\documentclass[
  reprint,
  superscriptaddress,
  amsmath,      
  amssymb,
  aps,
  prd,
  floatfix,
]{revtex4-2}

\usepackage[x11names,svgnames,table]{xcolor} 
\usepackage{graphicx}
\usepackage{array}
\usepackage{dcolumn}
\usepackage{float}
\usepackage{bm}            
\usepackage{booktabs}      
\usepackage{comment}

\usepackage{tikz}
\usetikzlibrary{shapes,positioning,arrows.meta,decorations.markings,calc,plotmarks}
\usepackage[compat=1.1.0]{tikz-feynman}
\tikzfeynmanset{warn luatex=false} 

\definecolor{CeruleanRef}{RGB}{12,120,180}
\usepackage[colorlinks=true,linkcolor=CeruleanRef,citecolor=CeruleanRef,urlcolor=black]{hyperref}
\usepackage{cleveref}
\usepackage{natbib}
\usepackage[utf8]{inputenc}   
\DeclareUnicodeCharacter{0301}{\'{}}  

\newcommand*{\SavedEqref}{}
\let\SavedEqref\eqref
\renewcommand*{\eqref}[1]{%
  \begingroup
    \hypersetup{linkcolor=CeruleanRef,linkbordercolor=CeruleanRef}%
    \SavedEqref{#1}%
  \endgroup
}

\definecolor{JeffersonBlue}{RGB}{35,45,75}
\definecolor{VirginiaOrange}{RGB}{248,76,30}
\definecolor{OxfordBlue}{rgb}{0,0.106,0.329}
\definecolor{umRed}{rgb}{0.73,0.09,0.19}
\definecolor{GoldDecorationlight}{RGB}{170,120,70}
\definecolor{ReddishBrown}{RGB}{153,0,76}
\definecolor{MainRed}{rgb}{.6,.1,.1}

\DeclareMathAlphabet{\mathbfit}{OML}{cmm}{b}{it}

\newcommand{\ReH}{\ensuremath{\mathrm{Re}\,\mathcal{H}}}
\newcommand{\ReE}{\ensuremath{\mathrm{Re}\,\mathcal{E}}}
\newcommand{\ReHt}{\ensuremath{\mathrm{Re}\,\widetilde{\mathcal{H}}}}
\newcommand{\xmi}{\ensuremath{\chi\mathrm{MI}}}

\begin{document}  

\preprint{APS/123-QED}

\title{Global Deep Neural Network Modeling of Compton Form Factors Constrained from Local $\chi^2$ Maps Fits} 

 \author{L. Calero Diaz}
\email{lc2fc@virginia.edu}
\email{liliet@lanl.gov}
\affiliation{Department of Physics, University of Virginia, Charlottesville, VA 22904, USA}
\affiliation{Los Alamos National Laboratory, Los Alamos, NM 87545, USA}

\author{D. Keller}
\email{dustin@virginia.edu}
\affiliation{Department of Physics, University of Virginia, Charlottesville, VA 22904, USA}

\date{\today}

\begin{abstract}

Over the past two decades, intense experimental efforts have focused on measuring observables that contribute to a three-dimensional description of the nucleon. Generalized Parton Distributions provide complementary insights into the internal structure and dynamics of hadrons, including information about the orbital angular momentum carried by quarks. The most direct process to access these distributions is Deeply Virtual Compton Scattering, in which the cross section can be expressed in terms of Compton Form Factors. These quantities are defined as convolutions of the Generalized Parton Distributions with coefficient functions derived in perturbative Quantum Chromodynamics.
We extract the Compton Form Factors from Deeply Virtual Compton Scattering data collected at Jefferson Lab, including the most recent measurements in Hall A, using a novel local fitting technique based on $\chi^2$ mapping to constrain the real parts of the Compton Form Factors $\mathcal{H}, \mathcal{E}$ and $\widetilde{\mathcal{H}}$. They are determined independently in each kinematic bin for the unpolarized beam-target configuration under the twist-2 approximation, following the formalism developed by Belitsky, Müller, and Kirchner. The extracted Compton Form Factors are then used to train and regularize a deep neural network, enabling a global determination of their behavior with minimal model dependence. This procedure is validated and systematically studied using pseudodata generated with kinematics matching those of the experimental measurements.

\end{abstract}

\maketitle

\section{\label{sec:introduction}Introduction}

Exploring the inner workings of strongly interacting systems through the analysis of quark and gluon structure is a cornerstone of particle physics. 
In the past few decades, comprehensive investigations of parton distribution functions (PDFs) have yielded detailed insights into the longitudinal momentum distribution of quarks and gluons, thereby offering a one-dimensional glimpse into the nature of hadrons. However, the comprehensive delineation of the multi-dimensional partonic structure of hadrons remains a paramount objective for ongoing experiments at esteemed facilities such as DESY, JLab, BNL, and CERN, as well as the prospective Electron-Ion Collider, where studying quantities revealing the transverse structure of hadrons is imperative.
Among those quantities are a new class of light-cone matrix elements, called Generalized Parton Distributions (GPDs) introduced in the 1990s \cite{Muller1994,Ji1_97,Ji2_97,Rad_96,Rad_97} and since then, they have been widely recognized as one of the key objects to explore the structure of hadrons. 
The GPDs seemingly unify different physical quantities, such as the PDFs and nucleon form factors (FFs), into the same framework. This makes them complicated multi-variable functions  with richer information than PDFs, encoding the largely unknown correlations between the transverse spatial structure of partons and their intrinsic longitudinal motion in the nucleon. This correlation provides a gateway to access the quark and gluon orbital angular momentum (OAM) and to elucidate the nucleon spin puzzle \cite{spinpuzzle}.  
In particular, they provide 3D number densities of quarks and gluons within the nucleon \cite{Burkardt,Diehl,Dupre}, and a connection to the matrix elements of the energy-momentum tensor, making it possible to evaluate the total angular momentum and “mechanical” properties of hadrons, like pressure and shear stress at a given point of space \cite{POLYAKOV1,POLYAKOV2}.

Experimental information on GPDs can be obtained from hard exclusive scattering processes such as deeply virtual Compton scattering (DVCS)~\cite{Ji2_97},  deeply virtual meson production (DVMP)~\cite{Rad_96}, time-like Compton scattering (TCS)~\cite{BergerTCS} and double deeply virtual Compton scattering (DDVCS)~\cite{Guidal_DDVCS}. All of them allow us to study the transitions of hadrons from one state to another, with a unique insight into changes taking place at the partonic level. At leading twist, the scattering is produced off a single parton of the nucleon with no other partons participating in the process and the nucleon structure is characterized by eight GPDs for each quark flavor: four GPDs conserving the helicity of the parton (chiral-even) and the other four flip the parton helicity (chiral-odd)~\cite{Diehl_GPDs,Hoodbhoy_GPDs}. There also exist gluon GPDs. Each GPD depends on 3 variables $x$, $\xi$ and $t$, where $x\pm \xi$ is the light-cone longitudinal momentum fraction of the struck quark before (“$+$” sign) and after (“$-$” sign) the scattering, resulting in a squared four-momentum transfer $t$ to the nucleon.
This gives rise to the division of the GPDs into two distinct regions, each carrying entirely different physical interpretations. An intuitive physical interpretation of the GPDs using light-cone coordinates for $x \in [\xi, 1]$ ($x \in [-1, -\xi]$) referred to as DGLAP~\cite{Dokshitzer:1977sg,Gribov:1972ri,Lipatov:1974qm,Altarelli:1977zs} region, can be attributed to the amplitude of hitting a quark (antiquark) in the nucleon with momentum fraction $x + \xi$ ($\xi - x$) and putting it back with a different momentum fraction $x - \xi$ ($-\xi - x$)  at a given transverse distance, relative to the transverse center of mass, in the nucleon. In the ERBL~\cite{Efremov:1979qk,Lepage:1980fj} region, $x \in [-\xi, \xi]$ the GPD resemble distribution amplitudes governing the emission or absorption of a parton-antiparton pair with squared momentum t.

GPDs are most directly accessible in the DVCS process where an incoming photon with high virtuality $Q^2$, emitted by a high-energy lepton beam, hits a parton of the nucleon which radiates a final real photon ($\gamma^*p\rightarrow \gamma p$). The possibility of studying GPDs in suitable exclusive scattering processes rests on factorization theorems, as does the usual extraction of parton densities from inclusive and semi-inclusive measurements. The proven QCD factorization theorems ~\cite{JCCollins_Fact,Ji3_fact} demonstrate that the DVCS amplitude can be factorized into a hard photon-quark Compton scattering calculable in perturbative QCD and a soft part which encodes the complex unknown non-perturbative dynamics of the quarks in the nucleon described, in the QCD leading order framework, by the four chiral-even GPDs ($H, E, \widetilde{H}, \widetilde{E}$),  in an expansion in inverse powers of $Q^2$ (twist-expansion) (Figure~\ref{Fig:1}).
This factorization has been shown to hold in the Bjorken limit for sufficiently large $Q^2>>M^2$ as well as a small net momentum transfer to the proton $|t|/Q^2 <<1$ at fixed $x_B = Q^2 /(2q \cdot P )$, where $q$ and $P$ are the virtual photon and initial proton four-momentum respectively. A consequence of the factorization is that GPDs are universal, as the differences between different processes are contained in the hard part. 
 At leading order, DVCS is not sensitive to the chiral-odd GPDs, but they can be measured in other exclusive processes such as pseudo-scalar meson production. 
 
\begin{figure}
\centering
\includegraphics[width = 1\linewidth]{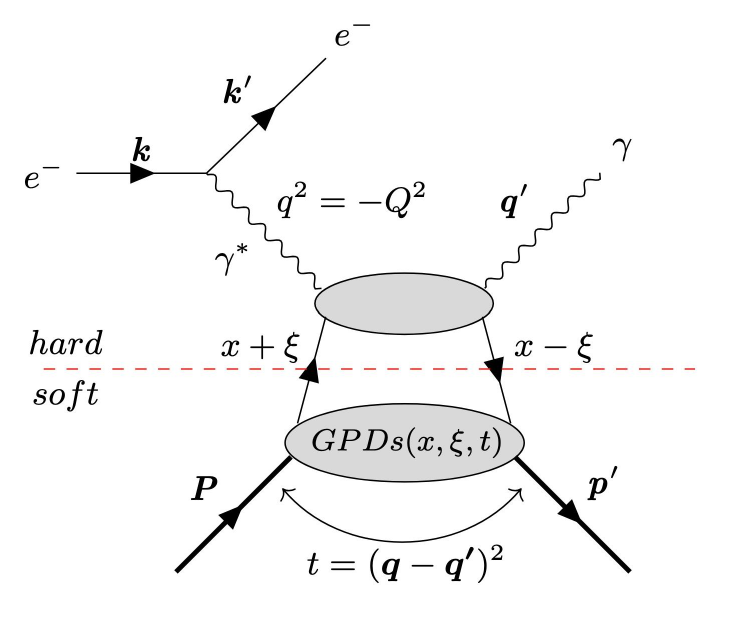}
\caption{Factorization of DVCS at leading twist and leading order.}
\label{fig:DVCS_factorization}
\end{figure}
Although DVCS is one of the cleanest channels to access GPDs, it requires high enough luminosities to measure small cross sections, and a large combination of experimental setups is needed in order to measure all the observables necessary to disentangle the contributions of the four GPDs. This includes different beam energies and beam charges, different kinematic coverages, DVCS on proton and neutron for quark GPDs flavor separation, and all possible unpolarized and polarized beam and target configurations. DVCS has been the focus of extensive research over the past decade by the DESY and JLab collaborations, with more recent studies conducted by the COMPASS experiment at CERN, which has now evolved into its successor, the AMBER experiment.
H1~\cite{Aktas2005,Aaron2009}  and ZEUS ~\cite{Chekanov2009} collider experiments at HERA measured the smallest $x_B \sim 10^{-4}$, and the largest $Q^2 \sim 100$ GeV$^2$ values where DVCS has been measured. Having access to both electron and positron beams, they published beam charge asymmetries as well as unpolarized cross sections at small $x_B$ regions dominated by sea quarks and gluons. As of today, HERMES ~\cite{Airapetian2001,Airapetian2007,Airapetian2008,Airapetian2009,Airapetian2010,Airapetian2011,Airapetian2012} experiment also at HERA, provided the most complete set of DVCS observables in $x_B$ between $0.04$ and $0.1$ and $Q^2$ up to $7$ GeV$^2$ using a fixed target which could be longitudinally or transversely polarized. 
At JLab ~\cite{Stepanyan2001,CLAS06_Pol,CLAS08_Asym,CLAS09_Asym,CLAS15_Asym,Camacho2006,JoCLAS:2015,Defurne2015,Defurne2017,Georges2022, HallAn}, the CLAS and Hall A collaborations also made a series of high statistics DVCS dedicated experiments using a fixed target. They worked at a larger value of $x_B$ of the order of $0.1$ to $0.7$ and $Q^2$ from $1$ to $10$ GeV$^2$, corresponding to the valence quarks region. In addition to unpolarized cross-sections, JLab has also provided asymmetry observables using a combination of longitudinally polarized beam and target. Among the available experimental data, only JLab has reported observables with high statistics and reasonably finely binned four-fold cross-sections and asymmetries. Hall A collaboration has also produced complementary neutron DVCS measurements that enable the separation of the $u$ and $d$ quark contributions to the leading GPDs~\cite{HallAn}.  DVCS cross-section measurements at $x_B = 0.056$ from the last runs taken by COMPASS \cite{Akhunzyanov2019} experiment at CERN with a fixed target, have been recently published.

Despite such a dedicated experimental program focused on DVCS significantly advanced worldwide, the intrinsic complexity of the GPDs needs more extensive and precise data to effectively constrain them in the multi-dimensional phase space of kinematic variables. Furthermore, one needs to cover this phase space with data collected for various processes and experimental setups, which is required to distinguish between many types of GPDs and contributions coming from various quark flavors and gluons. More data sensitive to GPDs will be delivered by the next generation of experiments. The upcoming DVCS experiment at COMPASS \cite{Gautheron2010}, which will employ a 160 GeV muon beam both positively and negatively charged, will collect data in a kinematic domain yet unexplored ($0.005 < x_B < 0.3$), between HERMES and the JLab experiments on the one hand, and the HERA collider experiments on the other hand, covering the region dominated by sea quarks.
The upgraded CLAS12 spectrometer \cite{Biselli2006} in Hall B at JLab, will allow for precise tuning of GPD parametrizations and improve the statistics of the previous CLAS measurements.  Approximately $85\%$ of the new data covers a phase space in the valence quark region that has never been probed with DVCS before. With the first CLAS12 measurements  on DVCS beam-spin asymmetries (BSA) published \cite{clas12}, measurements on unpolarized and polarised proton DVCS cross sections, as well as target-spin asymmetries (TSA) and double-spin asymmetries (DSA) with longitudinally and transversely polarized targets, are planned with a similar program for DVCS on the neutron. Additionally, new proposals are in preparation to utilize a future positron beam at JLab. 
Future electron-ion colliders like EIC\cite{EIC1,EIC2}, EicC~\cite{EicC}, and LHeC~\cite{LHeC}, are at the center of a lot of attention thanks to their promise of a high luminosity coverage over an extended region at relatively small $x_B$ and large $Q^2$ that will probe the gluon rich environment.

One major difficulty in the study of GPDs is that they appear in the DVCS amplitude as integrals over $x$ from $-1$ to $1$, as a consequence of the implied quark loop (Figure~\ref{Fig:1}). Since $x$ cannot be measured experimentally, the DVCS cross section is instead parametrized in terms of Compton Form Factors (CFFs) that are directly accessible. The factorization theorems allow us to express CFFs, as convolutions of GPDs with coefficient functions calculable at any order of perturbative QCD (pQCD). 
At leading order and leading twist, the CFF $\mathcal{H}$ associated to the GPD $H$ can be expressed as:
\begin{equation}
    \label{eq:1}
    \resizebox{0.78\hsize}{!}{%
    $
    \begin{split}
        \mathcal{H}(\xi,t) \equiv \sum_{q}\displaystyle\int_{-1}^{1} dxC_{0}^{q[-]}(x,\xi)H_q(x,\xi,t), \\     \end{split} 
    $
}
\end{equation}
summing over all the quarks flavors.
At leading twist, the coefficient function has no scale dependence and reads~\cite{BKM02}:
\begin{equation}
    \label{eq:2}
    \resizebox{0.72\hsize}{!}{%
    $
        C_{0}^{q[\mp]}(x, \xi) = e^2_q\Big(\frac{1}{\xi-x-i0} \mp \frac{1}{\xi + x - i0}\Big),\\ 
    $
}
\end{equation}
where $e_{q}^2$ is the charge of the quarks in the unit of proton charge. 
Using the residue theorem, it is possible to decompose the integral \eqref{eq:1} so that each CFF contains two real quantities,  for the CFF $\mathcal{H}$ for example:
\begin{equation}
    \label{eq:3}
    \resizebox{0.6\hsize}{!}{%
    $\mathcal{H}(\xi,t) = \Re e \mathcal{H}(\xi,t) + i\Im m \mathcal{H}(\xi,t) $
}
\end{equation}
where,
\begin{equation}
    \label{eq:4}
    \resizebox{0.9\hsize}{!}{%
    $\Re e \mathcal{H}(\xi, t) = \displaystyle\sum\limits_{q} \mathcal{P}\int_{0}^{1} dx C_q^- [H_q(x,\xi,t)-H_q(-x,\xi,t),  $
}
\end{equation}
\begin{equation}
    \label{eq:5}
    \resizebox{0.82\hsize}{!}{%
    $\Im m \mathcal{H}(\xi, t) = \displaystyle\pi\sum\limits_{q} e_q^2 \Big[H_q(\xi,\xi,t) - H_q(-\xi,\xi,t)\Big].   $
}
\end{equation}
$\mathcal{P}$ denotes the Cauchy principal value, the skewness $\xi$ is related to $x_B$ at leading twist as $\xi \approx x_B/(2 - x_B)$ and, 
\begin{equation}
    \label{eq:6}
    \resizebox{0.58\hsize}{!}{%
    $
        C_{q}^{[\mp]}(x, \xi) = e^2_q\Big(\frac{1}{\xi-x} \mp \frac{1}{\xi + x}\Big).\\ 
    $
}
\end{equation}
Similar expressions hold for the CFFs $\mathcal{E}, \mathcal{\widetilde{H}}, \mathcal{\widetilde{E}}$ associated with the GPDs $E, \widetilde{H}, \widetilde{E}$ respectively, where the top sign applies for the unpolarized GPDs ($H, E$) and the bottom sign is for the polarized GPDs ($\widetilde{H}, \widetilde{E}$). Equations \eqref{eq:4} and \eqref{eq:5} show that observables sensitive to the imaginary part of the CFFs will only contain information along the line $x = \pm \xi$, whereas the real part probes GPD integrals over the momentum fraction $x$. 
Given the complicated kinematic dependence of the GPDs, retrieving them from CFFs is therefore a major question of the field, known as the deconvolution problem~\cite{Deconv1}. Thus, the maximum model-independent information which can be extracted from the DVCS reaction at leading twist are 8 CFFs, which depend on two variables, $\xi$ and $t$, at QCD leading order. There is an additional $Q^2$ dependence in the CFFs (and in the GPDs) if QCD evolution is taken into account. 
Since the mapping from CFFs to GPDs is an ill-posed inverse problem, with uncertainties that amplify to the point of rendering direct inversion impractical~\cite{Deconv1,Deconv2}, theoretical inputs serve as essential and complementary constraints in determining GPDs, albeit at the cost of potential model-dependent bias.
Current parametrizations of GPDs suffer from model dependency of phenomenological GPD models, like KM~\cite{KM09,KM15,KM20}, GK~\cite{GK} and VGG~\cite{VGG}, which use similar Ans{\"a}tze with a rigid form and therefore cannot be considered as diverse sources for the estimation of model uncertainty.
Recently, efforts were conducted to produce GPD models using artificial neural network (ANN) techniques and fulfilling theory-driven constraints, keeping model dependency to a minimum and providing flexible parametrizations that give a better account of the systematic effects associated with the ill-defined extraction of GPDs from exclusive processes~\cite{ANNmodel}.
In addition to the experimental inputs, a new source of information regarding GPDs has emerged through lattice QCD simulations  where novel developments allow access to the nucleon structures from first principle calculation~\cite{LaMET1,LaMET2,LaMET3}.

\begin{figure*}[t]
\centering
\includegraphics[width = 1\linewidth]{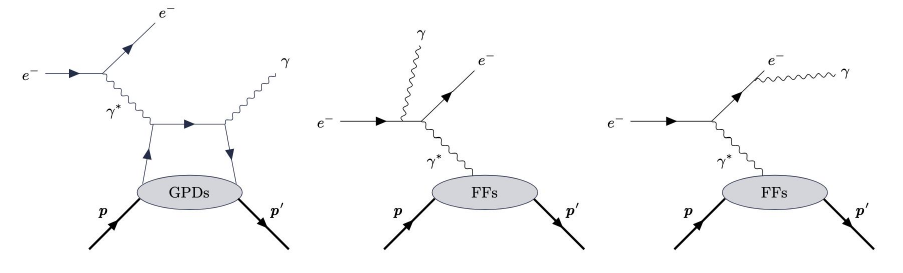}
\caption[Handbag diagram of the DVCS and the BH processes.]
{\label{fig:DVCS_BH}
Handbag diagram of the DVCS process at leading order and leading twist (left).
Bethe–Heitler diagrams where the real photon is emitted by the lepton either before (center) or after (right) the scattering off the nucleon.
}
\end{figure*}


A major challenge in extracting CFFs from exclusive measurements is the high dimensionality of the problem, as highlighted in the comprehensive review~\cite{Kumericki:review2016}.
Traditionally, the CFFs extraction technique falls into either a local fit (see Ref.~\cite{Guidal:fit08,Guidal-Mou:fitHermes09,Guidal:fitCLAS10,Guidal:Htilde10,BoerGuidal:fit14,Kumericki:fitHermes14,Moutarde:fit09,Dupre,GeorgesE12_2022}) or a global fit (see Ref.~\cite{Kumericki:DR3,KM09}) strategy. The analytical fit function is defined by the helicity amplitudes so the results can be specific to a particular DVCS formalism.
Global fits parameterize CFFs using experimental data across all kinematics, either at the CFF level or through GPD-based models like GK, VGG, and KM. However, these models introduce biases, particularly affecting orbital angular momentum extraction, which remains unexamined systematically. While global fits enable predictions in unmeasured kinematics, they come with model dependence. To reduce this bias, artificial neural networks (ANNs) have been developed to incorporate GPD properties without relying on specific parameterized models ~\cite{KMS11,Moutarde:ANN2019,KM20,GrigsbyKriesten:DNN}.
At present, the only known way to model-independently extract CFFs is locally in the kinematical points $\left(x_B, t, Q\right)$ at which the CFFs are fitted as free parameters on the experimental data. Therefore, this method introduces no bias on the general form of the CFF apart from the chosen framework of general approximation, like twist truncation, neglection, and dominance of some CFFs,... Local fits have however no ability to predict the result of measurements in previously unexplored kinematics with the shortcoming that GPDs can not be fitted themselves.

We focus our efforts on a novel local $\chi^2$ fit to extract the real Compton Form Factors (CFFs) \ReH, \ReE, and 
\ReHt\  using all available proton unpolarized DVCS data from JLab.
Since the unpolarized cross-section is the only physical observable used, it represents the least constrained scenario. Therefore, assessing the effectiveness of this extraction method will become increasingly valuable as more observables are incorporated. The extracted local results are then used to train a deep neural network (DNN) to achieve a model-independent global CFF fit, allowing for interpolation and extrapolation into kinematic regions where experimental data is unavailable.

\section{\label{sec:TheoryFramework}Theory Framework}
The DVCS process studies the virtual Compton scattering in deep inelastic
kinematics and at the leading order in perturbative QCD.  DVCS
is dominated by single-quark scattering, and therefore the amplitude can be expressed in
terms of the off-forward parton distributions. 
A virtual photon scatters from an unpolarized 
electron beam of energy $k$ off an unpolarized proton leading to the photon electroproduction cross section 4-fold differential cross-section:
\begin{eqnarray}
\label{eq:total_xs}
\frac{d^4\sigma}{d x_{Bj} d Q^2 d|t| d\phi} = \frac{\alpha^3 x_B y^2}{8 \pi Q^4 \sqrt{1 + \epsilon^2}} \frac{1}{e^6}
\big|\mathcal{T}\big|^2 \;.
\end{eqnarray}

The phase space of this process is parameterized by the Bjorken variable, $x_B=\frac{Q^2}{2pq}$, in terms of the momentum 
$q=k-k$ carried by the virtual photon of mass
$Q^2={-q}^2=-(k-k)^2$, the squared momentum 
transfer between the initial and final protons $t=\Delta^2$ with $\Delta=p'-p$ and the lepton energy loss $y=p·q/p·k$. The azimuthal angle $\phi$ between the leptonic and hadronic planes is defined in the Trento convention~\cite{Trento}. In the cross-section, $\alpha=e^2/(4\pi)$ is the fine structure constant, and $\epsilon=2x_BM/Q^2$ where $M$ is the proton mass.



The photon electroproduction cross-section is sensitive to the coherent interference of the DVCS amplitude with the Bethe-Heitler (BH) amplitude. The DVCS process (Figure~\ref{fig:DVCS_BH}, left), namely, the electroproduction of a photon off a proton with the exchange of a photon with virtuality, $Q^2$, much larger than the momentum transfer squared $t$, hypothetically provides one of the cleanest probes of the GPDs. The BH is an easily calculated electromagnetic process (Figure~\ref{fig:DVCS_BH}, center, right), where a real photon is emitted by the initial or the final lepton. Therefore both, DVCS and BH channels contain the same final states, which can not be distinguished experimentally, and the total photon electroproduction amplitude squared is given by the linear superposition of these two processes:
 \begin{equation}
    |\mathcal{T}|^2=\,|\mathcal{T}_{DVCS}|^2+|\mathcal{T}_{BH}|^2+\mathcal{I},
\end{equation}


The BH contribution is an undesirable contamination which was computed exactly from \cite{BKM02}. It is well known from QED calculations with great precision in terms of the proton electromagnetic form factors in the range of $-t < 0.4$ GeV$^2$ and will not be discussed in detail. The unpolarized BH amplitude is given by:
\begin{equation}
\resizebox{0.91\hsize}{!}{%
    $\big|\mathcal{T}_{BH}\big|^2 = \frac{e^6}{x^2_B y^2(1 + \epsilon^2)^2 t \mathcal{P}_1(\phi) \mathcal{P}_2(\phi)} \sum\limits_{n=0}^{2} c^{BH}_n \cos(n\phi)$ 
}\\%
\end{equation}

The harmonic terms $c^{BH}_n$ of the BH amplitude squared depend only upon bilinear combinations of the electromagnetic form factors $F_1(t)$ and $F_2(t)$,  which are computed from Kelly's parametrization ~\cite{Kelly:FFs}.  The factors $P_1(\phi)$ and $P_2(\phi)$ are the electron propagators in the BH amplitude.

The DVCS amplitude, and therefore the interference term  $\mathcal{I}$, encode the partonic structure of the nucleon given by the Compton tensor $T^{\mu\nu}$ ~\cite{Ji2_97} where the observables for the extraction of GPDs emerge. The proven QCD factorization theorem on DVCS \cite{Ji3_fact,JCCollins_Fact}, permits the application of the Compton tensor into perturbative coefficients and the long-distance dynamics which give rise to the GPDs and introduce the handbag diagrams twist expansion. At QCD leading twist and leading order approximation, there are four independent proton GPDs which can be accessed in the DVCS process: $\text{H}$, $\text{E}$, $\widetilde{\text{H}}$ and $\widetilde{\text{E}}$. These four GPDs depend on three variables: $x$, $\xi$ and $t$ where the latter two are experimental parameters with $\xi=x_B/(2-x_B)$ at leading order. The twist-2 CFFs are weighted integrals of these twist-2 GPDs over $x$ or combinations of GPDs at the line $x=\xi$.  The four  CFFs are:
\begin{equation}\label{eq:CFFs}
\begin{aligned}
  \mathcal{H}(\xi,t) &= \int_{-1}^{1} dx\,C^{q[-]}(x,\xi)\,H(x,\xi,t),\\[4pt]
  \mathcal{E}(\xi,t) &= \int_{-1}^{1} dx\,C^{q[-]}(x,\xi)\,E(x,\xi,t),\\[4pt]
  \widetilde{\mathcal{H}}(\xi,t) &= \int_{-1}^{1} dx\,C^{q[+]}(x,\xi)\,\widetilde{H}(x,\xi,t),\\[4pt]
  \widetilde{\mathcal{E}}(\xi,t) &= \int_{-1}^{1} dx\,C^{q[+]}(x,\xi)\,\widetilde{E}(x,\xi,t).
\end{aligned}
\end{equation}
where
\begin{equation}\label{eq:Ccoeff}
\begin{split}
  C^{q[\pm]}(x,\xi)
  &= -e_q^2\!\left(\frac{1}{x-\xi+i\varepsilon}\mp\frac{1}{x+\xi-i\varepsilon}\right).
\end{split}
\end{equation}
and $e_q$ is the charge of quarks in the unit of proton charge and summing over all quark flavors is implicit here.

This analysis is performed on unpolarized beam and unpolarized target cross-sections following the Belitsky-Kirchner-Muller DVCS formulation (BKM10)~\cite{BKM10} and we focus on the case where only twist-2 CFFs enter the cross-section.

At sufficiently large values of $Q^2$ and small values of $|t|$ and at leading order in the strong coupling constant $\alpha_S$, the squared DVCS amplitude is independent of the angle $\phi$ and a bilinear combination of the CFFs is contained in the coefficient $\mathcal{C}_{UU}^{DVCS}$ where $\mathcal{F} = \{\mathcal{H}, \mathcal{E}, \mathcal{\widetilde{H}}, \mathcal{\widetilde{E}}\}$ :

\begin{equation}\label{eq:dvcs2}
\bigl|\mathcal{T}_{UU}^{DVCS}\bigr|^2
= \frac{e^6}{y^2 Q^2}\,\frac{2-2y+y^2+\tfrac{\epsilon^2}{2}y^2}{1+\epsilon^2}\;
\mathcal{C}_{UU}^{DVCS}(\mathcal{F},\mathcal{F}^*).
\end{equation}

\par\noindent
\vspace{6pt}%

\begin{widetext}
\begin{equation}\label{eq:C_DVCS}
\begin{aligned}
\mathcal{C}_{UU}^{DVCS}(\mathcal{F},\mathcal{F}^*) =\;&
\frac{Q^2\bigl(Q^2 + x_B t\bigr)}{\bigl((2-x_B)Q^2 + x_B t\bigr)^2}
\Bigg\{ 
4(1-x_B)\,\mathcal{H}\mathcal{H}^*
+4\Bigl(1-x_B + \frac{2Q^2+t}{Q^2 + x_B t}\,\frac{\epsilon^2}{4}\Bigr)\widetilde{\mathcal{H}}\widetilde{\mathcal{H}}^*\\[6pt]
&\qquad
- \frac{x_B^2(Q^2+t)^2}{Q^2(Q^2 + x_B t)}\bigl(\mathcal{H}\mathcal{E}^* + \mathcal{E}\mathcal{H}^*\bigr)
- \frac{x_B^2 Q^2}{Q^2 + x_B t}\bigl(\widetilde{\mathcal{H}}\widetilde{\mathcal{E}}^* + \widetilde{\mathcal{E}}\widetilde{\mathcal{H}}^*\bigr)\\[6pt]
&\qquad
- \Bigg(\frac{x_B^2(Q^2+t)^2}{Q^2(Q^2+x_B t)} 
+ \frac{\bigl((2-x_B)Q^2 + x_B t\bigr)^2}{Q^2(Q^2 + x_B t)}\frac{t}{4M^2}\Bigg)\mathcal{E}\mathcal{E}^*
- \frac{x_B^2 Q^2}{Q^2 + x_B t}\frac{t}{4M^2}\,\widetilde{\mathcal{E}}\widetilde{\mathcal{E}}^*
\Bigg\}.
\end{aligned}
\end{equation}
\end{widetext}

\par\noindent
\vspace{6pt}%

The DVCS amplitude is contingent on the four twist-2 CFFs. These CFFs are complex-valued, consisting of two real magnitudes -- $\Re e\mathcal{F}$ and $\Im m \mathcal{F}$ -- each, leading to the emergence of eight distinct CFF parameters.  This intricate multiplicity of parameters underscores the complexity involved in their extraction process.

The unpolarized interference term is a linear combination of CFFs, with the following harmonic structure:
\begin{eqnarray}
\label{equ:Interference1}
\mathcal{I} = \frac{e^6}{x_By^3t{\mathcal{P}_1(\phi)\mathcal{P}_2(\phi)}} \sum\limits_{n=0}^{3} c^{\mathcal{I}}_n \cos(n\phi).
\end{eqnarray}

where the Fourier coefficients are:
\begin{eqnarray}
\label{equ:Interference2}
c^{\mathcal{I}}_n = C^n_{++}\Re e C^{\mathcal{I},n}_{++}(\mathcal{F})+C^n_{0+}\Re e C^{\mathcal{I},n}_{0+}(\mathcal{F}_{eff})\nonumber\\
+C^n_{-+}\Re e C^{\mathcal{I},n}_{-+}(\mathcal{F}_T)
\end{eqnarray}

  The double helicity-flip gluonic CFFs amplitudes, $C^{\mathcal{I},n}_{-+}(\mathcal{F}_{T})$,
  are formally suppressed by $\alpha_s$ and will be neglected here. $C^{\mathcal{I},n}_{++}(\mathcal{F})$ and  $C^{\mathcal{I},n}_{0+}(\mathcal{F}_{eff})$ are the helicity-conserving and the helicity-changing amplitudes respectively. For simplicity, similarly to \cite{Georges2022}, only the helicity-conserving amplitudes are considered which are given by:
\begin{eqnarray}
\label{equ:Interference3}
C^{\mathcal{I},n}_{++}(\mathcal{F}) = C^{\mathcal{I}}_{UU}(\mathcal{F})+\frac{C^{V,n}_{++}}{C^{n}_{++}}C^{\mathcal{I},V}_{UU}(\mathcal{F})\nonumber\\
+\frac{C^{A,n}_{++}}{C^{n}_{++}}C^{\mathcal{I},A}_{UU}(\mathcal{F}).
\end{eqnarray}

The complete expressions of the kinematic coefficients $C_{++}^n$, $C^{V,n}_{++}$and $C^{A,n}_{++}(n)$ are given in~\cite{BKM10}.  The $C^{\mathcal{I}}_{UU}$, $C^{\mathcal{I},V}_{UU}$ and $C^{\mathcal{I},A}_{UU}$ terms are a linear combination of the CFFs:

\begin{eqnarray}
\label{equ:Interference4}
C^{\mathcal{I}}_{UU}(\mathcal{F}) = F_1\mathcal{H}+\xi(F_1+F_2)\mathcal{\widetilde{H}}-\frac{t}{4M^2}F_2\mathcal{E}
\end{eqnarray}
\begin{eqnarray}
\label{equ:Interference5}
C^{\mathcal{I},V}_{UU}(\mathcal{F}) = \frac{x_B}{2-x_B+x_B\frac{t}{Q^2}}(F_1+F_2)(\mathcal{H}+\mathcal{E})
\end{eqnarray}
\begin{eqnarray}
\label{equ:Interference6}
C^{\mathcal{I},A}_{UU}(\mathcal{F}) = \frac{x_B}{2-x_B+x_B\frac{t}{Q^2}}(F_1+F_2)\mathcal{\widetilde{H}}
\end{eqnarray}

Under these conditions, the interference cross-section only depends on the CFFs $\Re e \mathcal{H}$, $\Re e \mathcal{E}$ and $\Re e \widetilde{\mathcal{H}}$.

\begin{figure}[ht!] 
  \centering
  \includegraphics[width = 0.92\linewidth]{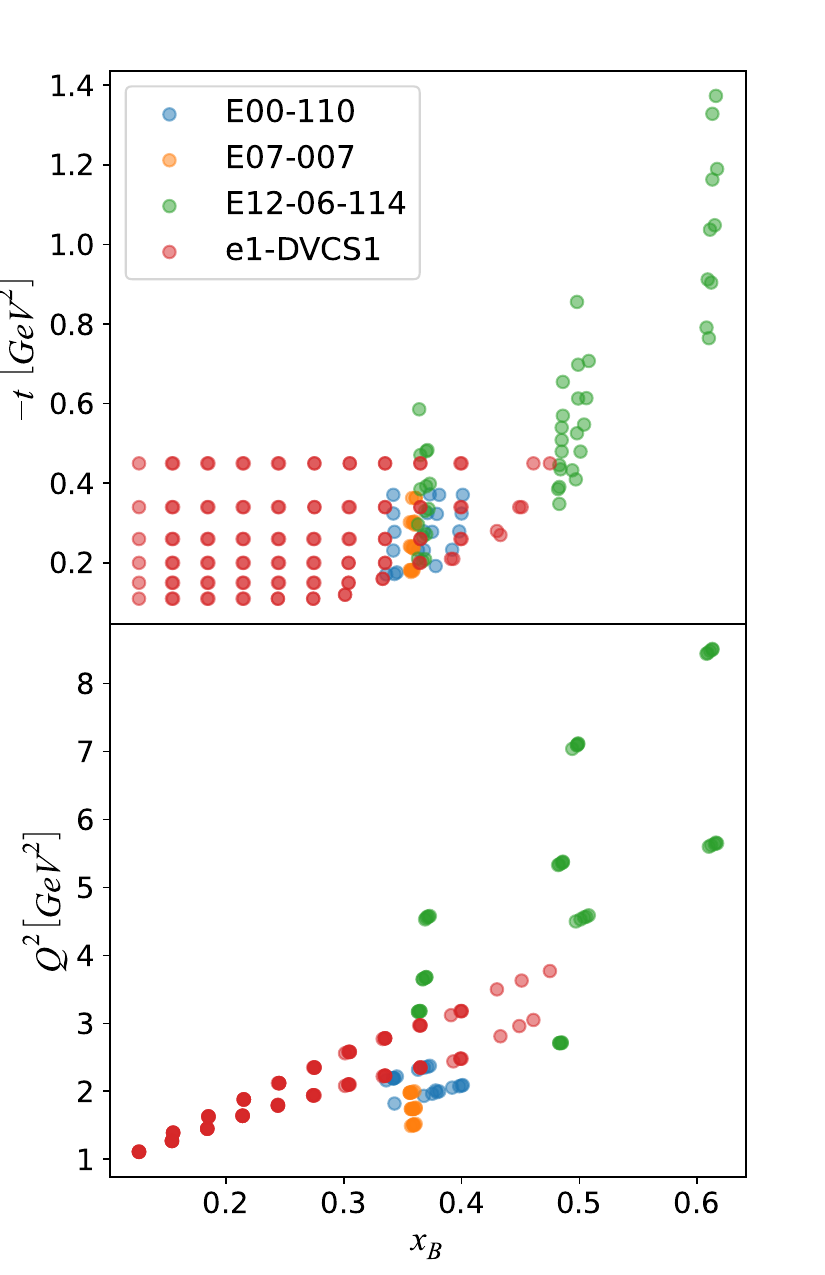}
  \caption[Kinematic region covered by the experimental data.]{Kinematical coverage of the experimental data listed in Table \ref{tab:data_summary} in $t$ vs $x_B$ (top) and  $Q^2$ vs $x_B$ (bottom) space.  }
  \label{fig:jlab_kinematics}
\end{figure}

\section{\label{sec:ExperimentalData}Experimental Data}

\begin{table*}
\caption{\label{tab:data_summary}Summary of the DVCS data from JLAB used in this analysis. The table includes information on the data source and the kinematic range covered by the helicity-independent cross-section.}
\begin{ruledtabular}
\begin{tabular}{lcccccc}
\multicolumn{1}{c}{Experiment} & Publication Year & $E_{beam}$ (GeV) & $Q^2$ (GeV$^2$) & $-t$ (GeV$^2$) & $x_B$ & Number of Points\\
\hline
Hall A E12-06-114 \cite{Georges2022} & 2022 & 4.487 - 10.992 & 2.71 - 8.51 & 0.204 - 1.373 & 0.363 - 0.617 & 1080\\
Hall A E07-007 \cite{Defurne2017} & 2017 & 3.355 - 5.55 & 1.49 - 2.00 & 0.177 - 0.363 & 0.356 - 0.361 & 404 \\
Hall A E00-110 \cite{Defurne2015} & 2015 & 5.75 & 1.82 - 2.37 & 0.171 - 0.372 & 0.336 - 0.401 & 468\\
Hall B e1-DVCS1 \cite{JoCLAS:2015} & 2015 & 5.75 & 1.11 - 3.77 & 0.11 - 0.45 & 0.126 - 0.475 & 1933\\ 
\end{tabular}
\end{ruledtabular}
\end{table*}

In this study, we analyze fixed target experimental DVCS data collected from Hall A~\cite{DefurneE00_2015,DefurneE07_2017,GeorgesE12_2022}  and Hall B~\cite{JoCLAS:2015}  at JLab. The datasets, detailed in Table~\ref{tab:data_summary}, comprises both helicity-independent and helicity-dependent cross-section measurements. However, for this specific analysis, the focus is solely on the least constrained helicity-independent cross-sections.
The data is finely binned into a fourfold differential cross-section format, characterized by the variables $Q^2$, $x_B$, t, and $\phi$. There are a total of  195 distinct kinematic conditions available for use in this experimental dataset defined by $Q^2$, $x_B$ and $t$. Figure~\ref{fig:jlab_kinematics} provides a visualization of the kinematic region covered by the utilized data. This data is a combination of results obtained from Jefferson Lab's 6 GeV and 12 GeV eras. Notably, the 12 GeV JLab data extends to higher values of $Q^2$, reaching up to 8.5 GeV$^2$. These measurements span moderate to high $x_B$ ranges, primarily probing the valence-quark region. The intermediate $Q^2$ values in these unpolarized fixed-target DVCS experiments significantly suppress higher-twist contributions, resulting in enhanced sensitivity to chiral-even GPDs.

\begin{figure}[h!] 
  \centering
  \includegraphics[width = 1\linewidth]{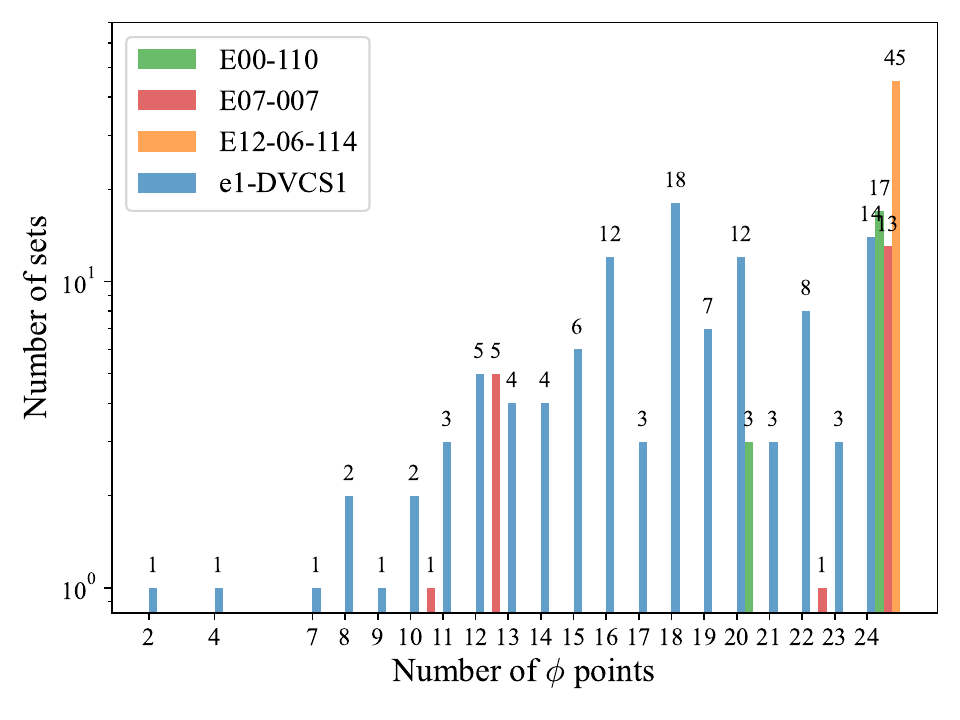}
  \caption[Number of $\phi$ points in the DVCS experimental data.]{Frequency of the number of $\phi$ points per kinematic sets of the DVCS data.}
  \label{fig:jlab_npoints}
\end{figure}

The number of $\phi$ points on each kinematic set of the experimental data is shown in Figure~\ref{fig:jlab_npoints}. The majority of Hall-A experiments have $24$ $\phi$ points per set, while the number of points in the Hall-B experiment is more variable. This accounts of a total of 1952 available data points from the HallA experiments and 1933 from the e1-DVCS1~\cite{JoCLAS:2015} experiment at Hall B, which operated at a fixed beam energy of 5.75 GeV and covered a $Q^2$ range from 1.0 to 4.6 GeV$^2$.

\section{\label{sec:Pseudo-data generation}Pseudo-data generation}

We employ simulated data, often referred to as pseudodata, of the total helicity-independent leptoproduction cross-section at twist-2, based on the theoretical framework detailed in Section~\ref{sec:TheoryFramework}. This pseudodata generation serves multiple purposes: to test and validate our extraction methods, to evaluate their reliability and robustness, and to systematically assess uncertainties and errors throughout the analysis pipeline, ultimately enhancing the reliability and accuracy of the extracted CFFs.

\subsection{\label{subsec:KM15*}GPDs PARAMETRIZATION}

\begin{figure*}[t!]
    \vspace{-2cm}
    \includegraphics[scale=0.65]{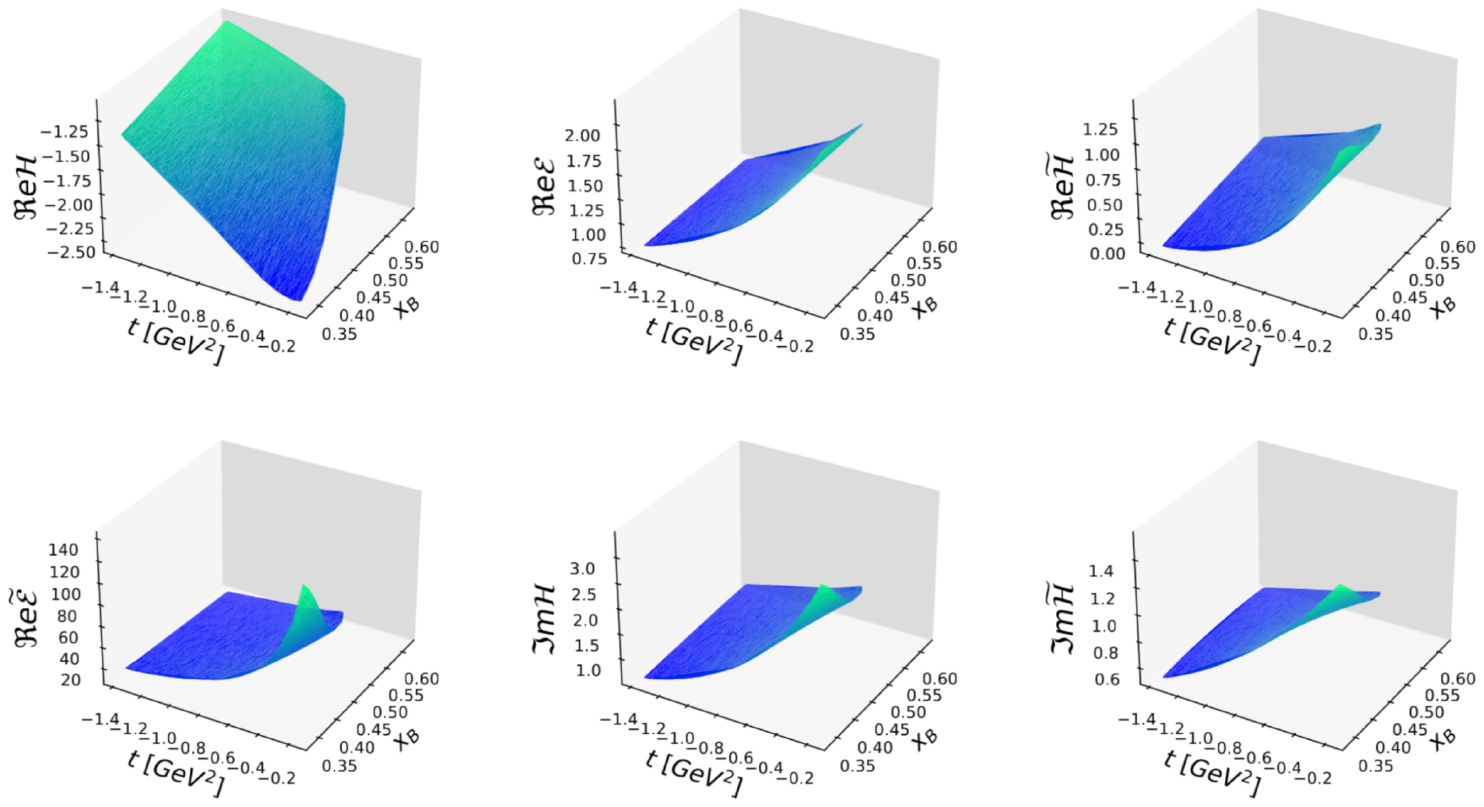}
    \vspace{-2.5cm}
    \caption[Parametrized $\mathrm{CFFs}$ from KM model]{$\mathrm{CFFs}$ obtained from KM model parametrizations in JLab kinematics range.}
    \label{fig:km15}
\end{figure*}

For a given kinematic set ($k, Q^2, x_B, t$), generating cross sections as a function of the azimuthal angle $\phi$
 requires the eight twist-2 CFFs entering the DVCS amplitude. The real parts of CFFs $\mathcal{H}$, $\mathcal{E}$, and $~\mathcal{\widetilde{H}}$ also appear in the interference term for the unpolarized target configuration.

To keep the problem realistic, given that GPDs have to fulfill a certain number of normalization constraints ~\cite{Diehl_GPDs,JI_GPDs,Guidal_2013}, the CFFs are generated by using a version of the GPD model originally developed by K. Kumericki and D. Mueller ~\cite{KM09}. 
In this approach, the valence and sea components are described by the simpler modeling of the GPD on the crossover line $x=\xi$ and the dispersion relation technique ~\cite{Teryaev:DR1,Diehl:DR2,Kumericki:DR3,Kumericki:DR4} is used to recover the remaining needed part. A similar modeling was used in the reference~\cite{KMS11} model dependent least-squares fit to the data. 
Overall, this approach allows to generate pseudodata that closely resemble the complexities and challenges of experimental data, where DVCS and BH processes are entangled, and GPDs must adhere to certain constraints. This synthetic data is then used for testing and validating our analysis techniques, as well as for systematic studies.

The adapted K. Kumericki and D. Mueller GPD model, which we designate as KM15$^*$, is used to generate the eight CFFs. The partonic decomposition of the imaginary component of $\mathfrak{I m} \mathcal{H}$ is:

\begin{equation}\label{eq:ImH}
\Im m \mathcal{H}\left(\xi, t\right)=\pi\left[\left(2 \frac{4}{9}+\frac{1}{9}\right) H^{\mathrm{val}}(\xi, \xi, t)+\frac{2}{9} H^{\mathrm{sea}}(\xi, \xi, t)\right].
\end{equation}

In the fixed target kinematics region and to LO accuracy, where the ${Q}^2$ virtuality is rather limited, the so-called scaling hypothesis is assumed i.e., on the assumption that the GPD does not evolve under the change of the photon virtuality. The functional form of the GPD $H$ for both, the sea and the valence contributions at the cross-over line, is motivated by a generic ansatz based on the double distribution (DD)~\cite{Muller1994,Radyushkin:DD1,Radyushkin:DD2} representation and a $t$-dependence inspired by a quark spectator model~\cite{Hwang:spectator}:

\begin{equation}
\label{eq:H_model}
H(\xi, \xi, t)=\frac{n r}{1+\xi}\left(\frac{2 \xi}{1+\xi}\right)^{-\alpha(t)}\left(\frac{1-\xi}{1+\xi}\right)^b \frac{1}{\left(1-\frac{1-\xi}{1+\xi} \frac{t}{M^2}\right)^p}
\end{equation}

Here $n$ is the residual normalization of PDF $q(x)=H(x, 0, 0)$ taken from PDF fits, $r$ is the skewness ratio at small $x$, i.e. the ratio of a GPD at some point on the cross-over trajectory and the corresponding $\mathrm{PDF}, \alpha(t)$ is the ``Regge trajectory'' borrowed from Regge phenomenology~\cite{collins_1977_Regge}, $b$ controls the large-$x$ behavior, and $M$ and $p$ control the $t$-dependence. 

A similar ansatz to equation~\eqref{eq:H_model} is used for $\widetilde{H}$ which introduces three additional parameters $\widetilde{M}, \widetilde{b}$ and $\widetilde{r}$:
\begin{equation}
\begin{split}
\Im \mathrm{m} \widetilde{\mathcal{H}}(\xi, t) &= \pi\left(2 \frac{4}{9}+\frac{1}{9}\right)  
\frac{\widetilde{n}^{\mathrm{val}} \hspace{1mm} \widetilde{r}^{\mathrm{val}}}{1+\xi}  
\left(\frac{2 \xi}{1+\xi}\right)^{-\alpha^{\mathrm{val}}(t)} \\
&\quad \times \frac{1}{\left(1-\frac{1-\xi}{1+\xi}  
\frac{t}{\left(\widetilde{M}^{\mathrm{val}}\right)^2}\right)^{\widetilde{p}^{\mathrm{val}}}}  
\left(\frac{1-\xi}{1+\xi}\right)^{\widetilde{b}^{\mathrm{val}}}
\end{split}
\end{equation}


 The valence quark parameters $n^{\mathrm{val}} = 1.35$, $\widetilde{n}^{\mathrm{val}} = 0.6$ and $p^{\mathrm{val}}=\widetilde{p}^{\mathrm{val}}=1$ are deduced from standard PDF parameterizations. For $\alpha^{val}(t)$ in the valence case, $\rho-\omega$ Regge trajectory is used,
\begin{equation}
    \alpha^{val}(t)=0.43+0.85 t / \mathrm{GeV}^2 .
\end{equation}

The remaining valence quark free parameters are taken from the state-of-the-art KM15~\cite{KM15} GPD parametrization version of the KM model that reproduces worldwide DVCS data.

The parameters of the sea-quark GPD $H^{\text {sea }}$ were taken to be as in~\cite{KM09} where the parameterization~\eqref{eq:H_model} is requested to reproduce their small $x_B$ fits  from H1/ZEUS data in the Mellin-Barnes representation~\cite{KM09,Mueller:Mellin} described within the parameters:
\begin{equation}
\begin{split}
  &\alpha^{\text {sea }}(t)=1.13+0.15 t / \mathrm{GeV}^2, n^{\text {sea }}=1.5, r^{\text {sea }}=1, \\
  &\quad b^{\text {sea }}=4.6, \left(M^{\text {sea }}\right)^2=0.5 \mathrm{GeV}^2, p^{\text {sea }}=2.  
\end{split}
\end{equation}
The CFFs are generally complex-valued, with their real and imaginary parts defined by equations \eqref{eq:4} and \eqref{eq:5}.
The GPD model provides the imaginary parts of the CFFs $\mathcal{H}$
and $~\mathcal{\widetilde{H}}$. By applying the Dispersion Relation (DR) \cite{Teryaev:DR1,Diehl:DR2,Kumericki:DR3,Kumericki:DR4}, we can then determine the real parts \ReH\ and \ReHt. 
The DR interconnects the $\xi$ dependencies of the real and imaginary parts of the CFFs, containing the same information up to a subtraction constant $\mathcal{C}_{\mathcal{F}}\left(t\right)$ independent of $\xi$ \eqref{DR}. This approach draws on a rich theoretical tradition of dispersion relations in hadron scattering, a method with roots tracing back to the 1950s, and is independent of perturbative QCD formalism.

\begin{equation}
    \begin{split}
    \label{DR}
         \Re e \mathcal{F}(\xi, t)&=\frac{1}{\pi} \mathcal{P} \int_0^1 \mathrm{~d} x\left(\frac{1}{\xi-x}\mp\frac{1}{\xi+x}\right) \Im m \mathcal{F}(x, t)\\
         &\quad+\mathcal{C}_{\mathcal{F}}\left(t\right).
    \end{split}
\end{equation}

The sign choice in the right-hand side of the DR equation \eqref{DR} corresponds to $\mathcal{F}=\{\mathcal{H}, \mathcal{E}\}$ for the minus sign and $\mathcal{F}=\{\mathcal{\widetilde{H}}, \mathcal{\widetilde{E}}\}$ for the plus sign.
The subtraction constant $\mathcal{C}_{\mathcal{F}}\left(t\right)$ is up to an opposite sign the same for $\mathcal{H}$ and $\mathcal{E}$, while it vanishes for the CFFs $\widetilde{\mathcal{H}}$ and $\widetilde{\mathcal{E}}$, and is perturbatively predicted to be zero for the combination $\mathcal{H}+\mathcal{E}$, see Ref.~\cite{Kumericki:DR3}. The non-vanishing subtraction constant is normalized by $C$ and $M_C$ which controls the $t$-dependence giving two additional free parameters. 
\begin{equation}
    \label{imh}
         \mathcal{C}_{\mathcal{H}}=-\mathcal{C}_{\mathcal{E}} = - \frac{C}{\left(1-\frac{t}{M_C^2}\right)^2} ; \quad \mathcal{C}_{\tilde{\mathcal{H}}}=\mathcal{C}_{\tilde{\mathcal{E}}}=0 .
\end{equation}

\begin{table}[t]
    $$
    \begin{array}{cccccccccc} 
        \toprule 
        M^{\text {val }} & r^{\text {val }} & b^{\text {val }} & C & M_C & \tilde{M}^{\text {val }} & \tilde{r}^{\text {val }} & \tilde{b}^{\text {val }} & r_\pi & M_\pi \\
        0.789 & 0.918 & 0.4 & 2.768 & 1.204 & 3.993 & 0.881 & 0.4 & 2.646 & 4  \\
        \midrule
    \end{array}
    $$
    \caption[GPDs model parameters of KM15 global fit]{Model parameters obtained by the global DVCS fit KM15~\cite{KM15}.} 
    \label{tab:km15pars}
\end{table}

In reference ~\cite{KM09}, the GPD $E$ is modeled solely in terms of the subtraction constant, $\mathcal{E}=\mathcal{C}_{\mathcal{E}}$, i.e., the $E$ contribution is only through the D-term~\cite{Polyakov:D-term} and therefore vanishes at $x = \xi$, resulting in $\Im m \mathcal{E} = 0$. The contribution of the GPD $\tilde{E}$ is described using the pion-pole~\cite{Penttinen:1999th} inspired effective ansatz and thus the amplitude in this channel is also purely real.
\begin{equation}\label{ReEt}
    \begin{aligned}
    & \Re e \widetilde{\mathcal{E}}(\xi, t)=\frac{r_\pi}{\xi} \frac{2.164}{\left(0.0196-\frac{t}{\mathrm{GeV}^2}\right)\left(1-\frac{t}{M_\pi^2}\right)^2}\hspace{1mm}, \\
    & \Im m \widetilde{\mathcal{E}}(\xi, t)=0,
\end{aligned}
\end{equation}
where $m_\pi^2=0.0196 \mathrm{GeV}^2$, while $M_\pi$ and $r_\pi$ are free parameters.

A summary of the GPD model free parameters taken from the global DVCS fit KM15 is shown in Table~\ref{tab:km15pars}.
For the rest of this work, we will refer to the described GPD parametrization in this section as KM$15^*$ model unless otherwise specified. Note that by employing the DR instead of directly modeling for example, $H(x, \xi, t)$, one can model the simpler functions $H(x, x, t)$ and $\mathcal{C}_{\mathcal{H}}(t)$, in a LO and leading-twist approximation, ignoring the effects of GPD evolution. These are all acceptable approximations when trying to describe presently available data in fixed-target kinematics where the $Q^2$ is relatively limited. However the assumption that the GPDs do not evolve when the photon virtuality changes, will be further scrutinized in this work.

The resulting model parametrization of the $\mathrm{GPDs}$, achieves a relatively good description of all helicity-independent JLab DVCS available data (Table~\ref{tab:data_summary}) with an overall normalized $\chi^2$ of $\approx 1.43$ for 195 data points. In Figure~\ref{fig:km15}, the obtained $\mathrm{CFFs}$ with the described KM$15^*$ model are shown as a function of $x_B$ and $t$ in the kinematic range of the HallA experiment.

\subsection{\label{subsec:pseudo_xs}PSEUDODATA CROSS-SECTION}

The unpolarized cross-section of the $e p \rightarrow e p \gamma$ process is 
generated as a function of the azimuthal angle $\phi$ using the same beam energy, and $\left(x_B, Q^2, t\right)$ kinematic bins as the experimental data described in Section \ref{sec:ExperimentalData}.
For each kinematic bin, the number of generated $\phi$ points matches the experimental data to properly account for statistical effects in the available measurements.

The calculation of the DVCS+BH amplitudes follows the BKM10 formulation described in Section \ref{sec:TheoryFramework}. The eight twist-2 CFFs entering in the cross-section are computed at each $x_B$ and $t$ values from the GPD model parametrization described in Section \ref{subsec:KM15*}. This parametrization is based on the established KM models, which satisfy most of the model-independent GPD normalization constraints and accurately represent the patterns observed in the experimental DVCS data used in this analysis.

\begin{figure}[t] 
  \centering
  \includegraphics[width = 1\linewidth]{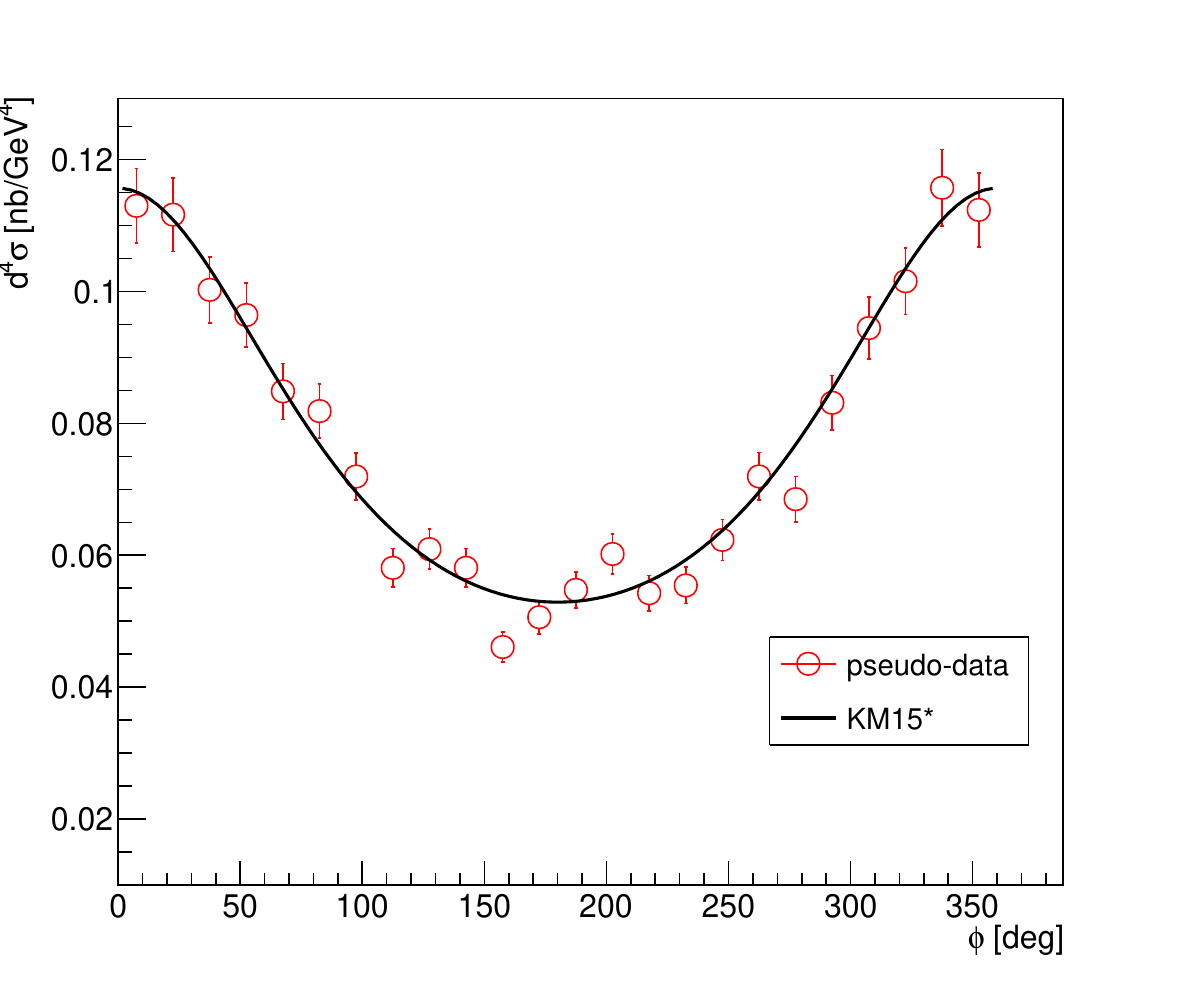}
  \caption[Generated pseudodata.]{Unpolarized generated pseudodata cross-section for the kinematics $\left(x_B, Q^2, t\right) = \left(0.34, 1.82 GeV^2, -0.17 GeV^2\right)$ corresponding to one set measured by the E00-100 HallA experiment. The solid line corresponds to the generating function using the KM15* model parametrization.}
  \label{fig:pseudo_example}
\end{figure}

An example of the generated pseudodata is shown on Figure~\ref{fig:pseudo_example} for the particular kinematics $(x_B, Q^2, t)=(0.34, 1.82$ $\mathrm{GeV}^2, -0.17$ $\mathrm{GeV}^2)$ and $5.75$ $\mathrm{GeV}$ beam energy.
This corresponds to a kinematic bin measured by the Hall-A E00-110 experiment. The $24$ generated points, corresponding to the $\phi$ binning of the experimental data, are superimposed on the theoretical curves. 
The uncertainties and the accessible $\phi$ regions vary for each $\left(x_B, Q^2, t\right)$ bin, and differ for the Hall-A and Hall-B experiments. The error bars added on the pseudodata correspond to the published experimental uncertainties of the JLab data and they average to about 5 $\%$ for the unpolarized DVCS cross-section.
The smearing of each $\phi$ point of the pseudodata has been done via a Gaussian distribution, centered at the theoretically computed value, with a standard deviation corresponding to the experimental uncertainties. 
The smeared pseudodata replicate the real-world experimental conditions that could influence the measured quantities.
It facilitates the assessment of how sensitive extraction algorithms are to these experimental conditions, offering valuable insights into optimizing the experiment for maximal CFF sensitivity.

Under these conditions, we deem that in the following we will perform the CFFs extraction in rather realistic conditions, taking into account the $\phi$-coverage of the data, their dispersion, and their uncertainties. By comparing the results from the pseudodata to the known input (theoretical predictions or actual CFFs), the accuracy and reliability of the analysis procedures can be accessed. This is essential for ensuring the robustness of the extraction process and enhancing the depth of understanding regarding systematic errors inherent in both the experimental procedures and analysis methodologies.

\section{\label{sec:CFFs_extraction} CFFs EXTRACTION}

Any attempt to extract GPDs from DVCS experimental data begins with the extraction of the corresponding CFFs. Over the past decade, the complete and precise determination of CFFs has been a major focus in the study of GPD-related processes, both theoretically and phenomenologically. A key challenge in extracting CFFs from exclusive measurements arises from the high dimensionality of the problem: multiple CFFs contribute simultaneously to DVCS cross-sections and asymmetries, making the individual determination of each CFF particularly difficult. This leads to a degeneracy problem, where different combinations of CFF values can produce nearly identical experimental signatures, resulting in inherent ambiguity in the interpretation of measurements.

\begin{figure}[t]
\includegraphics[scale=0.41]{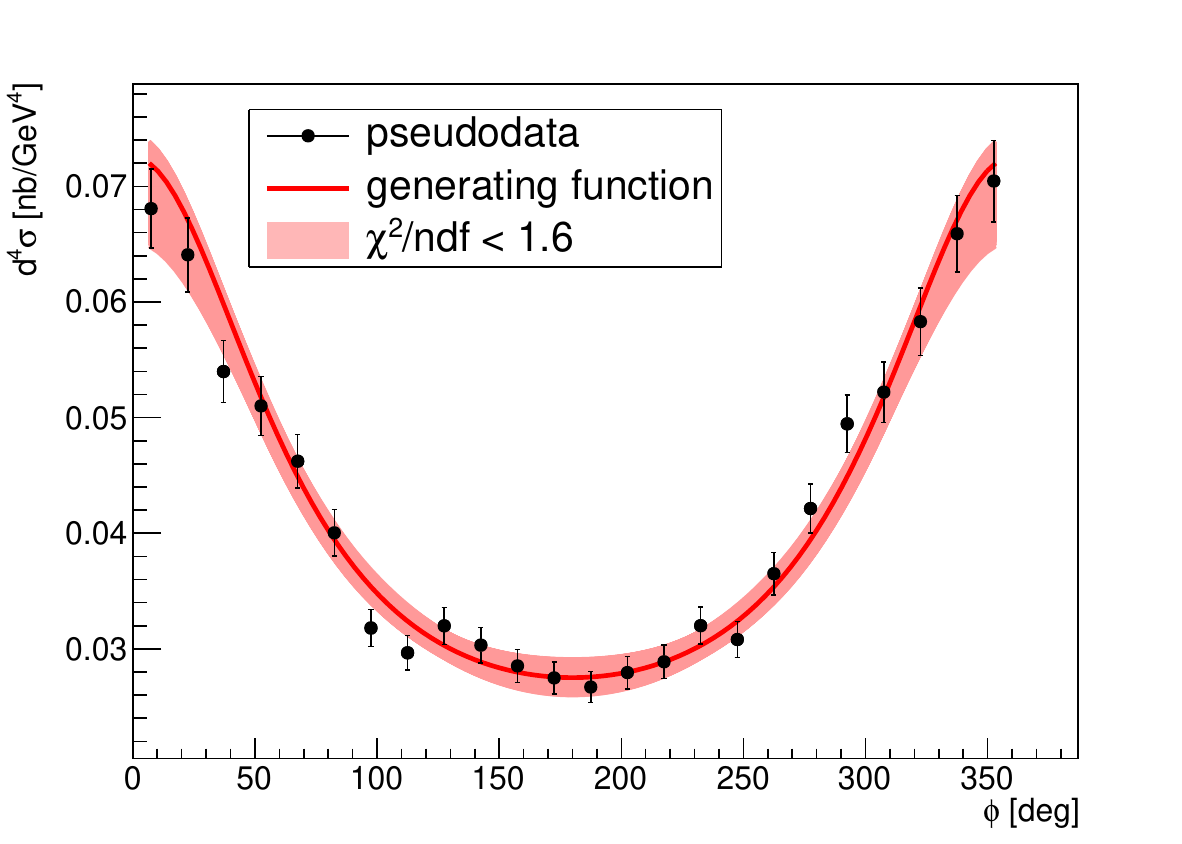}
\caption[Fits of one kinematic bin of the pseudodata cross-section with $\chi^2/ndf$ less than 1.6 and the corresponding range of CFF values that generated them.]{Fits of the pseudodata cross-section with $\chi^2/ndf$ values less than 1.6 (top) and the corresponding range of \ReH, \ReH\ and \ReHt\ that generated them (bottom) at the kinematic bin \mbox{k = 5.75 GeV}, \mbox{$Q^2 = 1.96$ GeV$^2$}, \mbox{$x_B = 0.37$}, \mbox{$t = -0.23$ GeV$^2$}. The true values generating the cross-section distribution are represented by solid lines.}
\label{fig:uncert}
\end{figure}

This behavior is illustrated in Figure~\ref{fig:uncert}, which is based on generated pseudodata. The generating function of the pseudodata cross-section (red line) is computed using the true values of the CFFs: \ReH\ (orange), \ReE\ (red) and \ReHt\ (green), shown as solid lines in the bottom panel. The red band in the top panel represents all fits with normalized $\chi^2 < 1.6$, yet the corresponding CFFs for these fits, shown in the bottom panel, span a wide range of values. 
This spread is particularly large for \ReE, whereas \ReH\ is more tightly constrained. Consequently, achieving a good fit to the cross-section does not necessarily guarantee an accurate extraction of the individual CFF values.

We choose to adopt an approach of extracting the CFFs by exploiting the $\chi^2$ distribution of randomly generated CFFs. As a matter of priority, this study is confined to the goal of locally extracting twist-2 CFFs directly from the least constraining and most challenging DVCS unpolarized observable in a model-independent way.
Additional observables can only improve the situation, and their inclusion is reserved for future study. We exploit the fact that at leading order and leading twist approximations the DVCS amplitude lacks dependence on the azimuthal angle, see (\ref{eq:dvcs2}). Therefore, the DVCS cross-section is taken as a free parameter in the fit function and the only contribution of CFFs comes from the Interference term which depends on \ReH, \ReE\ and \ReHt\ (\ref{equ:Interference4})-(\ref{equ:Interference6}). This way we reduce the  fitting problem from eight to four parameters.  
From the perspective that the real CFFs at leading twist are relevant to hadron tomography and helpful in the interpretation of the phenomenology of the DVCS process, we focus on an extraction schema that prioritizes both high accuracy and precision with minimal biases.  

Generically, no information can be reliably extracted from any CFFs unless several different observables measured at the same kinematics are studied simultaneously. Despite the limitations and approximation of this extraction method, having an extraction procedure capable of effectively constraining three of the CFFs out of one observable is a valuable contribution. In fact, it was observed in the pioneering local least squares minimization extraction~\cite{Guidal:fit08} at LO and LT that fitted two observables, unpolarized and beam-polarized Hall-A cross sections, resulted in a convergence of the fits for \ReH\ and $\Im m \mathcal{H}$ only while the other CFFs are left undetermined.
In this novel approach, we will show that the CFFs \ReH, \ReE, \ReHt, and the pure DVCS cross-section can, in fact, be constrained with a finite uncertainty with some insight into its $t$ and $x_B$-dependence, at Hall-A and at Hall-B kinematics.

The available data from JLab used in this analysis, Table ~\ref{tab:data_summary}, is limited to the valence region. This gives much more constraint on the CFF $\mathcal{H}$ than the other CFFs, as recognized by GPD models~\cite{GK,VGG,KM09}, while the CFF $\mathcal{E}$ suffers from scarcity and sizable statistical uncertainties, see Figure \ref{fig:uncert}. 
Kinematical twist-four and quantum loop corrections~\cite{BelitskyRad:2005,Ji3_fact} that are $\alpha_S$ suppressed, like the target mass and finite $t$ corrections~\cite{BMMP}, are omitted. Those corrections are related to leading-twist GPDs but involve different perturbative coefficients that get convoluted~\cite{BKM02}. Therefore, they lead to a different set of CFFs that can be considered independent of the twist-2 CFFs discussed here.

\begin{figure}[t]
\centering
\includegraphics[width=0.76\linewidth]{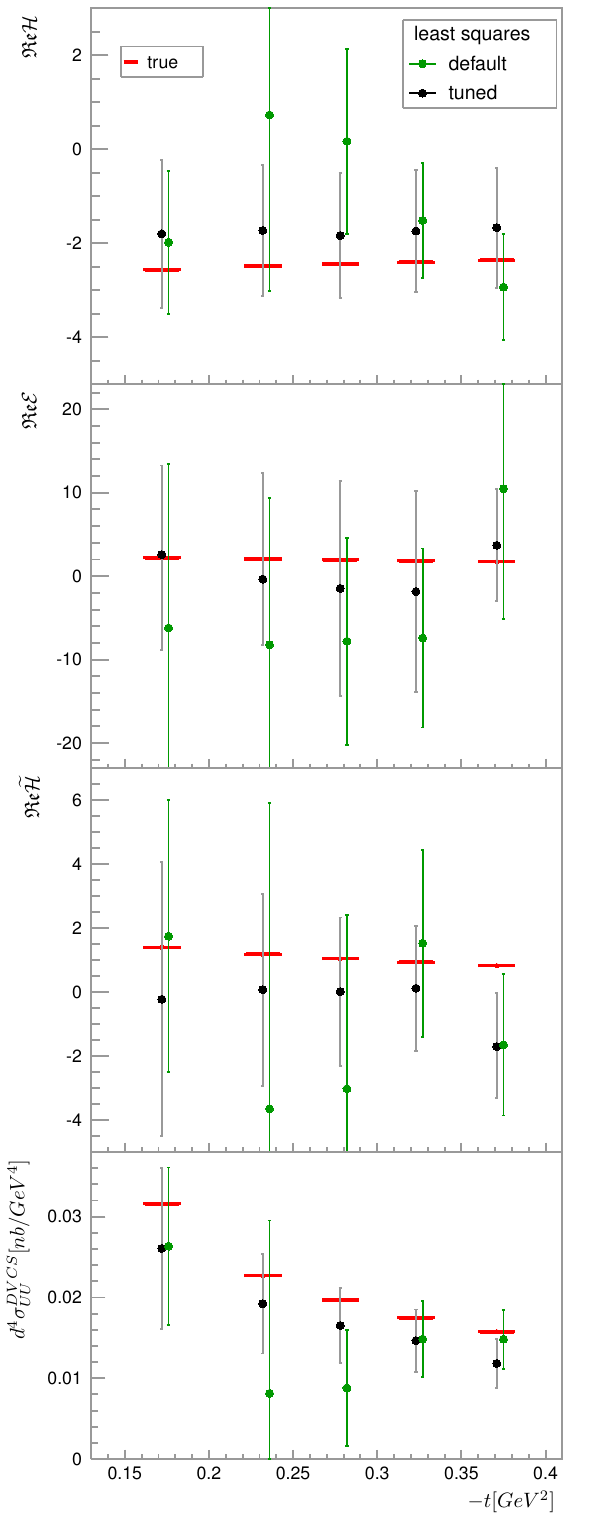}
\caption[Least squares fit optimization.]{Least squares extraction with MINUIT default parameters (green) and with the optimized algorithm parameters (black). The pseudodata from 5 kinematic points correspond to Hall-A experimental data with average \mbox{$Q^2 = 1.94$ $GeV^2$} and \mbox{$x_B=0.37$}. The parameters used to generate the pseudodata are shown in red, centered at the data $t$ value with an arbitrary bin width. The green points have been slightly shifted in $t$ for clarity on the visualization.}  
\label{fig:fit_opt_defl}
\end{figure}

\subsection{\label{subsec:least-squares fit}LEAST-SQUARES FIT}
For comparison, we perform an optimized local least-squares fit applied to the generated pseudodata. In this procedure, the best fit parameters are determined at each kinematic set by minimizing, via the least-squares method, the $\chi^2$ defined as:
\begin{equation}\label{eq:chi2}
\chi^2 = \sum_{i=1}^{n} \frac{(\sigma_i^{\text{fit}}-\sigma_i^{\text{data}})^2}{(\delta\sigma_i^{\text{data}})^2},
\end{equation}
where the fit function $\sigma^{\text{fit}}$ is the unpolarized photon electroproduction cross section described in Section~\ref{sec:TheoryFramework}. It consists of the BH term, the pure DVCS cross section (treated as a free parameter due to its constant behavior in $\phi$), and the interference term, which depends on three free parameters: \ReH, \ReE, and \ReHt. The quantities $\sigma^{\text{data}}$ and $\delta\sigma^{\text{data}}$ denote, respectively, the values and uncertainties of the pseudo- or experimental data. The index $i$ runs over all available $\phi$ points for a given $\left(x_B, Q^2, t\right)$ bin.
In the interest of comparison, the fit parameters are limited to the same range used to generate the $\chi^2$ maps described on the next section. 

We use the well-known MINUIT2 code from CERN~\cite{minuit2} with the MIGRAD minimization algorithm, which is generally considered the most robust minimizer for generic functions. Several internal parameters can be adjusted to improve the quality of the fit. A systematic pseudodata study was performed to optimize these parameters, resulting in extracted CFFs and DVCS cross sections that more closely reproduce the true values used in the pseudodata generation.
The two most relevant parameters tuned were the \verb|Tolerance|, which controls the convergence criterion of the minimization, and the \verb|Strategy| parameter, which can take the integer values 0, 1, or 2. Figure~\ref{fig:fit_opt_defl} shows the extracted values for five different kinematic sets using the default MINUIT2 configuration and the optimized settings which are both listed in Table~\ref{tab:fit_tuned_pars}. The deviation from the true values (red line) is significantly reduced after optimization.
The most accurate results were obtained when a looser \verb|Tolerance| of 0.01 was used together with the \verb|Strategy| set to 0, which is particularly effective in cases where many parameters are correlated. This optimization yields a notable improvement in the extraction of \ReE.
Point estimates from the fit should fluctuate statistically about the true values. The persistent displacement observed in the optimized solutions indicates a configuration-dependent systematic bias. A broad scan over minimizer settings (algorithm, tolerances, seeds, parameter bounds/penalties) shows that the configurations yielding the best nominal performance for \ReH\ and the DVCS term still exhibit the bias shown. Rather than applying an ad-hoc correction, the corresponding residual between fit and truth is assigned as a method-dependent systematic uncertainty of the least-squares estimator.


\begin{table}[]
    $$
    \begin{array}{ccc} 
        \toprule 
         & \verb|Strategy| & \verb|Tolerance| \\
         \hline\hline\addlinespace[0.5mm]
         \text{default value} & 1 & 0.01 \\
         \text{best fit} & 0 & 10 \\
        \midrule
    \end{array}
    $$
    \caption[MIGRAD minimization parameters]{MIGRAD minimization parameters.} 
    \label{tab:fit_tuned_pars}
\end{table}

We find that \ReHt\ and \ReE\ are more difficult to constrain compared to \ReH\ and the DVCS cross section. Nevertheless, the true values are generally encompassed within the estimated uncertainties.
After determining the optimal minimization parameters, the errors were estimated using the MINOS algorithm. This method is particularly well suited for non-linear problems or for cases where the $\chi^2$ profile deviates from a simple parabolic shape, as in our study. For proper error calculation, MINOS follows the function outward from the minimum to identify where it crosses $\chi^2_{\min}+1$, rather than relying on the curvature at the minimum and assuming a parabolic form. Consequently, the uncertainty on a given parameter corresponds to the parameter value at $\Delta \chi^2=+1$ above $\chi^2_{\min}$. In practice, MINOS evaluates $\chi^2$ at multiple points in the multi-dimensional parameter space and retains a smaller function value if encountered, thereby reducing the risk of trapping in local minima.
Although MINOS errors are computationally expensive, they are highly reliable since they account for non-linearities and parameter correlations. However, in some kinematic sets MINOS failed for certain parameters due to intrinsic instabilities in the fit, such as strong parameter correlations.
In such cases, the HESSE error matrix was used instead. This method evaluates the full second-derivative matrix of $\chi^2$ at the minimum (via finite differences) and inverts it:
\begin{equation}\label{eq:Hessian_matrix}
H_{i j}=\left.\frac{1}{2} \frac{\partial^2}{\partial \alpha_i \partial \alpha_j} \chi^2\right|_{\alpha=\alpha^0}.
\end{equation}
The corresponding uncertainty can then be propagated to any function $\mathcal{F}$ depending on the fit parameters via
\begin{equation}
\sigma_{\mathcal{F}}=T\left(\sum_{i, j=1}^n
\frac{\partial \mathcal{F}}{\partial \alpha_i} ,
H^{-1}_{ij} ,
\frac{\partial \mathcal{F}}{\partial \alpha_j}\right)^{1/2},
\end{equation}
where $T=\sqrt{\Delta \chi^2}$ is the tolerance factor. Setting $T=1$ yields the standard $68\%$ confidence interval.
The improved uncertainty estimates obtained from this procedure are shown in Figure~\ref{fig:fit_opt_defl} for the tuned fit.

\subsection{$\chi^2$ MAPS INFERENCE ($\chi$MI) }\label{sec:xMI_method}

\begin{figure*}[t] 
\centering
\includegraphics[width=1\linewidth]{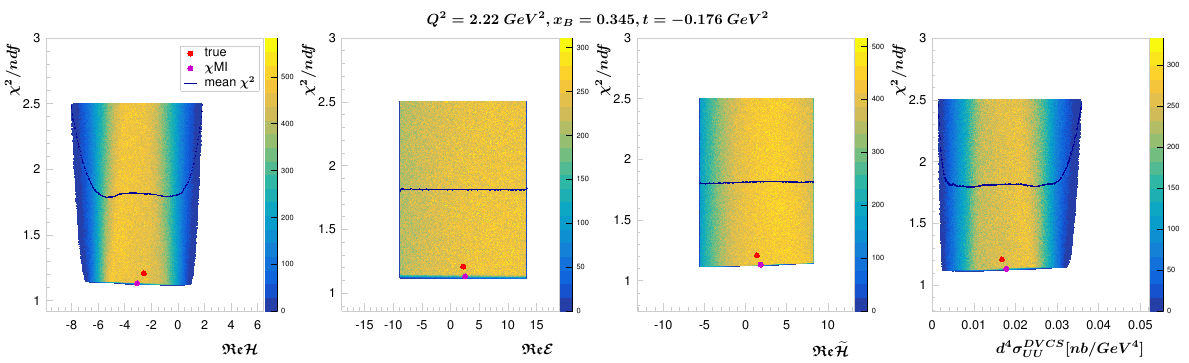}
\includegraphics[width=1\linewidth]{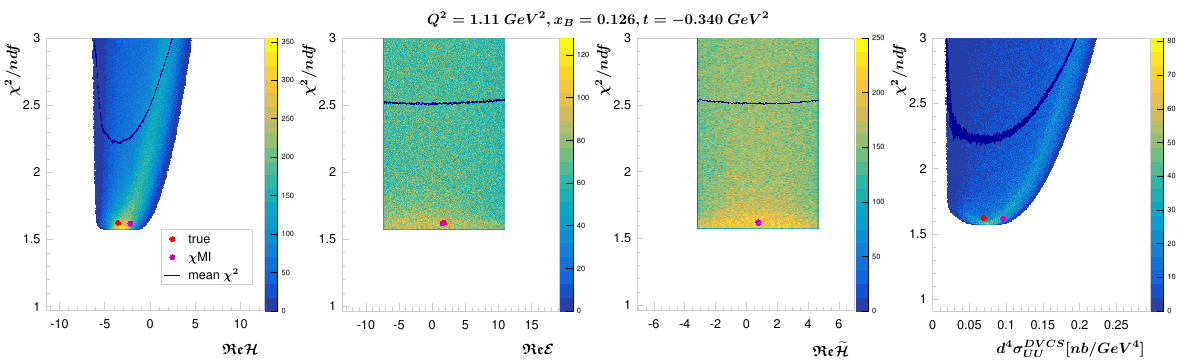}
\caption[Pseudodata $\chi^2$ maps at a fixed kinematic bin of Hall-A and Hall-B data.]{Pseudodata $\chi^2$ maps at fixed kinematic bins corresponding to Hall-A (top) and Hall-B (bottom) experiments.\label{fig:gen_maps_ex}}
\end{figure*}

DVCS observables receive contributions from several CFFs, which are strongly correlated, making their extraction an underconstrained problem. While it is often straightforward to obtain a good fit to experimental data, many different combinations of the real and imaginary parts of the CFFs can reproduce the measurements equally well, as illustrated in Figure~\ref{fig:uncert}. Traditional minimization approaches are often insufficient in this context, since they rely on local curvature information near the $\chi^2$ minimum and can fail to capture the full parameter correlations and degeneracies.

To address these challenges, we introduce a novel extraction method based on $\chi^2$ maps, referred to as $\chi^2$ Maps Inference ($\chi$MI), which allows us to effectively constrain three CFFs and the DVCS cross section in the helicity-independent case. The approach consists of evaluating $\chi^2$ at multiple points in the multi-dimensional hyperspace of the free parameters by generating Monte Carlo replicas of the eight CFFs. In this way, the full parameter space is sampled, and $\chi^2$ maps, i.e. the number density distributions of $\chi^2$ as a function of each parameter, are constructed. Based on the topology of these maps, additional selection criteria are applied. These selections are tuned with pseudodata in order to further improve the accuracy and precision of the extraction. The detailed procedure for generating the $\chi^2$ maps and the applied cuts are described in the following sections.

The best-extracted values of the three CFFs and the DVCS cross section are obtained as weighted averages over the $\chi^2$ confidence level (CL) distributions. For example, the extracted \ReH\ value at a fixed kinematic point is given by:
\begin{equation}\label{eq:wa}
  \overline{\Re e \mathcal{H}} = \frac{\sum\limits_{n=1}^{N} \Re e \mathcal{H}_n \cdot D_n(\chi^2_n, r)}{\sum\limits_{n=1}^{N} D_n(\chi^2_n, r)},
\end{equation}
and analogously for the other parameters. Here $n$ runs over all randomly generated CFF sets, and $D(\chi^2, r)$ denotes the complement of the cumulative distribution function of the $\chi^2$ distribution with $r$ degrees of freedom (also known in statistics as the survival function). Explicitly,
\begin{equation}\label{eq:D}
D(\chi^2,r) = \int_{\chi^2}^{+\infty} \frac{1}{\Gamma(r/2),2^{r/2}} , x'^{r/2-1} e^{-x'/2},dx',
\end{equation}
where $\Gamma$ is the incomplete gamma function. This corresponds to the upper-tail probability of the $\chi^2$ distribution between $\chi^2$ and $+\infty$, i.e., the probability of observing a $\chi^2$ value larger than the one obtained. A small value of $D(\chi^2, r)$ indicates that the corresponding set of parameters is unlikely to provide a satisfactory description of the data.

This method explores the global structure of parameter space without assuming that $\chi^2$ is quadratic in the vicinity of its minimum. This provides a more robust way of handling strong correlations among CFFs, albeit at the cost of increased computational requirements.
    
\subsubsection{Generation of $\chi^2$  maps}\label{subsubsec:maps_generation}

In the first stage, we generate the unpolarized cross-sections of the $ep\rightarrow ep\gamma$ process at the level of the eight twist-2 CFFs as a function of $\phi$. This generation is specific to a given kinematic bin characterized by $(Q^2, x_B, t)$ and a specific energy bin corresponding to the available experimental data kinematic points. For every uniform randomly generated set of eight CFFs, the $\chi^2$ is computed as:
\begin{equation}\label{eq:chi2_maps}
   \chi^2 = \sum_{i=1}^{n} \frac{(\sigma_i^{theo}-\sigma_i^{ data})^2}{(\delta\sigma_i^{ data})^2}, 
\end{equation}
where $\sigma_i^{theo}$ is the calculated cross-section at the point $\phi_i$, described in Section \ref{sec:TheoryFramework}. The quantities $\sigma^{\text {data }}$ and $ \delta \sigma^{\text {data }}$ are, respectively, the values and the uncertainties of the pseudo- or experimental data. As a result, we can construct the $\chi^2$ distribution of values as a function of the generated CFFs \ReH, \ReE\ and \ReHt. In the case of the pure DVCS cross-section parameter, it is calculated from the Monte-Carlo simulated CFFs using the theoretical calculation on \eqref{eq:dvcs2}. 

To keep the problem realistic, it was decided to limit, in a conservative and educated way, the range of variation of the CFFs while reducing the computing time and resources. 
The CFFs are confined within a bounded 8-fold hypervolume, with boundaries set at $\pm 5$ times the CFFs predicted by the KM15* model (described in Section \ref{subsec:KM15*}), which obeys most of the model-independent GPD normalization constraints. 
Centering the 8-CFFs hypervolume around the KM15* model and restricting it to a $\pm 5$ factor prevents us from exploring too unlikely cases. Generated values exceeding 3 times the KM15* model's value probably correspond to quite unrealistic CFFs. Given that GPDs have to fulfill a certain number of normalization constraints, such a strong deviation from the KM15* reference value is quite unlikely. However, the exploration of such a broad range of values enhances the robustness and credibility of the study. The only model-dependent input of this approach is the definition of the range of variation of the CFFs. 

Examples of the obtained maps are shown in Figure~\ref{fig:gen_maps_ex}.
We take the particular kinematics $\left(x_B, Q^2, t\right) = (0.34, 2.22$GeV$^2, -0.176$GeV$^2)$
and $\left(x_B, Q^2, t\right) = (0.13, 1.11$GeV$^2, -0.34$GeV$^2)$ with a 5.75GeV$^2$ beam energy.
This corresponds to a kinematic bin measured by the Hall-A and Hall-B experiments respectively.
The red point represents the true $\chi^2$ value obtained with the true CFFs at which the pseudodata was generated. The magenta point is at the the resulting $\chi^2$ value with the extracted parameters obtained from their weighted average \eqref{eq:wa}. We can see that several combinations of CFFs can give lower $\chi^2$ than the true one and therefore minimizing $\chi^2$ can lead to inaccurate results. 
A more striking limitation of the least-squares fit is revealed in the $\chi^2$ maps, which show an almost flat behavior near the minimum, particularly for \ReE\ and \ReHt. This flatness directly translates into large extraction uncertainties, as illustrated in Figure~\ref{fig:fit_opt_defl}.
A threshold is set at the maximum $\chi^2$ value of 50. This choice is informed by the observation that opting for higher $\chi^2$ values does not yield extra informative value as seen in Figure~\ref{fig:gen_maps_ex}.
The relevant topology primarily resides in the lower spectrum of normalized $\chi^2$, denoted as $\chi^2/ndf$, specifically those less than 2.  Here, $ndf$ represents the degrees of freedom calculated by subtracting the number of  free parameters i.e., four,  from the number of points in $\phi$ for a given kinematic configuration.

\subsubsection{Selection cuts on $\chi^2$ maps}\label{subsubsec:pre-selection}

Studying the $\chi^2$ maps topology in this 4-dimensional parameter space is a potential source of information on the CFFs correlations that can constrain the extraction of these parameters. The pseudodata mimicking the real data is a powerful tool to optimize and test the extraction method. We use the pseudodata to fine-tune the selection cuts applied on the $\chi^2$ maps to select the strongest $\chi^2$ values which allows to reduce the background from very unlikely parameters while finding more constrained high-density $\chi^2$ regions as a function of the parameters. This is of particular importance for the CFFs \ReE\ and \ReHt\ that pertain to a large homogeneous flat distribution.
    
The selection of the high-density areas is performed iteratively and the weighted average given in \eqref{eq:wa} is calculated at every step. Iterations are performed while the selected map region statistics is greater than 10000 points and the true value is within 1 standard deviation ($\sigma$) from the weighted average. In general, 2 iterations are performed on every kinematic set. The optimized selection is applied to the experimental data at the corresponding kinematic bin.
\begin{figure}[t] 
    \centering
    \includegraphics[width=1\linewidth]{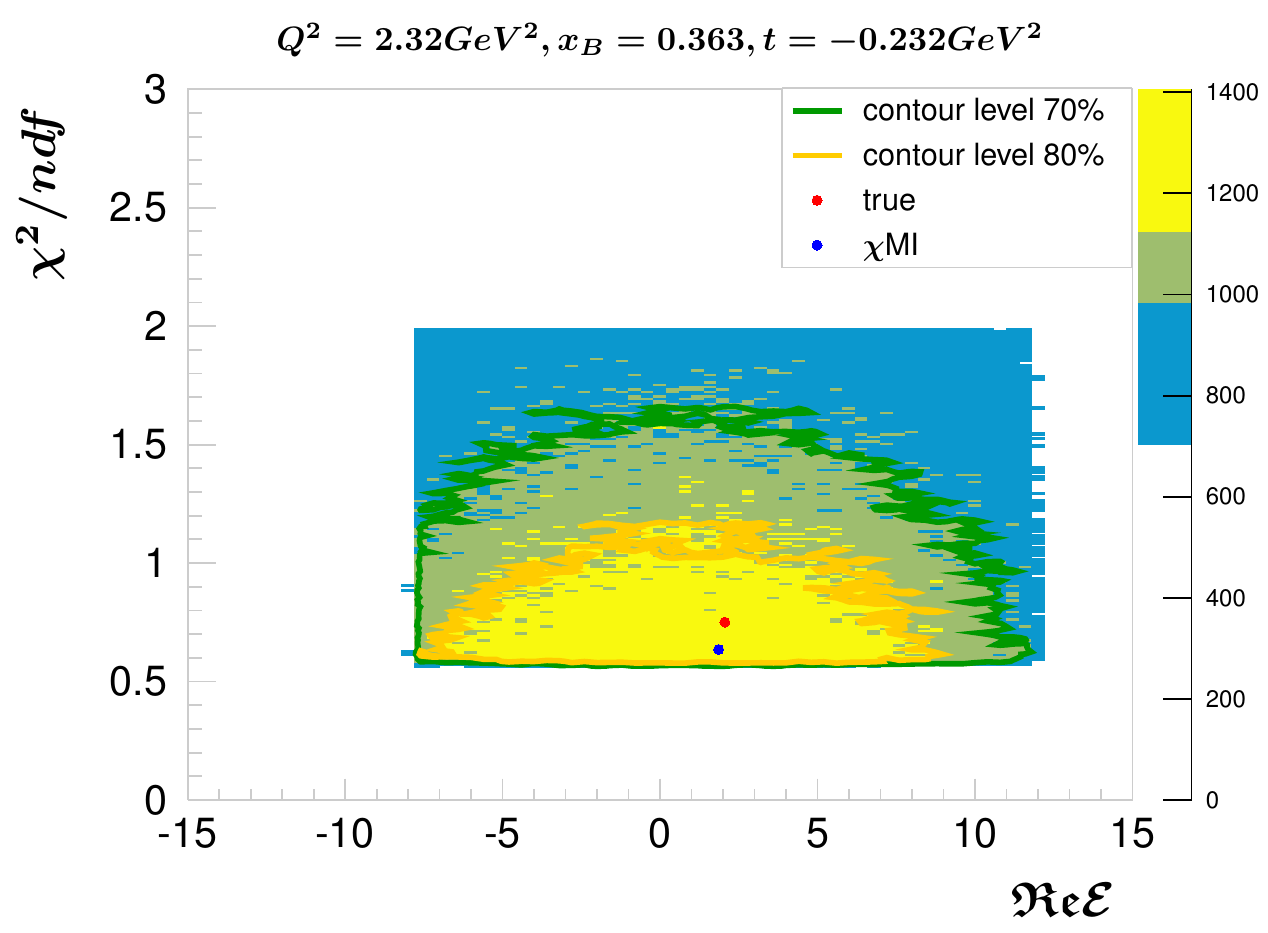}
    \caption[\ReE\ contour levels]{\ReE\ contour levels at the first iteration.}
    \label{fig:contours}
    \end{figure}
 
 \begin{figure*}[!ht] 
    \centering
    \includegraphics[width=1\linewidth]{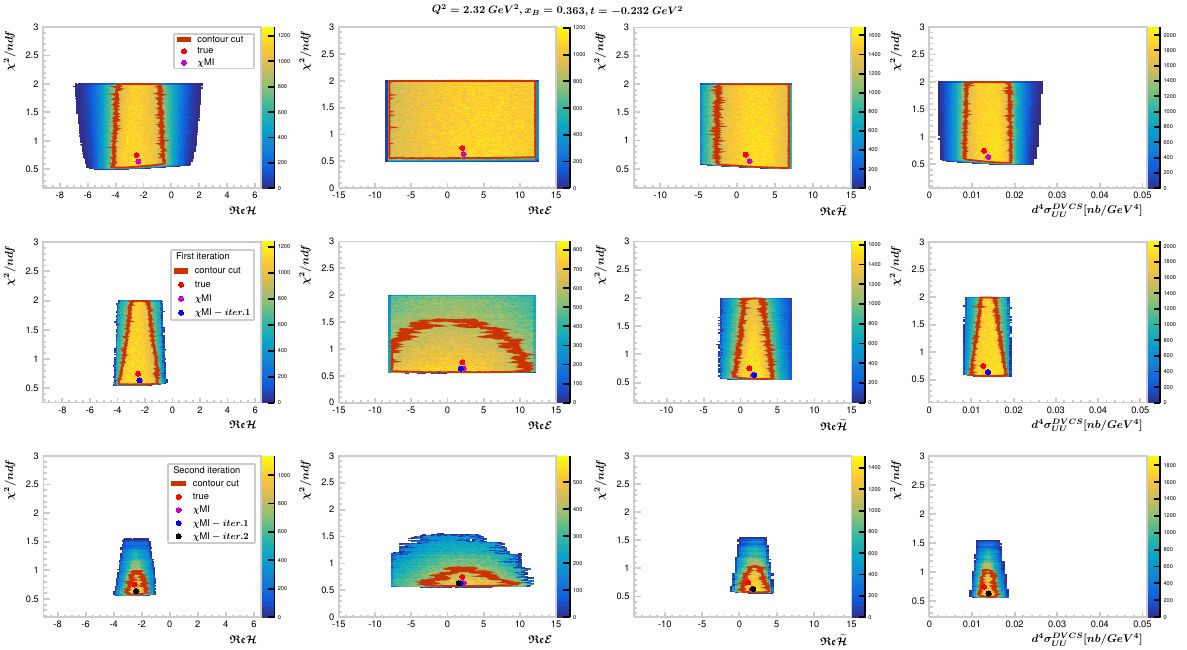}
    \caption[Pseudodata $\chi^2$ maps selections at a fixed kinematic bin of Hall-A.]{Selections on $\chi^2$ maps for the pseudodata at a fixed kinematic corresponding to a Hall-A experimental bin for all parameters. The top row shows the 2D maps without any selection cuts, in the center row the maps are shown after applying the contour cut (red, top) and the second iteration of selections is at the bottom after applying the contour cut (red, middle). }
    \label{fig:cuts_ex}
    \end{figure*}
    \begin{table*}[]
    $$
    \begin{array}{cccccc} 
        \toprule 
         & \mathfrak{Re}\mathcal{H} & \mathfrak{Re}\mathcal{E} & \mathfrak{Re}\mathcal{\widetilde{H}} & \sigma^{DVCS}_{UU}(nb/GeV^4) & \chi^2/ndf \\ \addlinespace[0.7mm]
         \hline\hline\addlinespace[0.5mm]
        \text{true} & -2.38 & 1.75 & 0.82 & 9.11\cdot10^{-3}  & 1.224  \\
        \text{no cuts} & -2.27\pm 1.15 & 1.10 \pm 4.85 & 1.45\pm 2.20 & (9.36 \pm 2.35)\cdot10^{-3} & 0.975  \\
        \text{iter. 1} & -2.20 \pm 0.67 & 0.97 \pm 4.52 & 1.34 \pm 1.64 & (9.22 \pm 1.33) \cdot10^{-3} & 0.978  \\
        \text{iter. 2} & -2.14 \pm 0.32 & 1.04 \pm 3.43 & 1.13 \pm 0.83 & (9.03 \pm 0.83)\cdot10^{-3}  & 0.986  \\
        \text{iter. 3} & -2.11 \pm 0.28 & 1.07 \pm 2.21 & 1.03 \pm 0.66 & (8.92 \pm 0.37)\cdot10^{-3}  & 0.990  \\
        \midrule
    \end{array}
    $$
    \caption[Extracted values of \ReH, \ReE, \ReHt\ and the DVCS cross-section from the pseudodata at one kinematic bin.]{Extracted values of \ReH, \ReE, \ReHt\ and the DVCS cross-section from the pseudodata at \mbox{$k =5.75$ $GeV$}, \mbox{$ Q^2 = 2.375$ $GeV^2$}, $x_B = 0.373$ and \mbox{$t = -0.372$ $GeV^2$} corresponding to a Hall-A kinematic set.} 
    \label{tab:uncert}
\end{table*}

Contour cuts are applied to the maps using a 70\% confidence level, defined relative to the highest-density bin of the 2D distribution. Figure~\ref{fig:contours}, for example, shows the confidence level contours at 70\% (green) and 80\% (orange) for \ReE\ after the first iteration. In this case, the next step selects only the region enclosed by the 70\% contour and recalculates the weighted average of \ReE. Repeated trials demonstrated that extending beyond two iterations, or varying the contour level by $\pm 10\%$ around 70\%, yields similar extracted parameter values. The main differences are confined to the precision, which is limited by the statistics within the contour, and to deviations from the true values that remain below one standard deviation.

Figure~\ref{fig:cuts_ex} shows the contour cuts for a fixed kinematic set over two iterations. At each step, the selected contour regions corresponding to the strongest $\chi^2$ zones must be satisfied simultaneously for all parameters. This iterative selection on the 2D maps enables the calculation of weighted averages in the regions most likely to contain the true values, thereby reducing the uncertainties on the extracted parameters. Notably, for \ReE\ and \ReHt\ the most probable $\chi^2$ values progressively concentrate near the true values after successive iterations, in contrast to the initially flat and homogeneous distributions. This demonstrates the capability of the $\chi$MI method to extract meaningful constraints even for the least sensitive CFFs.

\subsubsection{Uncertainty estimation}\label{subsubsec:stdev}

\begin{figure}[!ht] 
 \includegraphics[width=0.8\linewidth]{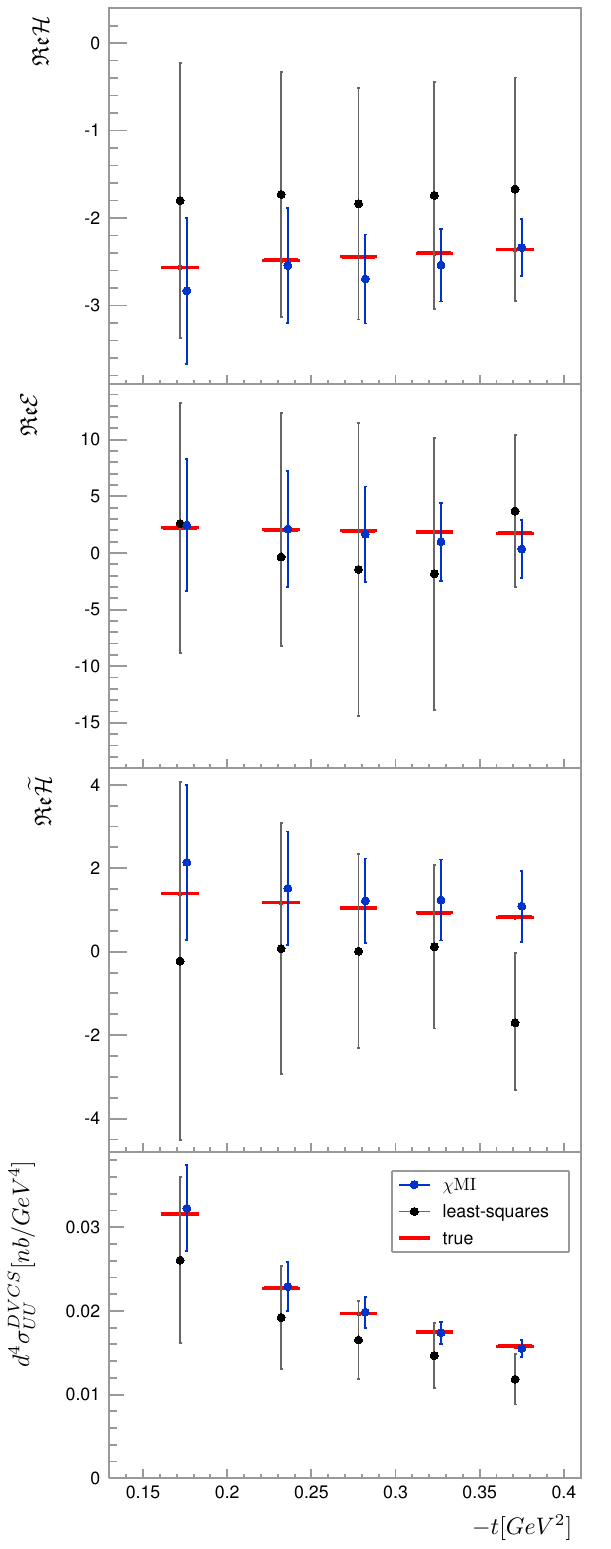}
        \caption[Extracted pseudodata parameters at Hall-A experiments kinematics.]{Extracted pseudodata parameters  corresponding to JLab Hall-A kinematics with average $Q^2=1.94\text{ GeV}^2, x_B = 0.369$. 
        The true values used to generate the pseudodata are represented with a red line centered at the experimental $t$ values with an arbitrary bin width. The results from the least-squares fit using the HESSE/MINOS technique are shown in black and the $\chi\text{MI}$ method is shown in blue. The points have been slightly shifted for visualization purposes.}
        \label{fig:pseudo_hallAB_extraction}
        \end{figure}

We recover the intuitive notion that when the likelihood is nearly flat near its maximum, the uncertainty on the parameter extraction is necessarily large. By applying selections on the 2D $\chi^2$ maps, we can significantly improve the precision of the extracted parameters—particularly for \ReE\ and \ReHt—while still ensuring that the true values remain contained within the selected regions. After iterative selections, the extracted values of the CFFs and the DVCS cross section are given by the weighted average, and the associated uncertainty is obtained from the standard deviation of the final map. For example, for \ReH:
\begin{equation}\label{eq:std}
    \sigma_{\mathfrak{Re}\mathcal{H}}(x_B,t,Q^2) = \sqrt{\frac{\sum_{n} \left(\mathfrak{Re}\mathcal{H}_n -\overline{\mathfrak{Re}\mathcal{H}}\right)^2}{N}},
\end{equation}
where $n$ runs over the $N$ points in the $\chi^2$ map, and $\overline{\mathfrak{Re}\mathcal{H}}$ is the weighted average defined in \eqref{eq:wa}. This statistical treatment ensures that the $\chi$MI method remains robust even in the presence of strong parameter correlations and near-flat likelihood regions.

The uncertainties and extracted values obtained after each iteration for one kinematic setting corresponding to the Hall A data are summarized in Table~\ref{tab:uncert}, together with the normalized $\chi^2$. In this specific set, by the third iteration the precision of the extracted values was improved by 75\%, 54\%, 70\%, and 84\% for \ReH, \ReE, \ReHt, and the DVCS cross-section, respectively, compared to the extraction without contour cuts, while the results remained consistent with the true values. This demonstrates the effectiveness of the $\chi$MI iterative selection procedure in reducing uncertainties without introducing bias.

\section{\label{Evaluation}$\chi$MI evaluation}
\begin{figure*}[!ht] 
        \centering
     \includegraphics[width=0.872\linewidth]{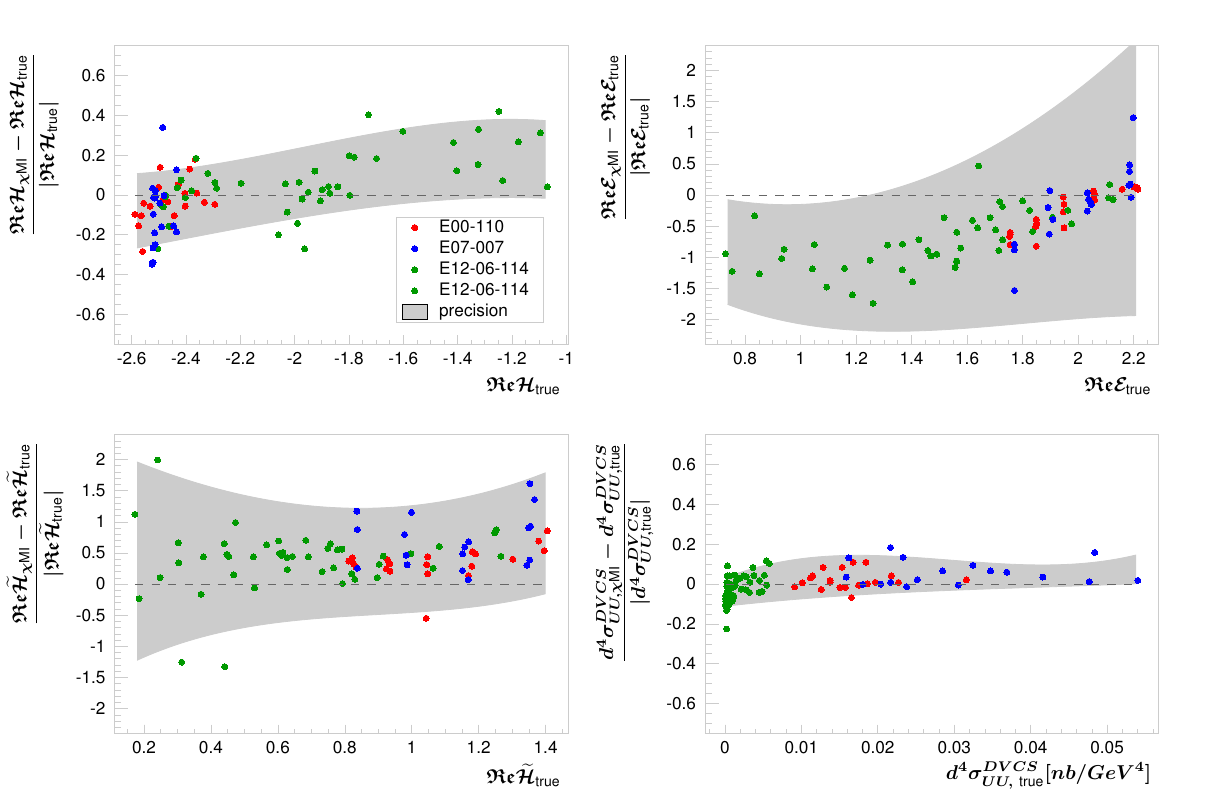}
        \caption[Pseudodata percentage difference from the \textit{true} CFFs at Hall-A kinematics.]{Pseudodata percentage difference of the extracted values (dots) and of their average error (gray band) from the \textit{true} parameters used to generate the pseudodata.}
        \label{fig:percent. diff}
        \end{figure*}
\begin{figure*}[!ht] 
        \centering
        \includegraphics[width=0.872\linewidth]{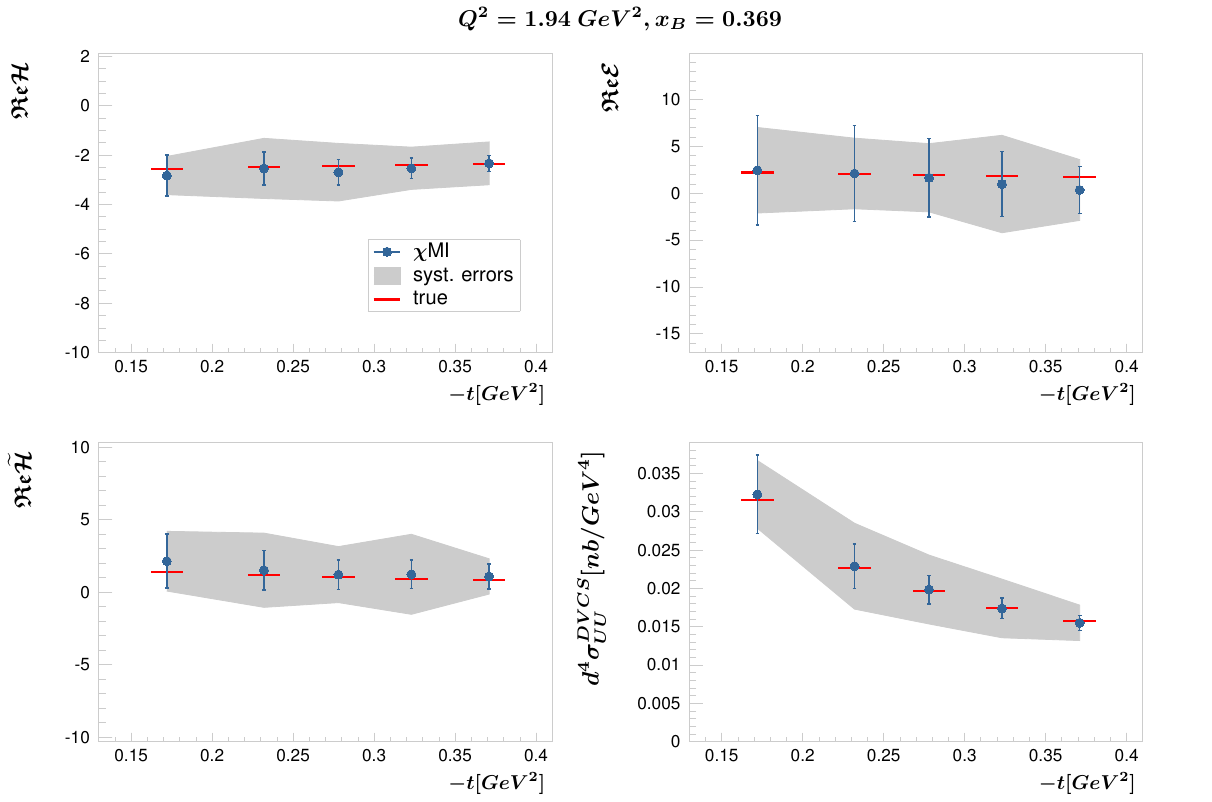}
        \caption[Systematic errors on the extraction parameters at kinematics bins corresponding to the Hall-A E00-110 experiment.]{Systematic errors on the extraction parameters at kinematics bins corresponding to the Hall-A E00-110 experiment with average \mbox{$Q^2 = 1.94$ GeV$^2$} and \mbox{$x_B = 0.369$}.}
        \label{fig:syst}
        \end{figure*}

Through our testing phase, the accuracy and precision are estimated using pseudodata generated at the same kinematics as the experimental data, with a smearing applied to mimic the experimental uncertainties. The goal of this study is to determine whether the $\chi$MI extraction method, applied to the pseudodata cross-sections, can accurately retrieve the eight original CFFs used to generate the pseudodata (i.e., the true CFFs) at the level of the CFFs \ReH, \ReE, \ReHt, and the DVCS cross-section under realistic experimental conditions. Additionally, the systematic uncertainties arising from deviations in the model used to generate the pseudodata are quantified.

\subsection{PSEUDO-DATA RESULTS}\label{subsec:pseudo_results}

The accuracy of the parameters extracted with the $\chi\text{MI}$ method is assessed by their proximity to the corresponding \textit{true} values. For a given parameter \mbox{$\mathcal{F} = \{ \Re e \mathcal{H}, \Re e \mathcal{E}, \Re e \widetilde{\mathcal{H}}, \text{DVCS}\}$}, the accuracy is defined as:
\begin{align}
    \label{eq:accuracy}
    \epsilon_{F}(x_B,t,Q^2) = \left( 1 - \left|\frac{\mathcal{F}_{true}-\mathcal{F}_{\chi\text{MI}}}{\mathcal{F}_{true}} \right|\right)\times 100 \%,
\end{align}
The precision of the extracted parameters is quantified by the standard deviation of the final $\chi^2$ map iteration, as given in \eqref{eq:std}.

The $\chi\text{MI}$ extraction method allows constraining the CFFs with increased precision compared to the regular least-squares fitting, particularly for the CFFs \ReE\ and \ReHt\ which normally carry significant uncertainties given their almost flat likelihood distribution near their maximum  - so with a small second derivative - the uncertainty on the parameter extraction is large with the Hessian method as well as with the MINOS method.
A pseudodata extraction comparison between the least-squares fit with HESSE/MINOS errors and the $\chi\text{MI}$ method for five different kinematic bins of JLab Hall-A experiments appears in Figure~\ref{fig:pseudo_hallAB_extraction}.
As previously noticed, the least-squares fits exhibit a stronger systematic bias relative to the input values, whereas the \xmi\ method produces fluctuations that more closely follow the expected systematic variations. This improvement arises because \xmi\ accounts for the local structure and correlations in the data, allowing each point to fluctuate naturally according to its uncertainty, rather than being constrained by a parametric fit, providing a more faithful representation of the underlying distribution.
This novel method gives the best achievable accuracy and precision for a local $\chi^2$ extraction technique in the least constrained DVCS observable scenario for the kinematic region covered by JLab experiments. 

        \begin{table}[]
        $$
        \begin{array}{ccccc} 
            \toprule 
             kinematics
             & \Re e \mathcal{H} & \mathfrak{Re}\mathcal{E} & \mathfrak{Re}\mathcal{\widetilde{H}} & \sigma^{DVCS}_{UU}(nb/GeV^4)  \\ \addlinespace[0.7mm]
             \hline\hline\addlinespace[0.5mm]
            \text{E00-110} & 91 (0.5)& 68 (4.1) & 60 (1.2)& 96 (1.9\cdot10^{-3}) \\
            \text{E07-007} & 87 (0.4) & 24 (2.2) & 48 (0.7) & 94 (0.2\cdot10^{-3})  \\
            \text{E12-06-114} & 85 (0.3)  & 58 (2.8) & 30 (0.7) & 94 (1.7\cdot10^{-3})  \\
            \text{e1-DVCS1} & 60(0.5) & 42(3.7) & 21 (1.1) & 60(6.5\cdot10^{-3})   \\
            \midrule
        \end{array}
        $$
        \caption[Average accuracy and precision of the $\chi\text{MI}$ method.]{Average accuracy and precision of the $\chi\text{MI}$ method. The accuracy \eqref{eq:accuracy} is given in percentage and the precision \eqref{eq:std}, shown in parenthesis,  is the standard deviation of the final iteration map.} 
        \label{tab:accuracy}
        \end{table}

Table~\ref{tab:accuracy} presents the obtained average accuracy from the pseudodata for the kinematics sets corresponding to each experimental data run. The average standard deviation is shown in parentheses.  The CFF \ReH\ and the DVCS cross-section are extracted with a high accuracy greater than 85\% and 94\% respectively at Hall-A kinematics. \ReE\ and \ReHt\ are the parameters harder to constrain with the best accuracy obtained at the Hall-A E00-110 experimental kinematic region which yields accuracies above 60\%. The accuracies for the \mbox{Hall-B} data are inferior; this is given by the lower $x_B$ region cover by this experiment where the CFFs were harder to constrain.

The percentage difference from the \textit{true} values for each kinematic bin of the \mbox{Hall-A} experiments is presented in Figure~\ref{fig:percent. diff} as a function of the \textit{true} values.  The gray band is obtained from the percent difference at $\pm \sigma$, e.g. for \ReH\ at a given kinematic set:       
    \begin{equation}\label{eq:perc.diff_sigma}
    \Delta_{\mathfrak{Re}\mathcal{H}} = \frac{(\mathfrak{Re}\mathcal{H}\pm \sigma_{\mathfrak{Re}\mathcal{H}})_{\chi\text{MI}}-\mathfrak{Re}\mathcal{H}_{\text{true}}}{\mathfrak{Re}\mathcal{H}_{\text{true}}}.     
  \end{equation}
  
Out of the four parameters, \ReH\ and the DVCS cross-section emerge with a quite well-nailed extraction and finite error bars in the order of $\approx20\%$ and $\approx10\%$ respectively for all the Hall-A kinematic range. \ReHt\ is extracted with differences from the \textit{true} values  below  $\approx50\%$ for most of the kinematic points. The CFF \ReE\ is extracted with deviations from the \textit{true} values  below $\approx50\%$ in the Hall-A experiments E00-110 and E07-007 kinematics. \ReE\ and \ReHt\ $\chi^2$ maps had very homogeneous topologies which make these parameters harder to constrain compared to \ReH\ and the DVCS cross-section parameters. The deviations of \ReE\ are larger as this CFF value decreases. This is consistent with the \ReE\ values in kinematic regions covered by the E12-06-114 experiment. 

\clearpage
\subsection{\label{subsec:Systematic} SYSTEMATIC UNCERTAINTIES}
To generalize the systematic uncertainty in the extraction method, it is imperative to modify the generated pseudodata to encompass all feasible  effects of the deviations of the parameters from the values employed in the evaluation and optimization of the extraction of the CFFs and the DVCS cross-section with the $\chi\text{MI}$ method. 
This process yields numerous samples of this kinematically dependent deviation. Aggregating all such samples provides direct means of estimating the systematic uncertainty associated with the extraction method itself. Incorporating this \textit{cumulative accuracy} measure into the quality metrics also ensures the robustness of the method, demonstrating its ability to consistently obtain extractions across a wide range of CFF studies. 

In this methodology, we generate multiple pseudodata samples by randomly generating sets of the eight CFFs within a defined region that spans up to two times the values predicted by the KM15$^*$ model. The purpose of this is to simulate a diverse range of possible scenarios and configurations of CFFs that are within a reasonable deviation from the KM15$^*$ model's predictions. The full extraction process with the $\chi\text{MI}$ method is then repeated on each sample to deduce the kinematic deviation given by the largest difference between the extracted and the true value among all the pseudodata samples. An example of the obtained systematic errors in the extracted parameters is shown in Figure~\ref{fig:syst}, as a function of $t$ with average \mbox{$Q^2 = 1.94$ GeV$^2$} and \mbox{$x_B = 0.369$} corresponding to the \mbox{Hall-A E00-110} experiment. The red lines are centered on the true values and the extracted values with the statistical errors bar are shown in blue. The obtained systematic error appears as a gray band. 

This step is crucial in assessing the systematic error of the CFF extraction from experimental data, as the uncertainty is contingent on both the kinematics and the magnitude of the CFFs.

\section{\label{sec:results}Results}
    We have evaluated the performance of the \xmi\ method and identified the optimal iterations and contour selections on the 2D $\chi^2$ maps on each kinematic set, as well as the expected effects of the experimental data resolution on the extraction accuracy and the systematic errors. Subsequently, we determine the CFFs \ReH, \ReE, and \ReHt, along with the DVCS unpolarized cross-section and their kinematical dependence using the described \xmi\ method at LO and LT approximations.  We use the helicity-independent photon electroproduction cross-section data from the \mbox{Hall-A} and \mbox{Hall-B} experiments presented in Table~\ref{tab:data_summary}, including the new Hall-A data after the 12 GeV JLab upgrade. 
    
    The description of the measured photon electroproduction cross-section by the \xmi\ method after extracting the CFFs and the pure DVCS cross-section that parametrizes it is discussed in the next section and compared with the published cross-section fits of the experimental datasets used.
We show that with our \xmi\ method we can extract the CFFs \ReH, \ReE, and \ReHt, along with the DVCS unpolarized cross-section unambiguously in the least constrained observable case. The sensitivity to the CFF \ReE\ and \ReHt\ with this approach is of great significance since they are much less well known experimentally.
The distribution of the extracted parameters as a function of the squared transverse momentum $t$ and $x_B$ are shown in the following sections and compared with results from other references when available.
The $Q^2$ evolution is also discussed which allows us to verify the applicability of the assumption that the GPDs do not evolve under the change of the photon virtuality i.e., $Q^2$-scaling hypothesis. 

\subsection{PHOTON ELECTROPRODUCTION}
\begin{figure*}[!ht] 
        \centering
        \includegraphics[width=0.45\linewidth]{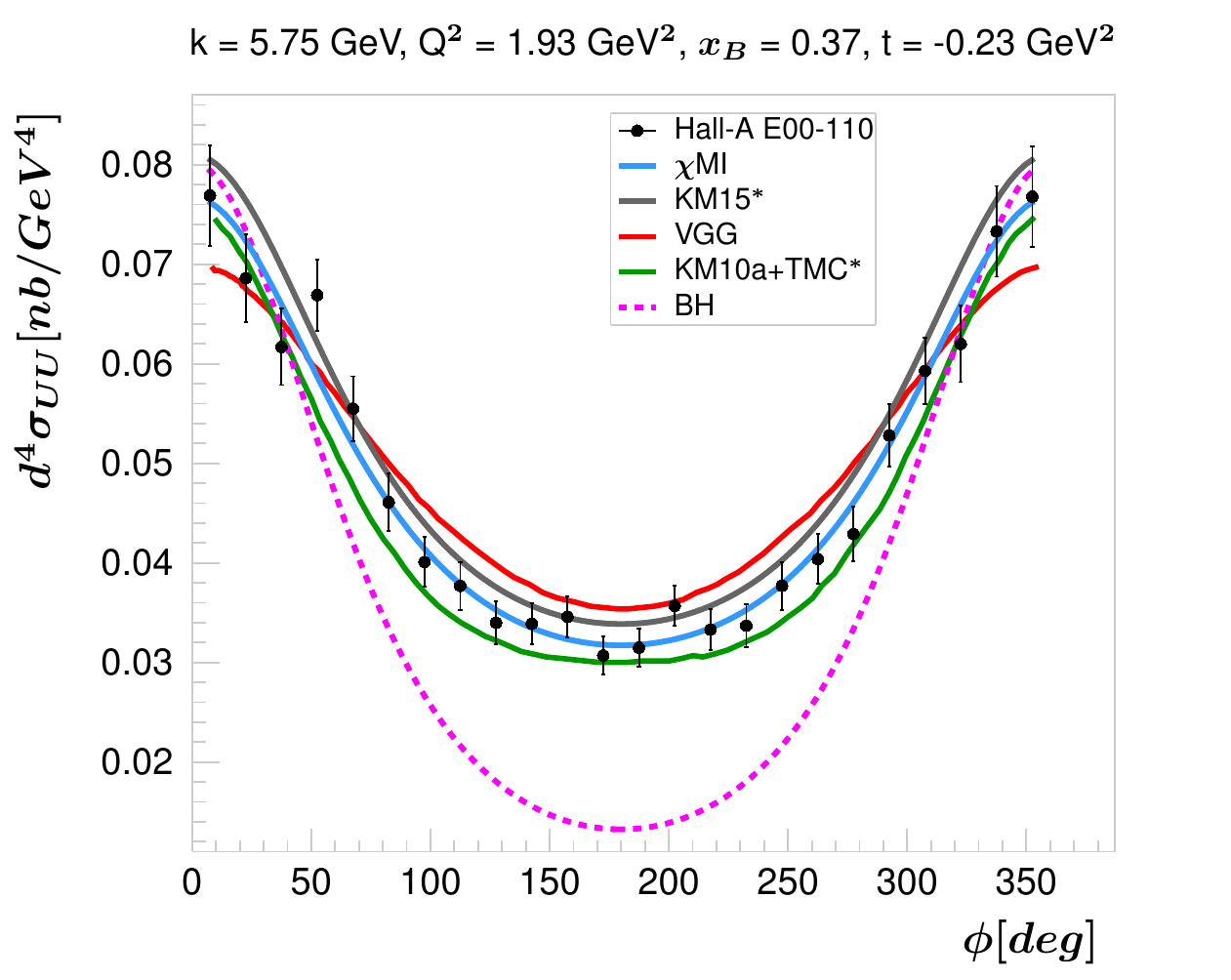}
        \includegraphics[width=0.45\linewidth]{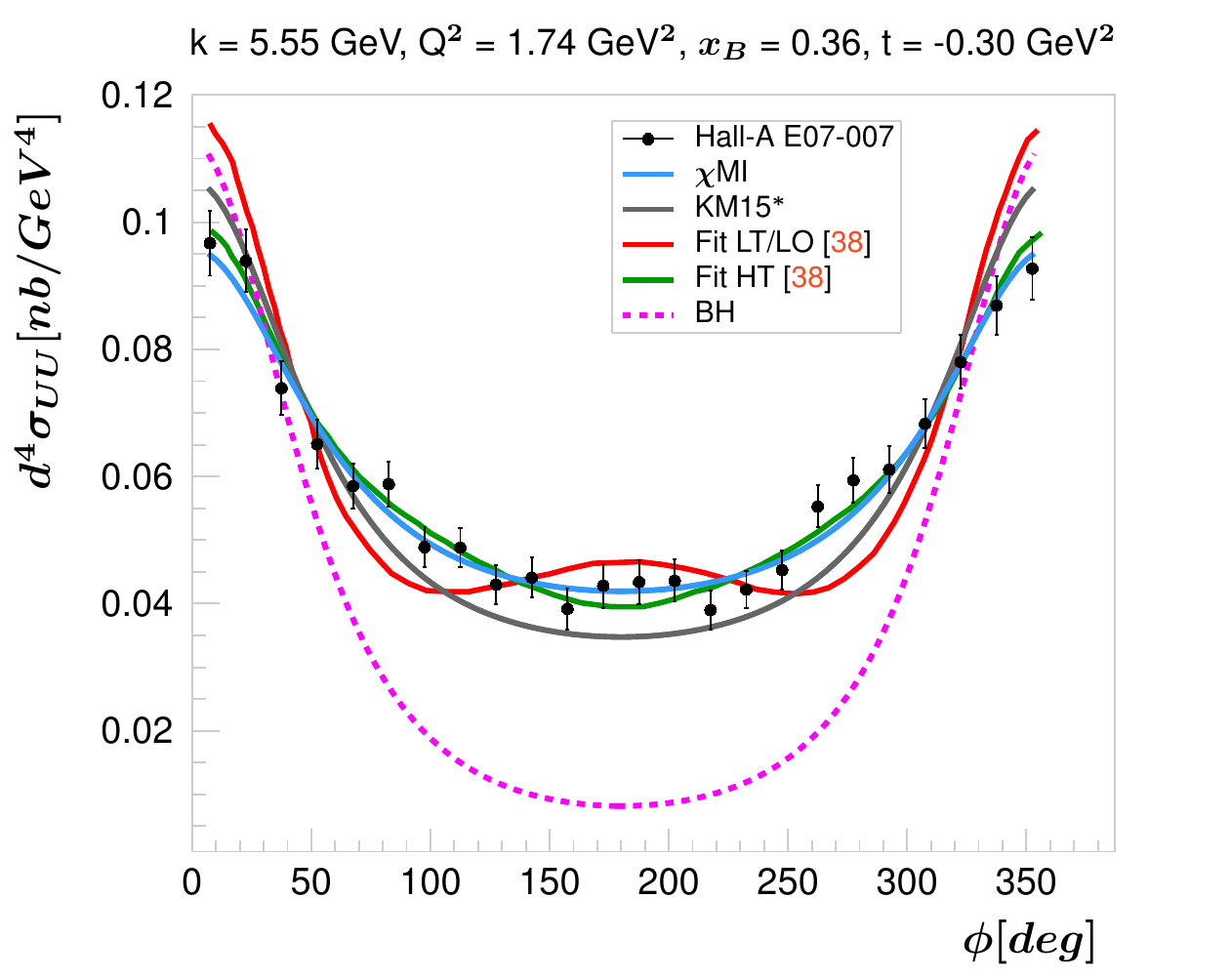}
        \includegraphics[width=0.45\linewidth]{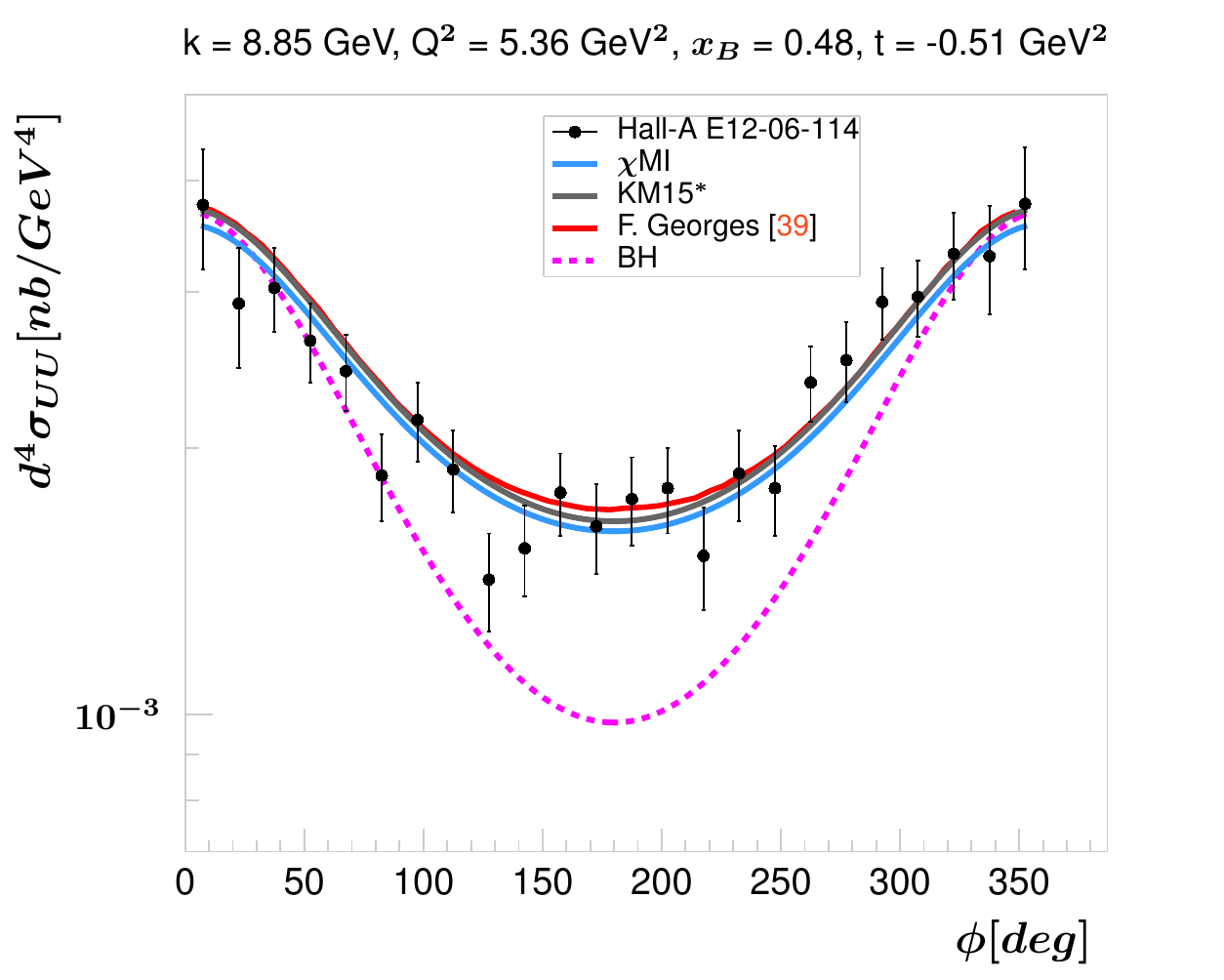}
        \includegraphics[width=0.45\linewidth]{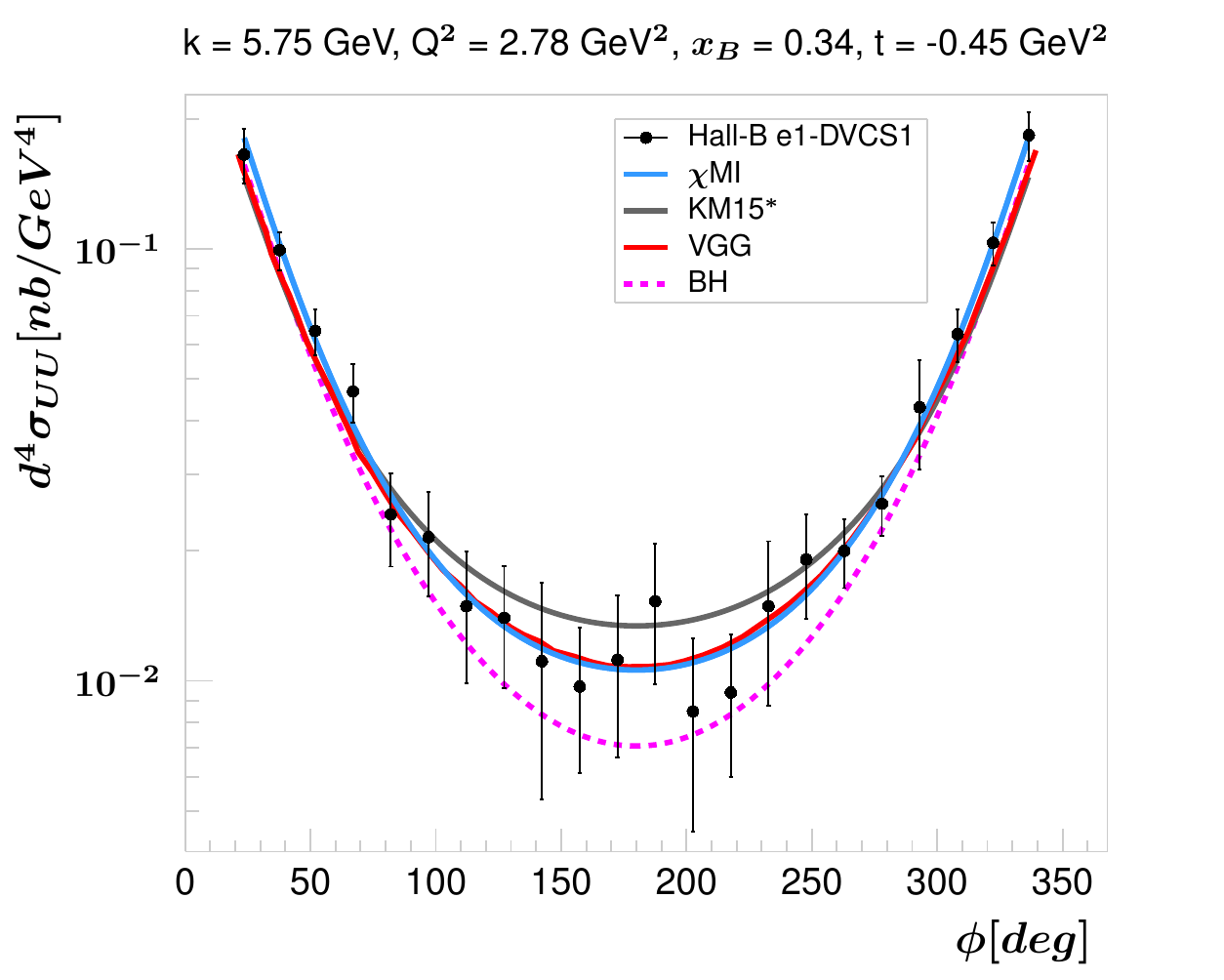}
        \caption[Helicity-independent photon electroproduction cross-section fits  for one kinematic set of the Hall-A and the Hall-B experiments.]{Helicity-independent photon electroproduction cross-sections (dots)  for one kinematic set of the Hall-A experiments \mbox{E00-110} (top-left), \mbox{E07-007} (top-right), \mbox{E12-06-114} (bottom-left) and of the Hall-B \mbox{e1-DVCS1} (bottom-right) experiment. The contribution of the BH process is shown on each plot as a dashed magenta line. The cyan line corresponds to the resulting cross-section distribution with the extracted parameters with the \xmi\ method. The theoretical KM15$^*$ model that was used to generate the pseudodata is shown in black.}
        \label{fig:xs_results}
        \end{figure*}

The total cross-section distributions as a function of the azimuthal angle $\phi$ 
obtained after the extraction the CFFs \ReH, \ReE, \ReHt\ and the pure DVCS cross-section with the \xmi\ method are shown in Figure~\ref{fig:xs_results} at one example kinematic set of each experimental dataset used in this work. The contribution of the BH process is shown on each plot as a dashed magenta line.  Notice that for the kinematic sets shown on the bottom panel of this figure, the BH contribution is large, and therefore the cross-sections are given in log-scale. Theoretical curves, including the KM15$^*$ model used to generate the pseudodata, and results from other references, are shown alongside for comparison. 

In this section we compare the cross-section results for each independent experiment. Overall, the \xmi\ method provides the best description of the photon electroproduction unpolarized cross-sections for the analyzed experimental data from Hall-A and Hall-B experiments, with the smallest $\chi^2$ per degree of freedom, giving the confidence that the extracted CFFs are realistic. This is remarkable for a local extraction with only one observable, the least constrained one, which brings high expectations on the extraction of CFFs with this technique when further observables are included or its implementation with an Artificial Neural Network (ANN)
napproach. Though we take notice of that, as pointed out in Figure~\ref{fig:uncert}, a good fit of the cross-section data is not sufficient for a reliable extraction of the CFFs. 

\subsubsection{Hall-A E00-110 experiment}
    
In Figure~\ref{fig:xs_results} (top-left), we show the experimental data from the E00-110~\cite{DefurneE00_2015} experiment carried out in Hall-A at Jefferson Lab for the Kin2 setting at \mbox{$t = - 0.23$ GeV$^2$} along with our \xmi\ predicted results (cyan) and the theoretical models KM15$^*$ (black), VGG~\cite{VGG} (red) and KM10a~\cite{KM09} (green) with target mass corrections (TMC) taken from~\cite{DefurneE00_2015}. The \xmi\ result is clearly very close to the helicity-independent data when compared to the other models considered. The information published in~\cite{DefurneE00_2015} for the experiment \mbox{E00-110} are the imaginary and real parts of  CFF combinations,  without a direct measurement of any of the CFFs independently.  These combinations were extracted simultaneously using a combined data-Monte Carlo $\chi^2$ minimization fit to determine the values that give the best agreement between the Monte Carlo predictions and the experimental data using the BKM formalism. The $\chi^2/ndf$ values they have obtained from the cross-section fits and the values resulting from the \xmi\ extraction method are given in Table~\ref{tab:E00_chi2} for the different kinematic settings of the \mbox{E00-110} experiment.  The \xmi\ extraction method offers improved accuracy with smaller $\chi^2/ndf$ values when compared to the Hall-A E00-110 cross-section fits from~\cite{DefurneE00_2015}.
\begin{table}[htbp]
\centering
\small
\setlength{\tabcolsep}{8pt}
\renewcommand{\arraystretch}{1.15}
\begin{tabular}{@{} l c c @{}}
\toprule
\textbf{Settings} & \textbf{Fit}~\cite{DefurneE00_2015} & $\boldsymbol{\chi^{2}_{\mathrm{MI}}}$ \\
\midrule
Kin2  & 1.16 & 0.89 \\
Kin3  & 0.99 & 0.84 \\
KinX2 & 0.82 & 0.61 \\
KinX3 & 1.28 & 1.21 \\
\bottomrule
\end{tabular}
\caption[Values of $\chi^2/\mathrm{ndf}$ for the kinematic settings of Hall-A E00-110.]
{Values of $\chi^2/\mathrm{ndf}$ for all kinematic settings of the Hall-A E00-110 experiment, from the extraction in~\cite{DefurneE00_2015} and from the \xmi\ method.}
\label{tab:E00_chi2}
\end{table}

\subsubsection{Hall-A E07-007 experiment}
The E07-007~\cite{DefurneE07_2017} experiment conducted at Hall-A, had the specific aim of isolating DVCS and DVCS-BH contributions to cross sections and demonstrating the sensitivity of high-precision DVCS data to twist-3 and/or higher-order contributions through a phenomenological study including kinematical power corrections. There are no direct values of the CFFs reported though.
This experiment measured cross-sections for three $Q^2$-values ranging from 1.5 to 2 GeV$^2$ at $x_B = 0.36$. Each kinematic setting was measured at two incident beam energies. 
The data were then fitted using a combined fit of two observables, the helicity-independent and the helicity-dependent cross-sections, using the BMMP~\cite{BMMP} formalism which incorporates leading-twist and leading-order contributions (LT/LO), higher twist contributions (HT), and next-to-leading order contributions (NLO). Those fits, along with our  \xmi\ results,  and the KM15$^*$ prediction are shown for one kinematic setting at \mbox{$t = -0.30$ GeV$^2$} in the top of the right panel of Figure~\ref{fig:xs_results}. 
The summary of $\chi^2/ndf$ values for the \mbox{Hall-A E07-007} fits utilizing the BMMP formalism from~\cite{DefurneE07_2017} and the \xmi\ method, at three different t settings, are presented in Table~\ref{tab:E07_chi2}. The $\chi^2/ndf$ values indicate that the \xmi\ method provides an improved description of the data when compared to the \mbox{Hall-A E07-007} fit with BMMP formalism from~\cite{DefurneE07_2017} in the LO/LT approximations and it is comparable to their fits with HT and NLO corrections.
\begin{table}[]
$$
\begin{array}{ccccc}
\toprule -t \text{ [GeV}^2] & \text { LO/LT } & \text { HT } & \text { NLO } & \chi\text{MI} \\\addlinespace[0.5mm]
\hline\hline\addlinespace[0.5mm] 0.18  & 1.20 & 0.98 & 0.99 & 0.77\\
0.24  & 1.76 & 0.99 & 1.0 & 0.99 \\
0.30  & 2.00 & 0.91 & 0.91 & 0.96 \\
\midrule
\end{array}
$$
 \caption[Values of $\chi^2/ndf$ for three $t$ kinematic bins of the \mbox{Hall-A E07-007} experimental data.]{Values of $\chi^2/ndf$ for three $t$ bins of the \mbox{Hall-A E07-007} experiment resulting from the extraction on~\cite{DefurneE07_2017} at LO/LT, HT, NLO and from the \xmi\ method.} 
        \label{tab:E07_chi2}
        \end{table}
\subsubsection{Hall-A E12-06-114 experiment}
The  E12-06-114 \cite{GeorgesE12_2022} experiment ran in Hall-A at Jefferson Lab after the 12 GeV upgrade. This experiment measured cross sections at fixed $x_B$ values over a broad range of $Q^2$, spanning from 2.7 GeV$^2$ to 8.4 GeV$^2$, using three distinct electron-energy settings. The aim of the $Q^2$-dependence measurements was to investigate the contribution of higher-twist terms relative to the leading-twist amplitudes. The cross sections were reported in nine different ($Q^2$,$x_B$) kinematic settings, with each setting comprising measurements at 3 to 5 different values of $t$. 
The cross sections reported in~\cite{GeorgesE12_2022} were fitted simultaneously using the BMMP formalism.  They present the first complete extraction of all 4 helicity-conserving CFFs as a function of $x_B$ averaged over $t$ appearing in the DVCS cross-section. We compare with their CFFs results in  Section~\ref{subsec:xB_dep}.
Figure~\ref{fig:xs_results}, left-bottom plot, displays their fits (red) alongside our fit using the \xmi\ method (cyan) for the specific kinematic setting Kin-48-3 at $t = -0.51$ GeV$^2$, shown for comparison. They both provide a good description of the data. Table~\ref{tab:chi2_E12} shows the obtained $\chi^2/ndf$ values of the E12-06-114 helicity-independent cross-section fit with our \xmi\ method for the nine different kinematic settings averaged over $t$.
\begin{table}
    \centering
    \begin{tabular}{cc|cc|cc}
    \toprule
        Settings & $\chi\text{MI}$ & Settings & $\chi\text{MI}$ & Settings & $\chi\text{MI}$\\\addlinespace[0.5mm]
        \hline\hline\addlinespace[0.5mm]
        Kin-36-1 & 1.27 & Kin-48-1 &  1.07 & Kin-60-1 & 1.78 \\
        Kin-36-2 & 1.30 & Kin-48-2 &  1.38 & Kin-60-2 & 1.30 \\
        Kin-36-3 & 1.31 & Kin-48-3 &  1.15 &  & \\
                 &  & Kin-48-4 &  0.97 &  & \\
    \midrule
    \end{tabular}
    \caption[Values of $\chi^2/ndf$ for the kinematic settings of the \mbox{Hall-A E12-06-114} experimental data.]{Values of $\chi^2/ndf$ resulting from the E12-06-114 helicity-independent cross-section fit with the \xmi\ method for the nine different kinematic settings averaged over $t$.}
    \label{tab:chi2_E12}
\end{table}

\subsubsection{Hall-B e1-DVCS1 experiment}

The e1-DVCS1~\cite{JoCLAS:2015} experiment ran in Hall-B at Jefferson Lab, with 110 finely bins in ($Q^2$, $x_B$ and $t$), nevertheless in the same 
$(Q^2 , xB , t)$ bin limits as those used for the Hall-A E00-110 analysis
 \mbox{($Q^2 =2.3$ GeV$^2$}, \mbox{$xB =0.36$}, and $-t =0.17$, 0.23, 0.28 and 0.33 GeV$^2$)  there are larger statistical uncertainties and lack of $\phi$-coverage around $\phi = 180^{\circ}$. 
The e1-DVCS1 experiment accesses the lowest $x_B$ region of all Jefferson Lab data with $0.12<x_B<0.47$. 
In~\cite{JoCLAS:2015}, well-defined minimizing values for $\mathfrak{Im}\mathcal{H}$ and \ReH\ are found. We compare their results for \ReH\ in Section~\ref{subsec:t_dep}. They use the local-fitting procedure at leading-twist and leading-order where the two observables, unpolarized and beam-polarized cross sections, are fitted simultaneously.
Figure~\ref{fig:xs_results} (right-bottom) shows the \xmi\ fit to the cross sections for fixed values of $Q^2$ = 2.78 GeV$^2$ and $x_B$ = 0.34 at $t$ = -0.45 GeV$^2$. The BH contribution in this kinematic set is significantly large and the total cross-section is best described by the \xmi\ method.  Over the 110 Hall-B data kinematic bins, the average $\chi^2$ per degree of freedom was the smallest for the \xmi\ method (0.76) compared to the theoretical models listed in Table~\ref{tab:chi2_e1}.

\begin{table}
    \centering
    \begin{tabular}{cccccc}
    \toprule
        VGG  & KMS  & KM10a & KM10 & KM15$^*$  & \xmi \\\addlinespace[0.5mm]
        \hline\hline\addlinespace[0.5mm]
        1.91 & 1.85 & 1.46 &  3.92 & 1.17 & 0.76 \\
        \midrule
    \end{tabular}
    \caption[Average $\chi^2/ndf$ over the \mbox{Hall-B e1-DVCS1} experimental data.]{Average $\chi^2/ndf$ over 110 kinematic bins of the \mbox{Hall-B e1-DVCS1} helicity-independent experimental data resulting from the theoretical models VGG, KMS, KM10a, KM10 taken from~\cite{JoCLAS:2015} and the KM15$^*$ model and our \xmi\ method.}
    \label{tab:chi2_e1}
\end{table}


       \subsection{\mbox{\large{$\mathbfit{|t|}$}} DEPENDENCE}\label{subsec:t_dep}

\begin{figure}[!h] 
        \centering
        \includegraphics[width=1.05\linewidth]{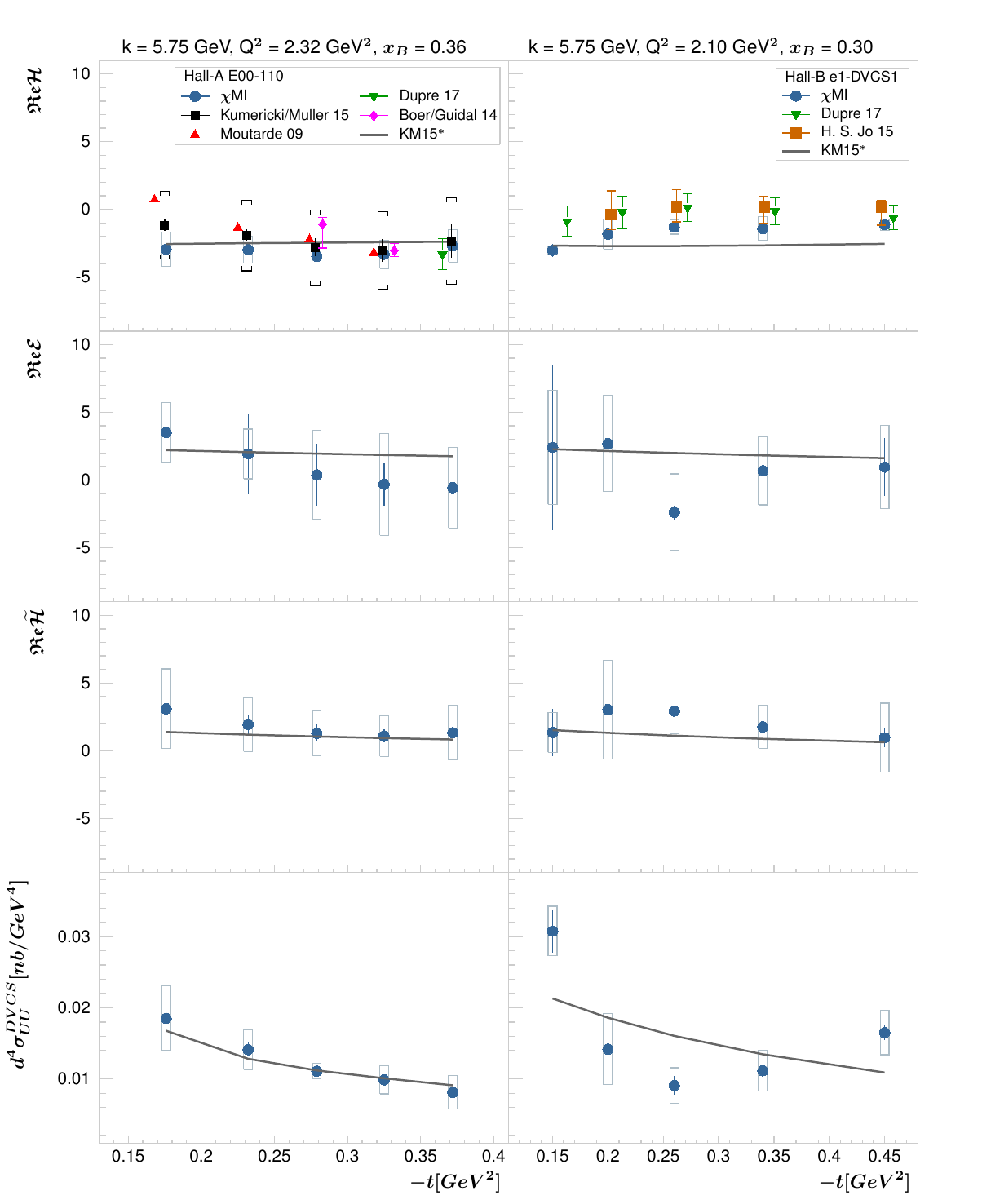}
        \caption[Extracted CFFs and the unpolarized DVCS cross-section as a function of $t$ for the experiments Hall-A \mbox{E00-110} and Hall-B e1-DVCS1.]{Extracted CFFs \ReH, \ReE, \ReHt\  and the DVCS cross-section as a function of $t$ for the experiments Hall-A \mbox{E00-110} (left panel) at $Q^2 = 2.32$ GeV$^2$, $x_B = 0.36$ and Hall-B e1-DVCS1 (right panel) at $Q^2 = 2.10$ GeV$^2$, $x_B = 0.30$. A black line shows the KM15$^*$ model. The results from \mbox{Kumeri\v{c}ki} and M\"uller~\cite{KM15}, Moutarde~\cite{Moutarde:fit09}, Bo\"er and Guidal~\cite{BoerGuidal:fit14}, Dupr\'e et. al.~\cite{Dupre} and H. S. Jo et. al.~\cite{JoCLAS:2015} are presented for comparison. Some points have been slightly shifted for visibility.}
        \label{fig:t-dep}
        \end{figure}

Studying the t-dependence of the CFFs is of great importance since it provides access to the transverse profile of the proton. Figure~\ref{fig:t-dep} shows the CFFs \ReH, \ReE, and \ReHt\  and the unpolarized DVCS cross-section as a function of $t$ for the Kin3 kinematic setting of the Hall-A E00-110 experiment (left) and for the Hall-B e1-DVCS1 experiment (right) at $Q^2 = 2.10$ GeV$^2$ and $x_B = 0.30$. From this point onward, all plots display statistical uncertainties as vertical lines and systematic uncertainties as boxes surrounding the extracted values. The predicted behavior as a function of $t$ by the KM15$^*$ model is given by a black line. The KM15$^*$ model, which has been shown to reproduce DVCS data reasonably well, is in good agreement with our extracted parameters using the \xmi\ method. In particular, we have been able to predict the CFFs \ReHt\ and \ReE, which, to the best of our knowledge, have not previously been determined locally without ANN techniques in the kinematic range accessed by the Jefferson Lab DVCS data, except in the work of F. Georges et al.~\cite{GeorgesE12_2022}. The limited sensitivity, especially for \ReE, arises from the $H$ dominance in the DVCS cross-section. This is reflected in the 2D $\chi^2$ maps of \ReHt\ and \ReE, which appear as flat distributions, but can nevertheless be constrained after applying contour cuts.

        \begin{figure}[!h] 
        \centering
        \includegraphics[width=0.9\linewidth]{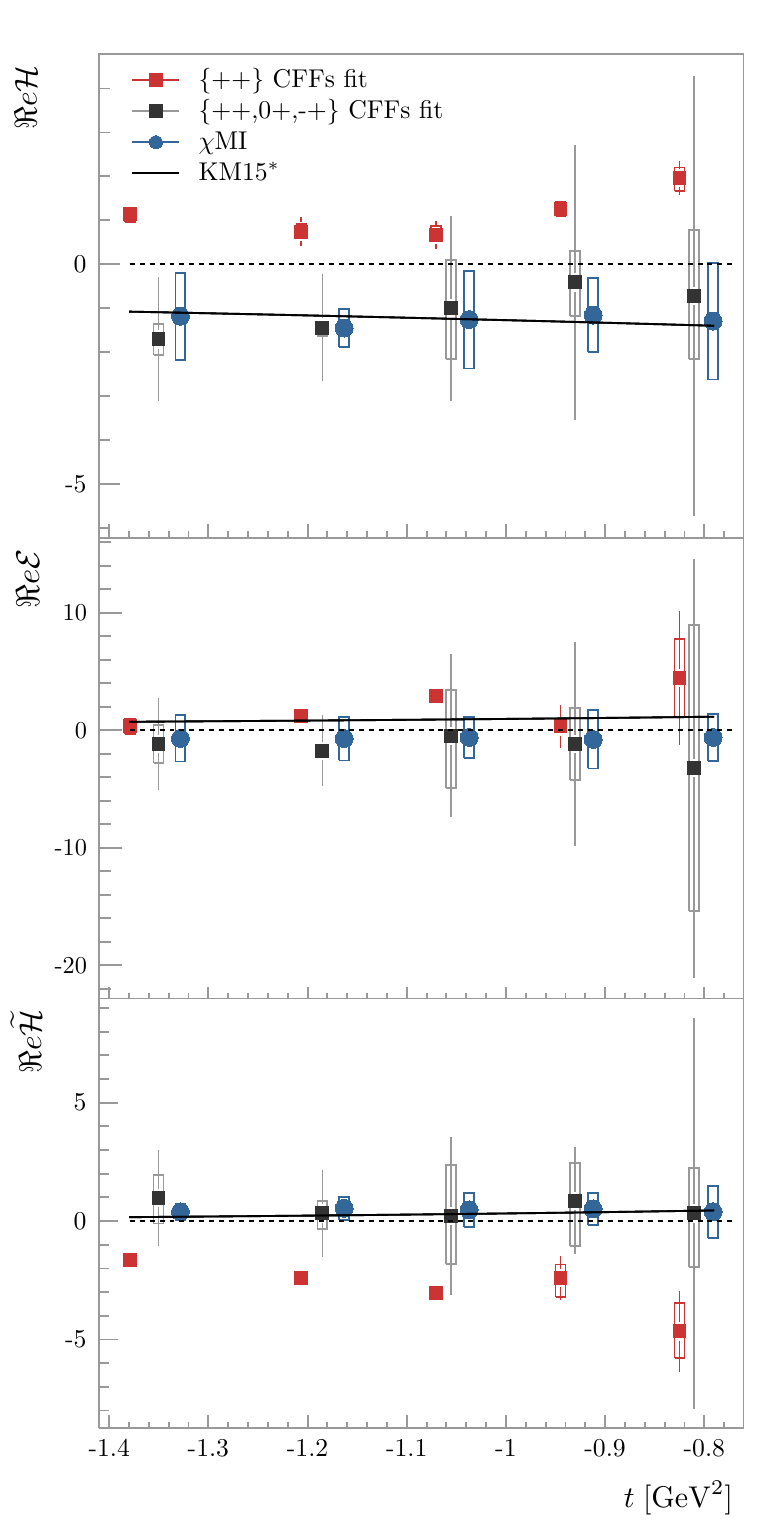}
        \caption{Values of the CFFs \ReH, \ReE, \ReHt\ at  \mbox{$Q^2 = 8.48 \,\mathrm{GeV}^2$}, $x_B = 0.60$ obtained with the \xmi\ method (blue circles). Results from F. Georges et. al~\cite{GeorgesE12_2022} are shown along side as squares from a fit including only the helicity-conserving CFFs (red) and a fit including both helicity-conserving and helicity-flip CFFs (black).  Bars around the points indicate statistical uncertainty and boxes show the total systematic uncertainty. The solid lines show the KM15$^{*}$ model. Some points have been slightly shifted for visibility.}
        \label{fig:t-dep_FG}
        \end{figure}

        We compare the extracted CFF \ReH\ with other available local fit results. \mbox{Kumeri\v{c}ki} and M\"uller~\cite{KM15}, performed a local fit of the Kin3 kinematic of the Hall-A E00-110 data (black squares). Their extracted values are supported by the extraction with the \xmi\ method within the errors. 
        Dupr\'e et. al.~\cite{Dupre} extracted the CFFs \ReH\ and $\mathfrak{Im}\mathcal{H}$ based on a simultaneous least-squared fitting of the unpolarized and beam-polarized observables by generating an ensemble of fits with randomly distributed start values of CFFs' multipliers.
   Our analysis allows for the determination of the CFF \ReH\ across the full $t$ range of the Hall-A data (left panel). In contrast, the results from Dupré et al.~(green triangles) provide a constraint at a single point, $t = -0.32$ GeV$^2$, which shows good agreement with the value extracted using our \xmi\ method. On the Hall-B data (right panel) \ReH\ is constrained in the entire $t$ range and they are consistent with the \xmi\ values except for the lowest $t = -0.12$ GeV$^2$.
Moutarde~\cite{Moutarde:fit09} also provides \ReH\ values extracted with the earlier E00-110~\cite{Camacho:2006} data that contained four $t$ bins (red triangles) between -0.33 and -0.17 GeV$^2$. Their local fit results are reproduced by the \xmi\ method excluding the low $t = -0.17$ GeV$^2$ although, the systematic errors of Moutarde's extraction have not been plotted. The same previous experimental data was used in the work of Bo\"er and Guidal~\cite{BoerGuidal:fit14} with a similar least-squared method followed by Dupr\'e et. al. In~\cite{BoerGuidal:fit14}, they were able to constrain \ReH\ at two kinematic sets (magenta diamonds) that are in agreement with the values we obtained with the \xmi\ method. They were not able to constrain though \ReH\ with the Hall-B dataset. Within uncertainties, the results for the CFF \ReH\ from the Hall-B e1-DVCS1 publication~\cite{JoCLAS:2015} (orange squares), obtained through a simultaneous fit of the unpolarized and beam-polarized cross sections, are also consistent with our extraction. The consistency observed in Figure~\ref{fig:t-dep} between similar kinematics in the Hall A and Hall B experiments, indicates that both data sets are compatible and can be reliably combined in a global fit. This agreement holds across their common kinematic range.

The first complete extraction of all helicity-conserving CFFs contributing to the DVCS cross section was performed in the Hall A E12-06-114 experiment by F. Georges et al.\cite{GeorgesE12_2022}. In their analysis, two observables were fitted simultaneously: the unpolarized and beam-polarized cross sections, using the BMMP formalism\cite{BMMP}, which includes target-mass corrections. All kinematic bins in $Q^2$ and $\phi$ at constant ($x_B$, $t$) are fit simultaneously. Given the larger $Q^2$ coverage of the E12-06-114 experiment, the sensitivity to the CFFs $\mathcal{E}$ and $\mathcal{\widetilde{E}}$ arises from the $Q^2$-dependent kinematic factors weighting these terms relative to the contribution of $\mathcal{H}$ and $\mathcal{\widetilde{H}}$. They provide the only source of comparison for the CFFs \ReE\ and \ReHt\ from a local least-squares fit at the Hall-A kinematics.
A comparison with their results is presented in Figure~\ref{fig:t-dep_FG} for the kinematic setting $Q^2 = 8.48 \,\mathrm{GeV}^2$ and $x_B = 0.60$. In their extraction, the values of the CFFs as a function of $t$ are shown for two scenarios: one where only helicity-conserving CFFs are included in the fit (red squares), and another where both helicity-conserving and helicity-flip CFFs are allowed to vary (black squares).
Our results obtained with the $\chi$MI method are consistent with the Georges et al. extraction that incorporates both helicity-conserving and helicity-flip CFFs in the fit, while their fitting using only
helicity-conserving CFFs significantly underestimates their uncertainties.
Importantly, $\chi$MI provides an independent validation: in this highly unconstrained parameter space, the weighted-$\chi^2$ mapping procedure and the contour-based selections inherent to $\chi$MI successfully recover trends for \ReH, \ReE, and \ReHt\ that are consistent with both the KM15$^*$ expectations (black line) and the extraction by Georges et al., despite relying solely on the unpolarized observable.

        \begin{figure}[t] 
        \centering
        \includegraphics[width=1\linewidth]{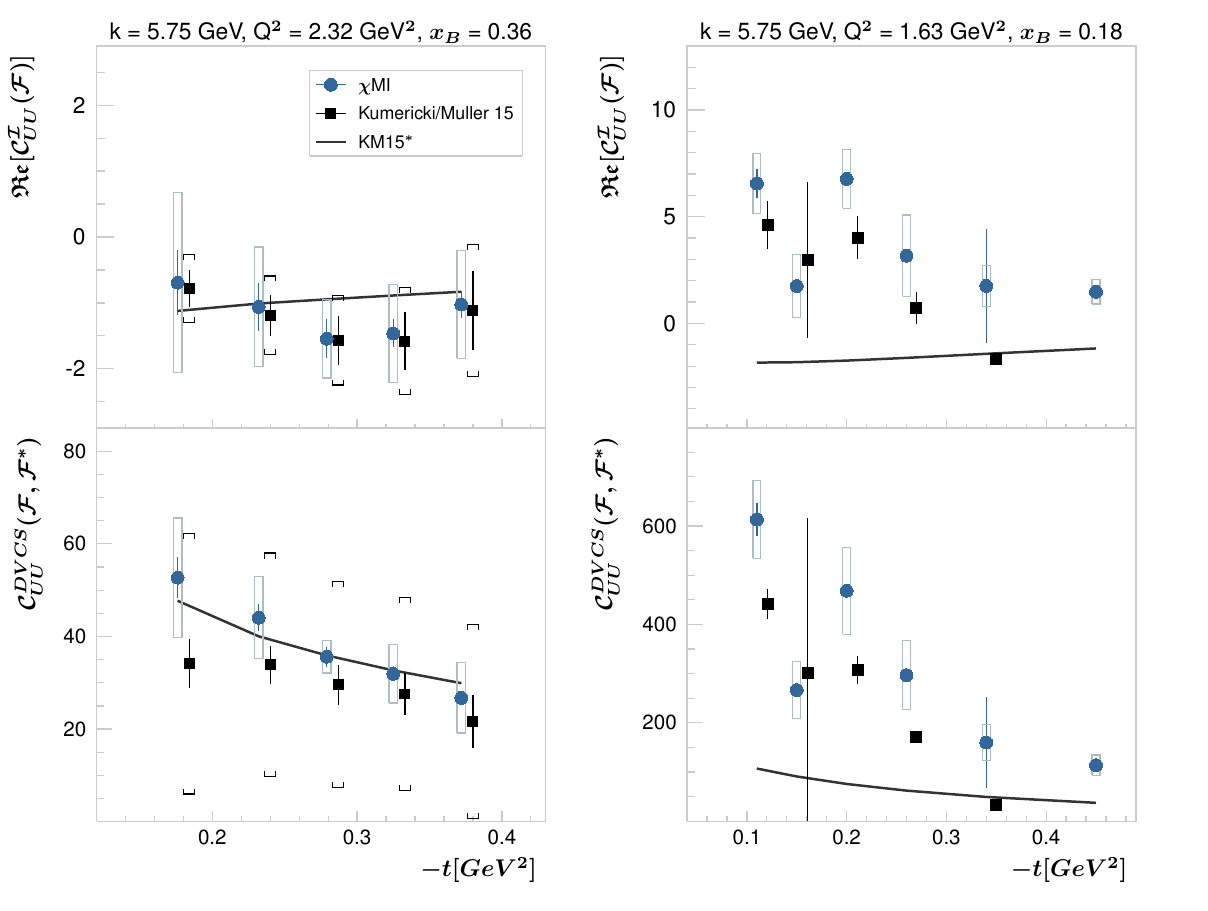}
        \caption[CFFs combinations $\mathfrak{Re}C^{\mathcal{I}}_{UU} (\mathcal{F})$ and  $C^{DVCS}_{UU} (\mathcal{F,F^*})$ as a function of $t$ for the experiments Hall-A \mbox{E00-110} and Hall-B e1-DVCS1.]{Combination of CFFs $\mathfrak{Re}C^{\mathcal{I}}_{UU} (\mathcal{F})$ \eqref{equ:Interference4} (top panel), and  $C^{DVCS}_{UU} (\mathcal{F,F^*})$ \eqref{eq:C_DVCS} (bottom panel), as a function of $t$ for the experiments Hall-A \mbox{E00-110} (left) at $Q^2 = 2.32$ GeV$^2$, $x_B = 0.36$ and the Hall-B e1-DVCS1 (right) at $Q^2 = 1.63$ GeV$^2$, $x_B = 0.18$. The results from \mbox{Kumeri\v{c}ki} and M\"uller~\cite{KM15} (black squares) have been slightly shifted for visibility. The KM15$^*$ is shown alongside (black line).  }
        \label{fig:CI_CDVCS}
        \end{figure}


         \begin{figure*}[t] 
        \includegraphics[width=1.04\linewidth]{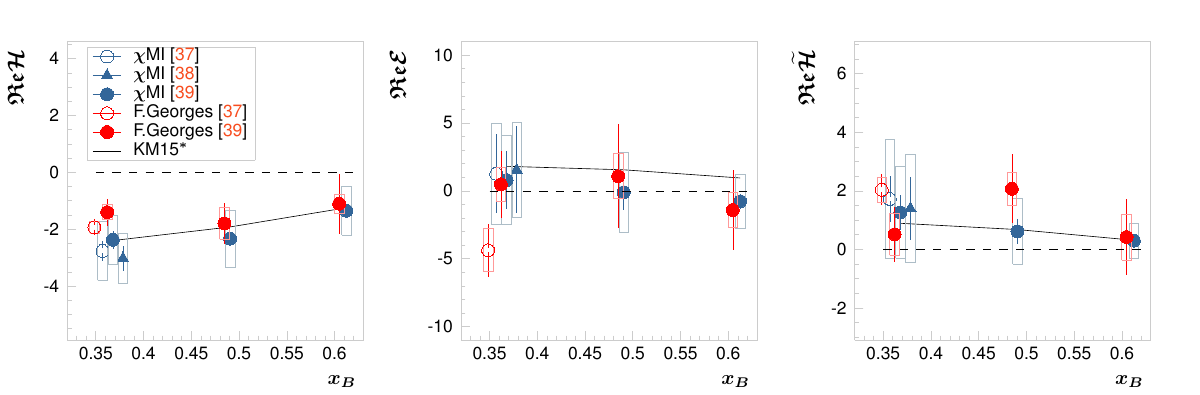}
        \caption[Extracted of the CFFs as a function of $x_B$ averaged over $t$.]{Extraction of the CFFs \ReH\ (left), \ReE\ (center) and \ReHt\ (right) from the \xmi\ method (blue) and from F. Georges et.al.~\cite{GeorgesE12_2022} (red) as a function of $x_B$ averaged over $t$ and $Q^2$. The results are extracted from the Hall-A datasets E00-110~\cite{DefurneE00_2015} (empty circle), E07-007~\cite{DefurneE07_2017} (triangle) and E12-06-114~\cite{GeorgesE12_2022} (circle) with the \xmi\ method and from the Hall-A datasets E00-110~\cite{DefurneE00_2015} (empty circle) and E12-06-114 (circle)~\cite{GeorgesE12_2022} in F.Georges et. al. The average $t$ values are $-0.281$ and $-0.258$ GeV$^2$ for the experiments E00-110 and E07-007 respectively at $x_B = 0.36$. The average $t$ values
        for the experiment E12-06-114 are -0.345, -0.702, -1.050 GeV$^2$ at $x_B = 0.36$, 0.48, 0.60, respectively. Some points have been slightly shifted for visibility. }
        \label{fig:xB-dep}
        \end{figure*}

         The underconstrained nature of the full CFF determination motivates the extraction of the imaginary and real parts of the CFF combinations $C^{\mathcal{I}} (\mathcal{F})$ and the bilinear form $C^{DVCS}(\mathcal{F,F^*})$. In the unpolarized case, we compute $\mathfrak{Re}C^{\mathcal{I}}_{UU} (\mathcal{F})$ from \eqref{equ:Interference4} substituting the extracted CFFs \ReH, \ReE\ and \ReHt\ and compare it with the results from \mbox{Kumeri\v{c}ki} and M\"uller~\cite{KM15} on the top panel of Figure~\ref{fig:CI_CDVCS}. This corresponds to the Kin3 setting of the Hall-A E00-110 (left) and for the Hall-B e1-DVCS1 experiment (right) at $Q^2 = 1.63$ GeV$^2$ and $x_B = 0.18$. With the extracted unpolarized DVCS cross-section parameter, we can compute $C^{DVCS}_{UU} (\mathcal{F,F^*})$ from \eqref{eq:dvcs2}. The resulting values are shown on the bottom panel of Figure~\ref{fig:CI_CDVCS} for Hall-A (left) and Hall-B (right) at the same kinematics as the top panels. 
        Our results for the CFFs combination forms are consistent in the Hall-A kinematic shown (left). There are larger deviations between our results with~\cite{KM15} for the Hall-B kinematic set (right) however they mention these fit results on the Hall-B data are underestimated.

        These agreements with previous local fit results, demonstrates that $\chi$MI can robustly extract information on subleading CFFs in regions where traditional local minimization methods either struggle with strong correlations or rely heavily on specific fit assumptions, when sufficient constraints are not available.

\subsection{\texorpdfstring{$\mathbfit{Q^2}$ EVOLUTION}{Q^2 EVOLUTION}}
\label{subsec:QQ_dep}
    
    In Figure~\ref{fig:xB-dep}, we compare our extracted parameters (blue) with the results from F. Georges et. al.~\cite{GeorgesE12_2022} (red) as a function of $x_B$ averaged in $t$, including the poorly known \ReE. The theoretical prediction from the KM15$^*$ model is shown as a black line, which shows a reasonable agreement with both results. The resulting CFFs with the Hall-A E00-110~\cite{DefurneE00_2015} were also calculated at the available $x_B = 0.36$ with average $t = -0.281$ GeV$^2$ (empty circles). We have also included the results using the Hall-A E07-007~\cite{DefurneE07_2017} that was also measured at $x_B = 0.36$ with an average $t = -0.258$ GeV$^2$ (triangles). We find the extracted CFFs \ReH, \ReE\ and \ReHt\ to be consistent with their results. These measurements also demonstrate that the full extraction of the CFFs \ReH, \ReE\ and \ReHt\ is within reach with the \xmi\ method at leading order and leading twist in the least constrained  helicity-independent cross-section observable. 

The first-moment sum rules relate the GPD 
$\widetilde{H}$ 
  (summed over quark flavor $q$) to the proton axial form factor $G_A$:
\begin{equation} \label{eq:GA}
  \sum_q \int_{-1}^{1} \widetilde{H}_q(x,\xi,t)\,dx = G_A(-t).
\end{equation}
The GPD 
$\widetilde{H}$
  is interpreted as a momentum decomposition of the corresponding form factor $\widetilde{H}$. In this context, the present measurement of the CFF \ReHt\
  provides complementary information to other determinations of the axial form factor 
$G_A$
 , which remains less well known experimentally than the electromagnetic form factors $G_E$
  and $G_M$
 . These results therefore supply additional constraints on the quark momentum distribution that underlies the corresponding form factors within the studied $x_B$	
  range.

    \subsection{$\mathbfit{Q^2}$ EVOLUTION}\label{subsec:QQ_dep}

        For fixed target kinematics, where $Q^2$ lever arm is rather limited, the $Q^2$ evolution on the GPDs is generally dropped since one may rely on the so-called scaling hypothesis, i.e., on the assumption that the GPD does not evolve under the change of the photon virtuality.  Here we check the veracity of this assumption in the Hall-A kinematics. 
        
\begin{figure}[t] 
        \centering
        \includegraphics[width=1\linewidth]{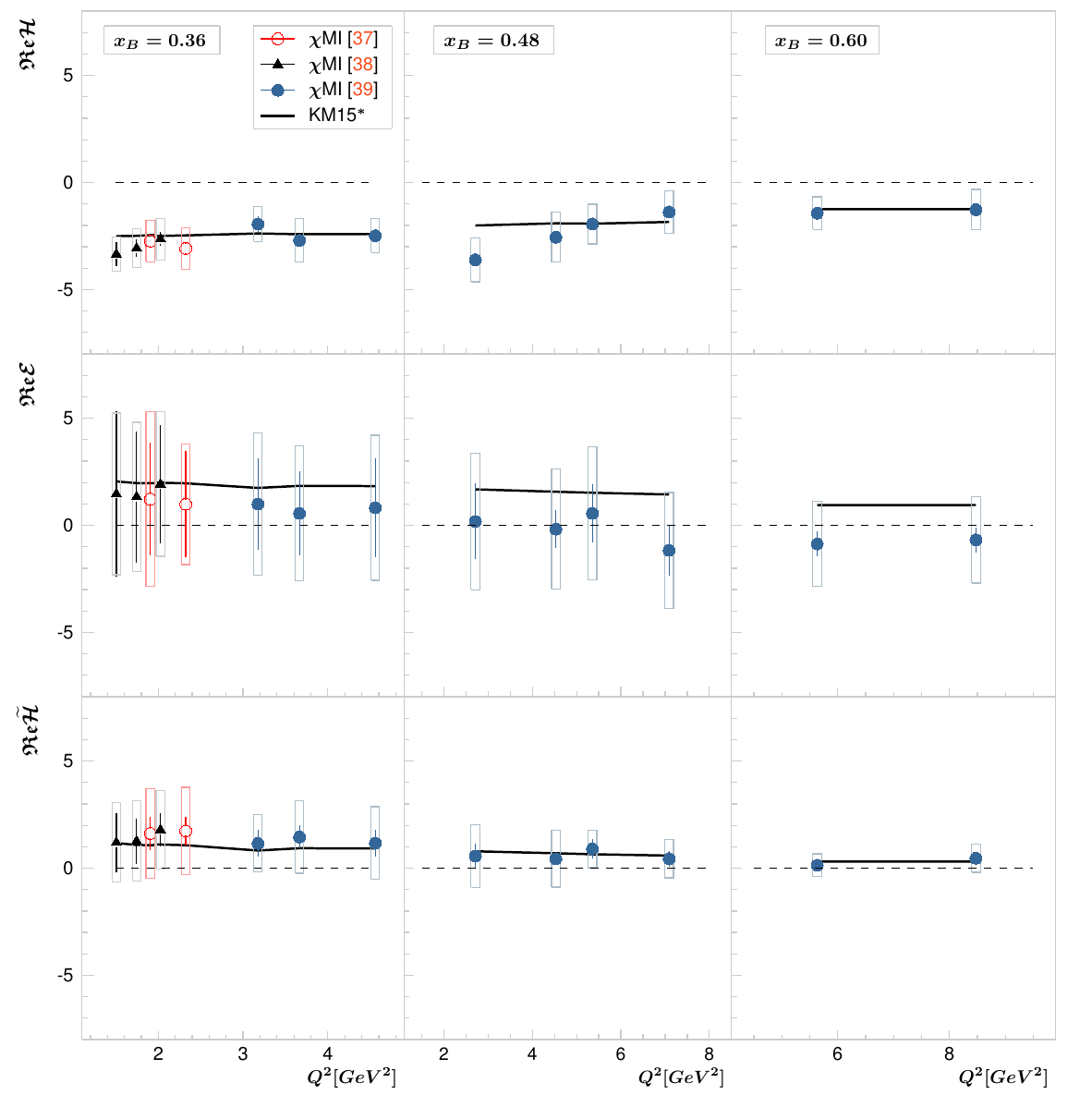}
        \label{fig:QQ-dep}
        \end{figure}

        Figure~\ref{fig:QQ-dep}, displays the $Q^2$ dependence of the extracted CFFs averaged in $t$ for the Hall-A experiments E00-110 (red), E07-007 (black) and E12-06-114 (blue) for the $x_B$ values 0.36 (left), 0.48 (center) and 0.60 (right).   
        In typical Jefferson Lab kinematics, there is a correlation between $Q^2$ and $x_B$, so that even after selecting small multidimensional bins, some variation of $x_B$ with $Q^2$ remains. For the $x_B = 0.36$ bin, the E00-110 experiment covers $x_B$ values from 0.343 to 0.373, while the E07-007 and E12-06-114 experiments span narrower ranges of 0.356 to 0.361 and 0.363  to 0.373, respectively. The $x_B$ bins of 0.48 and 0.6, only covered by the E12-06-114 experiment, extend over $[0.482,0.508]$ and $[0.608,0.617]$, respectively.
        With the new Hall-A data after the Jefferson Lab 12 GeV upgrade, we can gain insight into the $Q^2$ dependence in a larger region. The KM15$^*$ model which does not include $Q^2$ evolution on the GPDs is shown alongside for reference and it is computed for each experimental $Q^2$ value at the corresponding average $t$.

    \begin{figure}[]
        \centering        \includegraphics[width=1\linewidth]{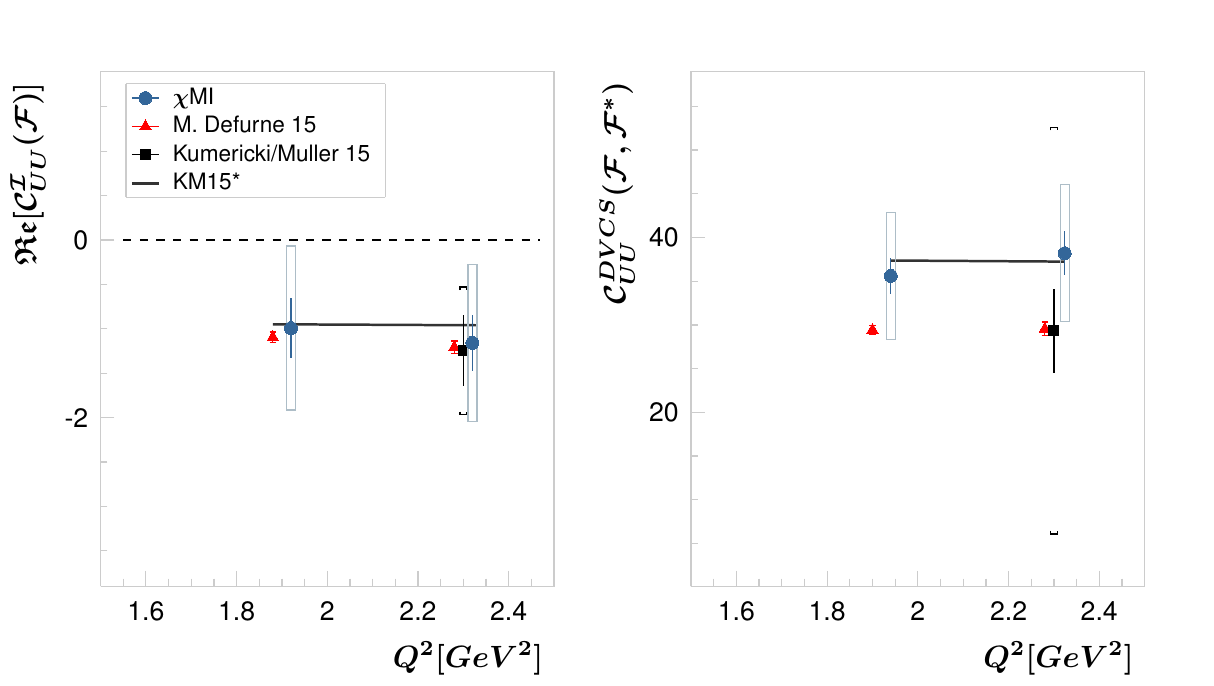}
        \label{fig:CI_CDVCS_QQ}
        \end{figure}

        Overall, the extracted CFFs exhibit little or no dependence on $Q^2$ within the range of $Q^2$ values shown in the figure.
        This slow $Q^2$-evolution of the extracted CFFs indicates that this kinematic region is suitable for an analysis under the scaling hypothesis. Nevertheless, we observe a sizeable scaling deviation on \ReH\ extracted from the HAll-A E12-06-114 data for $x_B = 0.48$ which expands the largest $Q^2$ range but is compatible with the scaling hypothesis within the errors.

        For comparison with the results from M. Defurne et. al.~\cite{DefurneE00_2015}, we show in Figure~\ref{fig:CI_CDVCS_QQ}, the CFFs combinations $\mathfrak{Re}C^{\mathcal{I}}_{UU} (\mathcal{F})$ and  $C^{DVCS}_{UU} (\mathcal{F,F^*})$ as a function of $Q^2$ at xB = 0.36 and integrated over t. The result from \mbox{Kumeri\v{c}ki} and M\"uller~\cite{KM15} for the Kin3 setting of the Hall-A E00-110 experiment is also shown along with the KM15$^*$ model prediction. No $Q^2$ dependence is observed for these CFFs combinations and the logarithmic $Q^2$–evolution can safely be neglected within this $Q^2$ lever arm at this $x_B$.

\section{\label{sec:mglobal}DNN Global model }
Deep neural networks (DNNs) are powerful universal function approximators, a property guaranteed by the Universal Approximation Theorem~\cite{NN_1,NN_2}. Their ability to represent highly nonlinear and complex mappings makes them especially effective for modeling and information extraction compared to other machine learning methods. Crucially, DNNs can capture the behavior of a function directly from data without requiring prior knowledge of its analytic form. This abstraction enables analysis and inference even in situations of extreme complexity where explicit functional descriptions are inaccessible.

The CFFs are highly nontrivial functions of energy and momentum transfer. Deep neural networks (DNNs) provide a powerful framework for modeling these quantities, as they can capture nonlinear dependencies directly from data without requiring an explicit analytic form. Once trained, DNNs can generalize beyond the training set, enabling reliable predictions of CFFs in unexplored kinematic regions. To predict the values of the extracted CFFs 
\ReH, \ReE, and \ReHt\ at arbitrary kinematic points, we perform a global fit by training a DNN on the CFFs extracted with the \xmi\ method across all available kinematic bins.

\begin{figure}[!t] 
        \centering
        \includegraphics[width=1\linewidth]{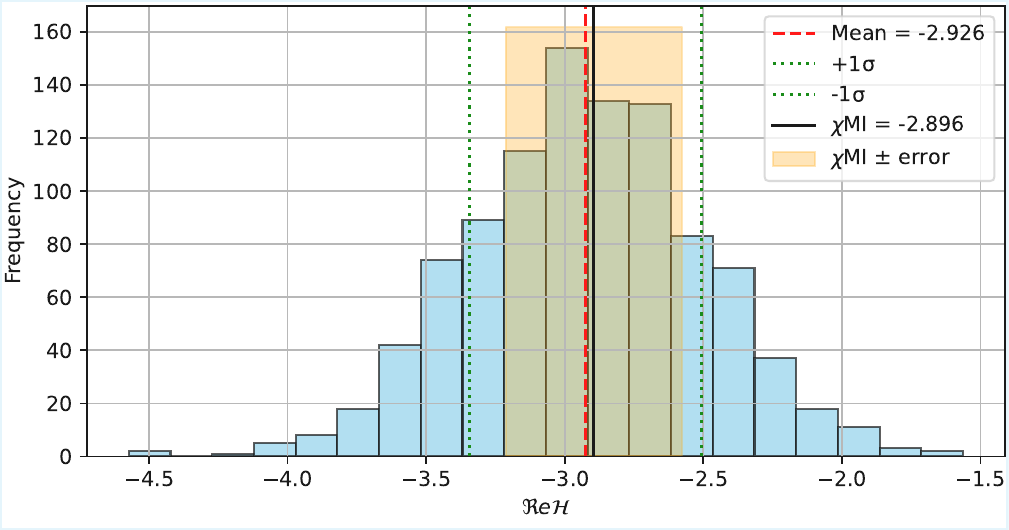}
        \caption{\ReH\ replica models prediction distribution for the kinematic bin \mbox{($Q^2 = 1.82$, $x_B = 0.343$, $t = -0.172$)}.}
        \label{fig:ReH_rep_set1}
\end{figure}

\subsection{\label{subsec:Bootstrap} Bootstrap approach}

The bootstrap is a resampling technique in which synthetic datasets are generated by perturbing the original data within its experimental uncertainties. While true Monte Carlo error propagation would require full covariance information, bootstrap resampling provides a practical alternative under the assumption of normally distributed errors. For each kinematic point, a replica dataset of CFFs is constructed by sampling from Gaussian distributions centered on the \xmi-extracted values with widths set by their uncertainties. Each replica is used to train a DNN, and the ensemble of resulting models encodes the propagated uncertainties in the extracted CFFs. The robustness of this approach depends on the statistical precision of the input data, the neural network architecture, and the fidelity of the training procedure.

After running 1000 bootstrap replicas, a distribution of predicted outputs is obtained. The standard deviation of these predictions represents the combination of propagated experimental uncertainty and additional algorithmic variability. For a fixed kinematic set, it is given by
\begin{align}
\label{eq:precision_DNN}
\sigma (x_B,t,Q^2) =
\sqrt{\frac{1}{N-1}\sum_{j=1}^{N} \left(\mathcal{F}^{\hspace{0.5mm}j}_{\text{DNN}}-\overline{\mathcal{F}}_{\text{DNN}}\right)^2},
\end{align}
where $\mathcal{F}^{\hspace{0.5mm}j}_{\text{DNN}}$ is the predicted CFF \ReH, \ReE\ or \ReHt\ from the j-th replica dataset, and 
$\overline{\mathcal{F}}_{\text{DNN}}$ is the mean prediction across all replicas. The final CFF distribution as a function of kinematics is given by $\overline{\mathcal{F}}_{\text{DNN}}$, with the precision quantified by \eqref{eq:precision_DNN}. 
An example of the \ReH\ replica models prediction distribution is shown in Figure~\ref{fig:ReH_rep_set1} for the kinematic bin \mbox{($Q^2 = 1.82$, $x_B = 0.343$, $t = -0.172$)}. The mean of the distribution, indicated by the dashed red vertical line, represents the DNN extracted value of \ReH\, while the dashed green lines indicate the prediction uncertainty given by the standard deviation of the distribution. The solid black vertical line denotes the 
\xmi\ value, with its associated error represented by the yellow band.
\begin{table*}[htbp]
\centering
\footnotesize
\begin{ruledtabular}
\begin{tabular}{l l}
\textbf{CFF} & \textbf{Architecture (layer sequence)} \\ \midrule
$\ReH$ &
\begin{minipage}[t]{0.78\textwidth}\raggedright
Input(13) $\rightarrow$ Dense(292, relu) $\rightarrow$ BatchNorm $\rightarrow$ Dropout(0.1) $\rightarrow$
Dense(148, tanh) $\rightarrow$ BatchNorm $\rightarrow$ Dropout(0.4) $\rightarrow$
Dense(292, leaky\_relu) $\rightarrow$ BatchNorm $\rightarrow$
Dense(52, tanhshrink) $\rightarrow$ BatchNorm $\rightarrow$
Dense(532, leaky\_relu) $\rightarrow$ BatchNorm $\rightarrow$
Dense(556, tanh) $\rightarrow$ BatchNorm $\rightarrow$ Dropout(0.4) $\rightarrow$
Dense(268, relu6) $\rightarrow$ BatchNorm $\rightarrow$
Dense(652, tanh) $\rightarrow$ BatchNorm $\rightarrow$ Dropout(0.5) $\rightarrow$
Dense(700, leaky\_relu) $\rightarrow$ Dropout(0.3) $\rightarrow$
Dense(196, tanhshrink) $\rightarrow$ BatchNorm $\rightarrow$ Dropout(0.1) $\rightarrow$
Dense(268, tanhshrink) $\rightarrow$ BatchNorm $\rightarrow$ Dropout(0.1) $\rightarrow$
Dense(700, tanhshrink) $\rightarrow$ Dense(1)
\end{minipage} \\
$\ReE$ & \begin{minipage}[t]{0.78\textwidth}\raggedright
Input(13) $\rightarrow$ Dense(292, relu6) $\rightarrow$ Dropout(0.3) $\rightarrow$
Dense(164, tanh) $\rightarrow$ Dropout(0.4) $\rightarrow$
Dense(388, tanh) $\rightarrow$ BatchNorm $\rightarrow$
Dense(36, leaky\_relu) $\rightarrow$ BatchNorm $\rightarrow$ Dropout(0.1) $\rightarrow$
Dense(68, relu6) $\rightarrow$ BatchNorm $\rightarrow$ Dense(1)
\end{minipage} \\
$\Re\widetilde{H}$ & \begin{minipage}[t]{0.78\textwidth}\raggedright
Input(13) $\rightarrow$ Dense(228, tanh) $\rightarrow$ BatchNorm $\rightarrow$
Dense(228, relu) $\rightarrow$ BatchNorm $\rightarrow$ Dropout(0.5) $\rightarrow$
Dense(196, tanhshrink) $\rightarrow$ BatchNorm $\rightarrow$
Dense(132, tanhshrink) $\rightarrow$ BatchNorm $\rightarrow$
Dense(324, leaky\_relu) $\rightarrow$ Dropout(0.5) $\rightarrow$
Dense(228, leaky\_relu) $\rightarrow$ Dense(1)
\end{minipage} \\
\end{tabular}
\end{ruledtabular}
\caption{Summary of DNN architectures and training batch sizes for each CFF. ...}
\label{tab:DNN_architectures}
\end{table*}

\subsection{\label{subsec:Bootstrap}  Data preprocessing and models architecture}

The DNN input consists of 195 experimental kinematic sets 
\mbox{($Q^2, x_B$, $t$)} from the Hall-A and Hall-B experiments described in Section~\ref{sec:ExperimentalData}.
Replica datasets are generated by Gaussian sampling around each extracted CFF  at the corresponding kinematic set with the \xmi\ method, using their associated uncertainties. Each replica dataset serves as the target for training an independent DNN.

Prior to training, the input and target variables underwent feature engineering steps and normalization. Engineered features were created from the original kinematic variables \((Q^2, x_B, t)\), including cross-terms \((x_B t, Q^2 x_B, Q^2 t)\), squared terms \((Q^4, x_B^2, t^2)\), and the inverse of \(Q^2\) to capture possible non-linear dependencies.  
Smoothing was applied by transforming \(Q^2\), \(x_B\), and \(t\) using logarithmic scaling, which compresses their dynamic ranges and reduces the impact of statistical fluctuations. Specifically, the log-transformed features were computed as:
\begin{equation}
\label{eq:log_transform}
\log \nu = \log\left( \max\left( \nu - \nu_{\min} + \epsilon, \epsilon \right) \right),
\end{equation}
where \(\nu\) represents any of the kinematic variables, \(\nu_{\min}\) is the minimum observed value in the dataset, and \(\epsilon = 10^{-6}\) prevents undefined operations.  

Input features were normalized using the \mbox{\textit{StandardScaler}}~\cite{scikit-learn-standardscaler}
that rescales each feature by subtracting its mean and dividing by its standard deviation. This transforms the feature to have zero mean and unit variance, which helps stabilize and speed up the training of neural networks by ensuring that all input features have similar scales and contribute equally during optimization. This is especially effective when the original features have distributions close to Gaussian.
Target variables (CFFs) were normalized using the \mbox{\textit{RobustScaler}}~\cite{scikit-learn-robustscaler}, which subtracts the median and scales by the interquartile range. This method is less sensitive to outliers and skewed distributions, making it better suited for CFF targets that can exhibit long-tailed distributions.  
These preprocessing steps improve numerical stability, accelerate convergence, and enhance the network’s ability to generalize to unseen data.

The DNN models were optimized using Keras Tuner~\cite{KerasTuner}, which automatically searched for the best architecture by tuning the number of hidden layers, the number of neurons per layer, and the choice of activation functions. The Adam optimizer~\cite{kingma2017adam} was selected for training due to its adaptive learning rate and robust convergence properties. For each DNN replica model the loss function is defined as:
\begin{equation}\label{eq:loss_global}
L\left(\nu, \mathcal{F}\right)
= \frac{1}{N} \sum_{i=1}^N\left(\mathcal{F}^{\hspace{0.5mm}i}_{\text{DNN}}-\mathcal{F}^{\hspace{0.5mm}i}_{\chi \text{MI}}\right)^2,
\end{equation}
where $\nu$ denotes the input kinematics features and $\mathcal{F}$ denotes the DNN output variables (\ReH, \ReE\ or \ReHt). This function computes the mean squared difference between the inferred CFF from the DNN $(\mathcal{F}^{\hspace{0.5mm}i}_{\text{DNN}})$ and the CFF extracted via the \xmi\ method $(\mathcal{F}^{\hspace{0.5mm}i}_{\chi\text{MI}})$ over the N kinematic sets.

\begin{figure}[!t] 
        \centering
        \includegraphics[width=1\linewidth]{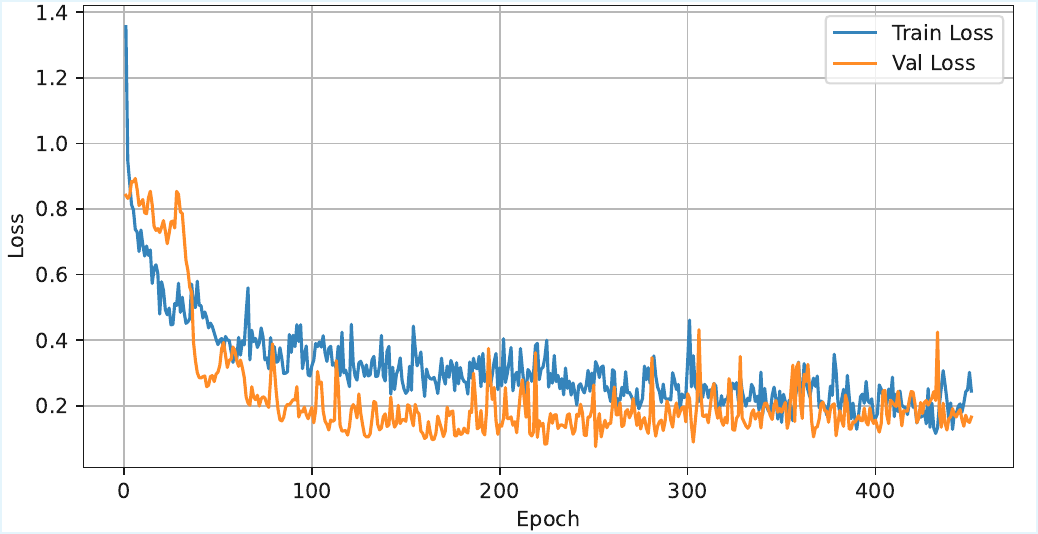}
        \caption{Training and validation loss for one DNN replica model of \ReH.}
        \label{fig:ReH_loss}
\end{figure}

Once the optimal hyperparameters were identified, a benchmarking procedure was carried out to determine the most suitable batch size for stable and efficient training. For all three CFFs models the best batch size found was \texttt{None}. 
The performance and stability of the optimized model were assessed using 5-fold cross-validation~\cite{refaeilzadeh2009cross}, ensuring that the results were robust against variations in the training data. For the final training, the dataset was randomly split into 90\% training and 10\% testing subsets. The model was trained for a maximum of 5000 epochs, with an early stopping callback monitoring the validation loss and a patience of 200 epochs to prevent overfitting while allowing for sufficient convergence time. 
An example of the training and validation loss for a single replica of the \ReH\ model is shown in Figure~\ref{fig:ReH_loss}.

Various standard nonlinear activation functions are employed in the models, including the Rectified Linear Unit (\texttt{ReLU}), \texttt{Leaky ReLU} and hyperbolic tangent (\texttt{tanh}) which are discussed in comprehensive surveys of activation functions~\cite{Dubey2022ActivationSurvey,Nwankpa2021ActivationTrends}. \texttt{ReLU6}~\cite{Howard2017MobileNets,Sandler2018MobileNetV2}, and the \texttt{tanhshrink} ~\cite{PyTorchTanhshrink} function are used as well.
The standard \texttt{ReLU} activation has become widely adopted due to its simplicity, computational efficiency, and ability to mitigate the vanishing gradient problem. \texttt{Leaky ReLU} extends this by allowing a small, nonzero slope for negative inputs, improving gradient flow through inactive neurons.
The \texttt{tanh} function outputs values in the range $(-1, 1)$, producing zero-centered activations that can facilitate optimization.
\texttt{ReLU6}, a capped variant of ReLU with an upper bound of 6, is particularly effective in quantized or resource-constrained settings, improving numerical stability.

Finally, the \texttt{tanhshrink} activation, defined as 
$x-tanh(x)$, blends linear and nonlinear characteristics, promoting smoother gradient transitions while preserving representational flexibility. It has been previously determined to enhance network performance for similar architectures ~\cite{Sipper2022NeuralNetworksALaCarte}. 
Furthermore, batch normalization layers~\cite{Ioffe2015Batch} are incorporated to normalize intermediate layer inputs, reducing internal covariate shift and accelerating convergence. Dropout layers~\cite{Srivastava2014Dropout} are applied as a regularization technique to prevent overfitting by randomly deactivating units during training.
Table~\ref{tab:DNN_architectures} summarizes the resulting architectures of the DNNs employed for each CFF.



\subsection{\label{subsec:results}  CFFs projections in selected kinematics}

\begin{figure}[!t] 
        \centering
        \includegraphics[width=0.95\linewidth]{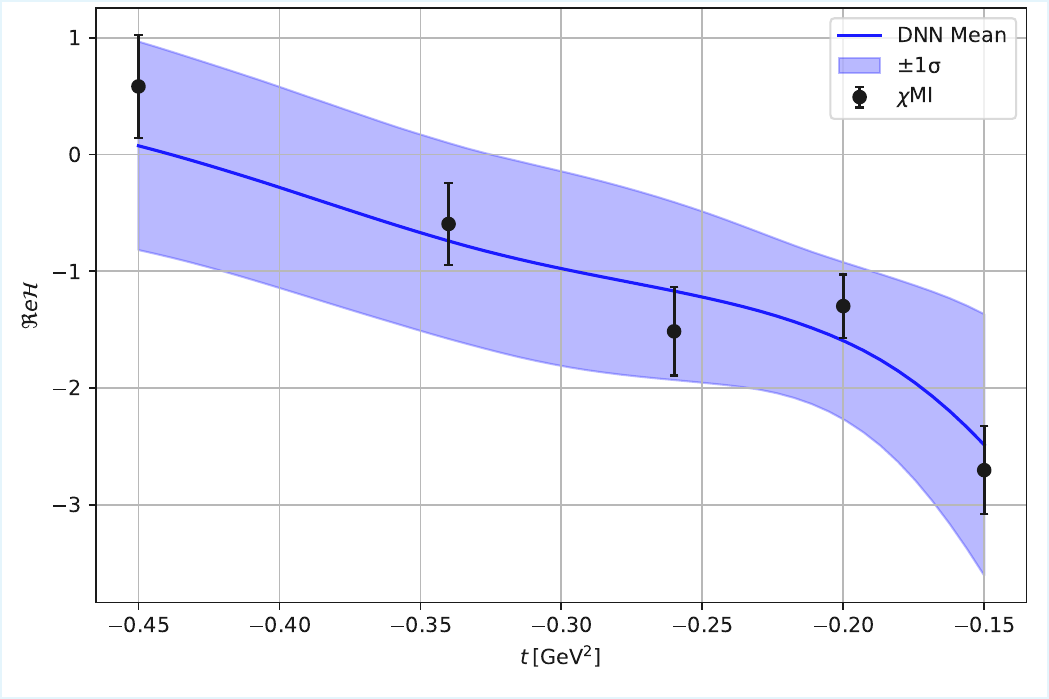}
        \includegraphics[width=0.95\linewidth]{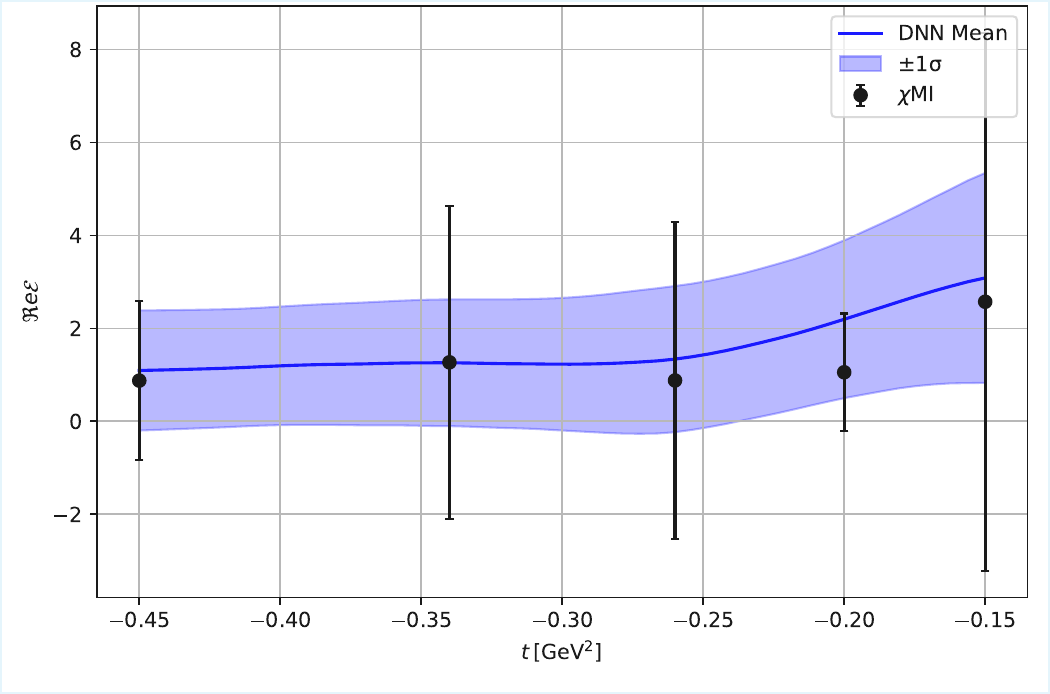}
        \includegraphics[width=0.95\linewidth]{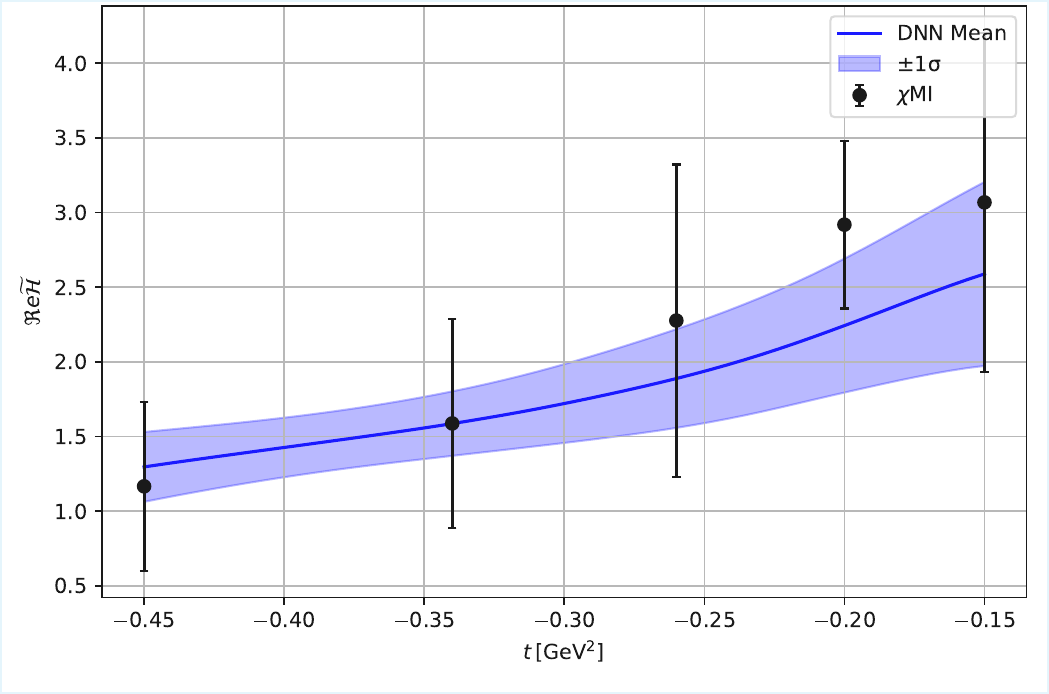}
        \caption{Global DNN predictions for \ReH\ (top), \ReE\ (center), and \ReHt\ (bottom) versus $t$ at fixed $Q^2 = 1.94~\mathrm{GeV}^2$ and $x_B = 0.274$. Blue curves show the replica-averaged predictions, with shaded bands indicating the $1\sigma$ uncertainty. Black points are locally extracted \xmi\ results at the same kinematics.}
        \label{fig:CFFs_vs_t}
    \end{figure}

Figure \ref{fig:CFFs_vs_t} presents the resulting global DNN models for \ReH, \ReE\ and \ReHt\ as functions of $t$, evaluated at fixed values of $Q^2 = 1.94~\mathrm{GeV}^2$ and $x_B = 0.274$.
The locally extracted $\chi\mathrm{MI}$ CFFs for the same kinematics are shown as black points.
The blue solid curves represent the DNN model predictions averaged over all replicas, while the shaded bands indicate the corresponding one–standard-deviation uncertainty.
These global DNN models capture the kinematic dependence of the CFFs, enabling both interpolation and extrapolation across the phase space. This continuity and smoothness in the CFFs help ensure that the GPDs derived from them remain stable and physically meaningful. Furthermore, the DNN framework simplifies the treatment of error propagation, allowing uncertainties from local fits to be directly mapped onto the global CFF predictions.

\begin{figure}[!t] 
        \centering
        \includegraphics[width=0.95\linewidth]{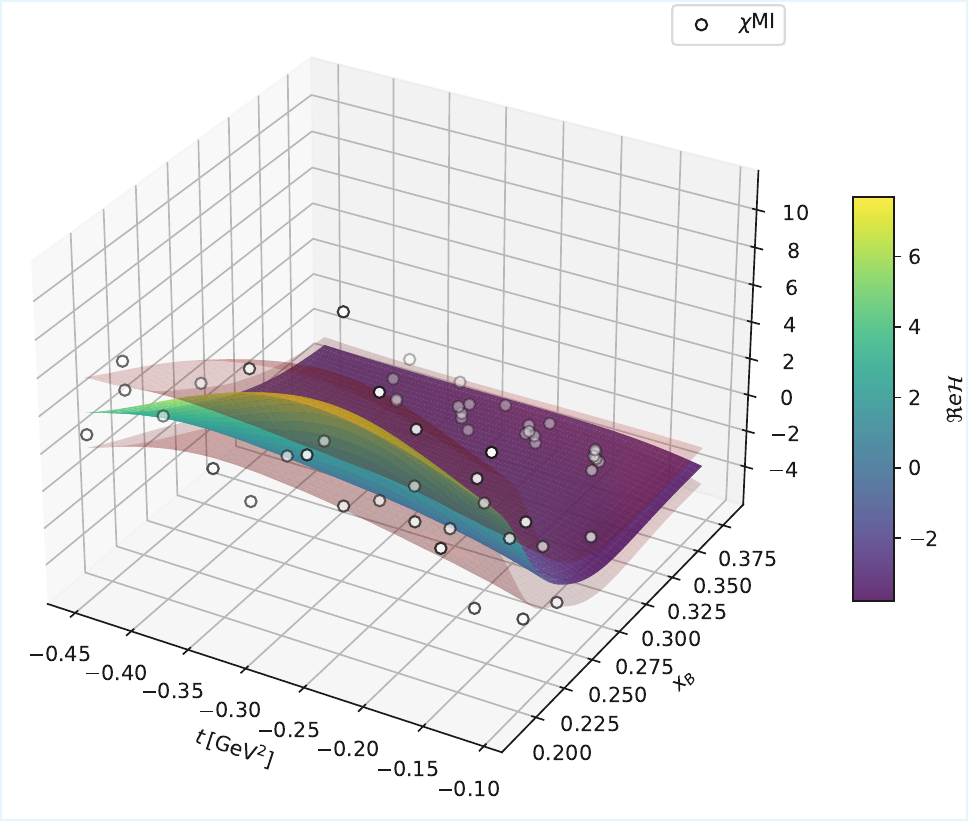}
        \includegraphics[width=0.95\linewidth]{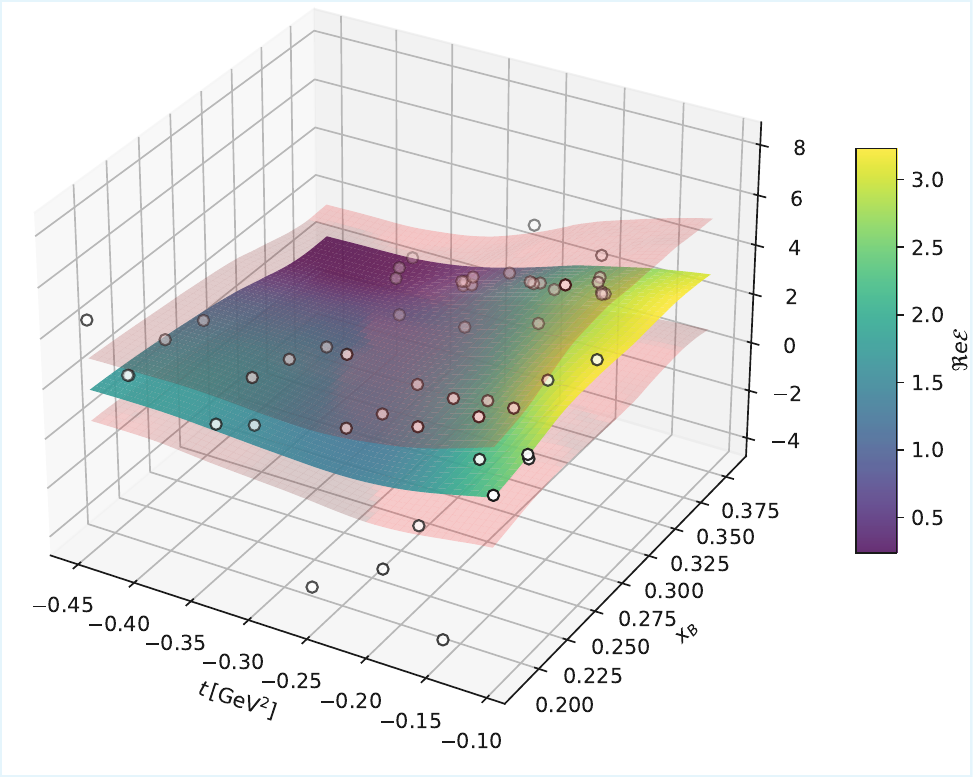}
        \includegraphics[width=0.95\linewidth]{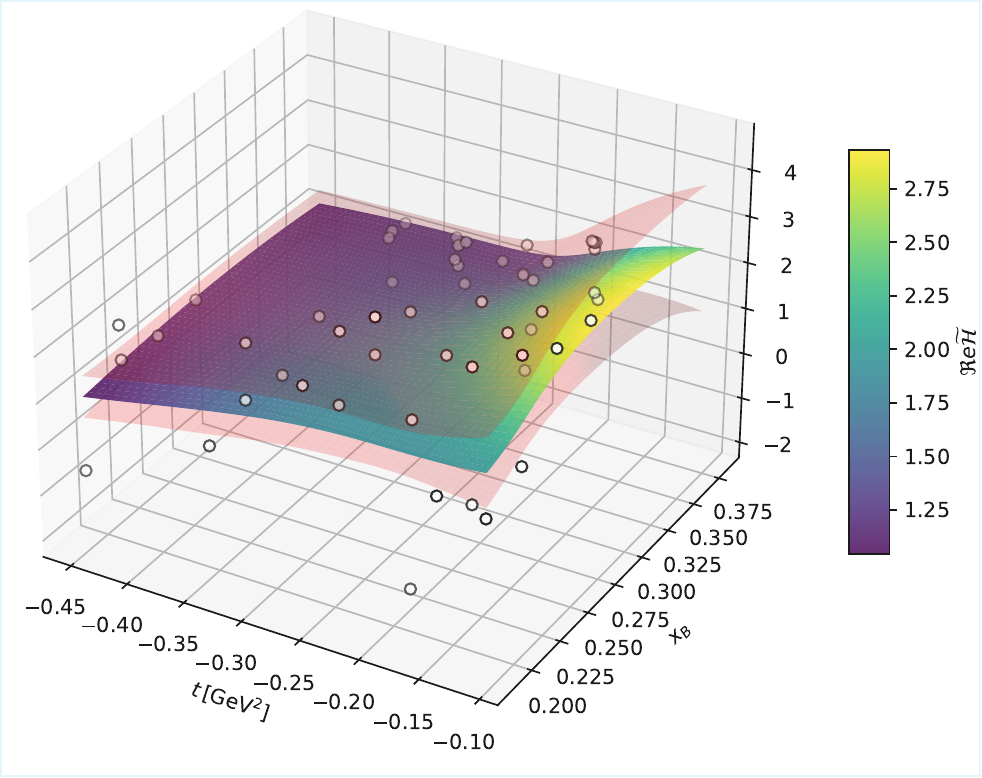}
        \caption{DNN-predicted \ReH\ (top), \ReE\ (center), and \ReHt\ (bottom) surfaces versus $t$ and $x_B$ for $Q^2 = 1.78~\mathrm{GeV}^2$. The central surface corresponds to the mean prediction, while the outer surfaces bound the 68\% confidence level. White circles denote the locally extracted \xmi\ CFFs within $Q^2 \in [1.50, 2.00]~\mathrm{GeV}^2$.}
        \label{fig:CFFs_vs_xB_t}
    \end{figure}

Figure \ref{fig:CFFs_vs_xB_t} shows the predicted DNN model surfaces as functions of $t$ and $x_B$ for $Q^2 = 1.78~\mathrm{GeV}^2$, corresponding to the mean $Q^2$ of the locally extracted $\chi\mathrm{MI}$ CFFs within the range $Q^2 \in [1.50, 2.00]~\mathrm{GeV}^2$. The white circles represent these extracted values. The central surface denotes the predictions mean, while the outer surfaces indicate the 68\% confidence-level bounds.

Deep neural networks are poised to become a standard and highly effective tool for global CFF modeling, offering strong adaptability to new datasets and predictive power across diverse kinematic regions. Unlike traditional fitting methods, the DNN approach does not require assuming specific functional forms for partonic distributions or nonperturbative contributions, thereby reducing a significant class of systematic uncertainties and minimizing potential biases in the predictions.

     \begin{figure}[!t] 
        \centering
        \includegraphics[width=0.95\linewidth]{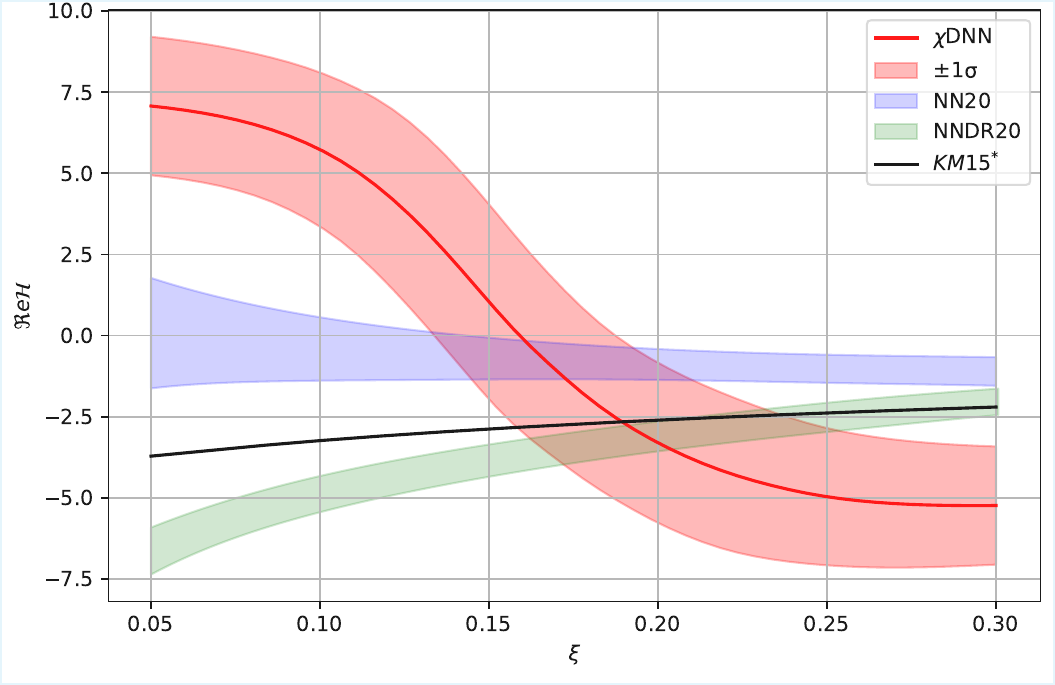}
        \includegraphics[width=0.95\linewidth]{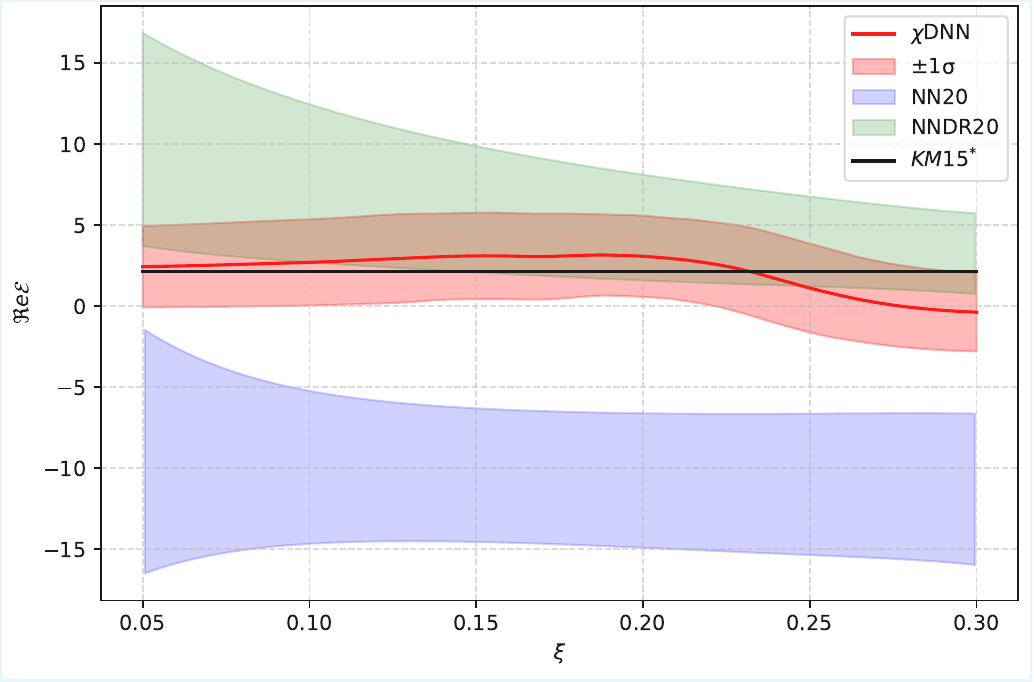}
        
        \caption{Global DNN predictions for \ReH\ (top) and \ReE\ (bottom) versus $\xi$ at fixed $Q^2 = 4~\mathrm{GeV}^2$ and $t = -0.2~\mathrm{GeV}^2$. Comparison with results from \cite{KM20}.}
        \label{fig:CFFs_vs_xi_KM20_1}
    \end{figure}

\subsection{\label{subsec:results}  Comparison with previous DNN analyses}

A central result of this work is the demonstration that a 
global Deep Neural Network fit constrained solely by 
unpolarized cross section data can achieve a level of precision comparable to---and in some cases better than---earlier 
global extractions that combined cross sections with beam-spin 
asymmetry observables. Previous DNN-based global fits, such 
as those presented in Refs.~\cite{KM20,Moutarde:ANN2019}, made use of both unpolarized cross sections and beam asymmetries, 
providing additional kinematic sensitivity to the real and 
imaginary parts of the Compton Form Factors (CFFs). In 
contrast, the present analysis represents the least-constrained 
scenario, relying exclusively on cross sections, yet it achieves competitive precision.

\begin{figure}[!t] 
        \centering
        \includegraphics[width=0.95\linewidth]{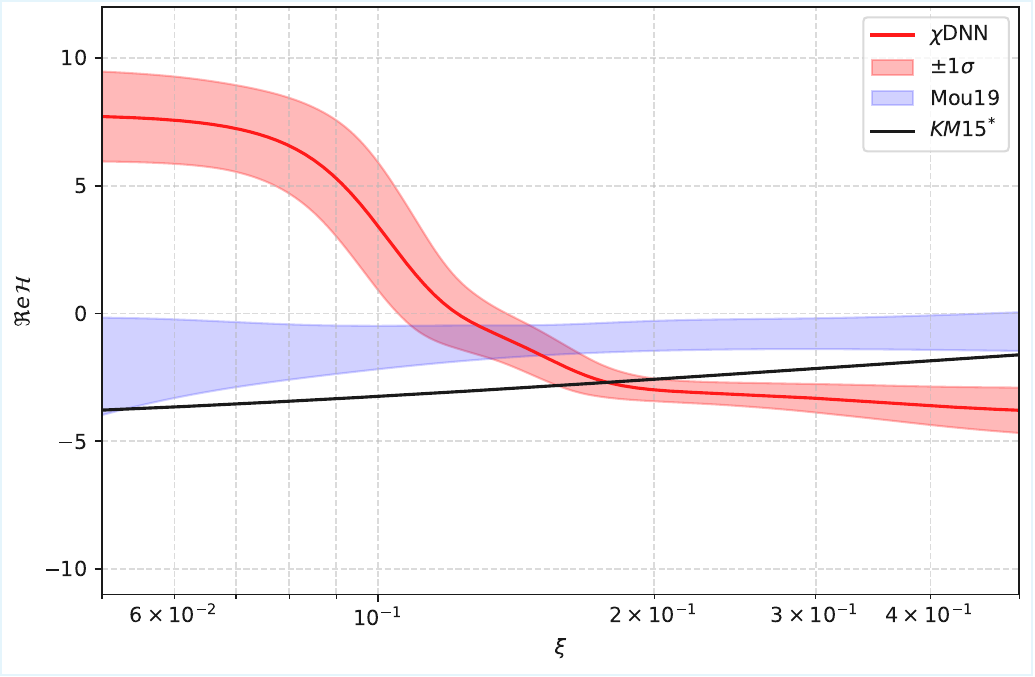}
        \includegraphics[width=0.95\linewidth]{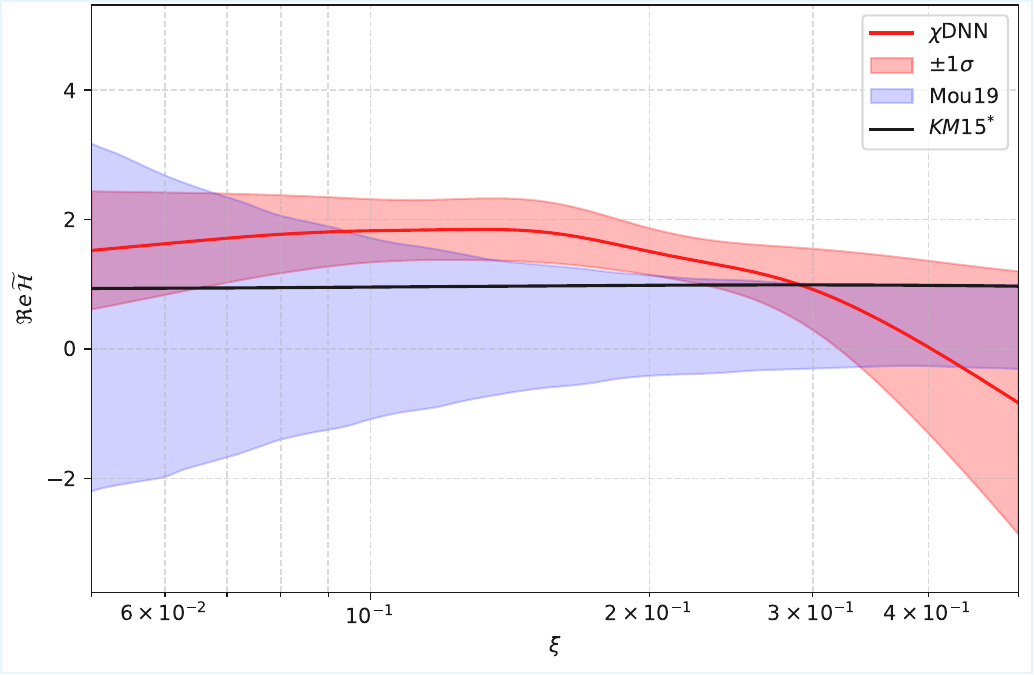}
        
        \caption{Global DNN predictions for \ReH\ (top) and \ReHt\ (bottom) versus $\xi$ at fixed $Q^2 = 2~\mathrm{GeV}^2$ and $t = -0.3~\mathrm{GeV}^2$. Comparison with results from \cite{Moutarde:ANN2019}.}
        \label{fig:CFFs_vs_xi_KM20_2}
    \end{figure}

Figure~\ref{fig:CFFs_vs_xi_KM20_1} illustrates the comparison for \ReH\ and \ReE\ with the previous deep neural network model of Ref.~\cite{KM20}. Their neural network results were trained using Jefferson Lab $6\,\mathrm{GeV}$ measurements~\cite{JoCLAS:2015,Defurne2015,Defurne2017} of the unpolarized cross section, various beam and target asymmetries, as well as the helicity-dependent cross sections. Along with their neural network fit NN20 (blue), the NNDR20 fit (green), which additionally imposes the dispersion relation constraint~\eqref{DR}, is also displayed.  
The comparison with the deep neural network model of Ref.~\cite{Moutarde:ANN2019} is shown for \ReH\ and \ReHt\ in Figure~\ref{fig:CFFs_vs_xi_KM20_2}, where the $\xi$ axis is displayed on a logarithmic scale. Their model (blue) uses differential cross sections, beam-spin asymmetries, and beam-charge asymmetries from H1~\cite{Aktas2005,Aaron2009}, ZEUS~\cite{Chekanov2009}, HERMES~\cite{Airapetian2001,Airapetian2007,Airapetian2008,Airapetian2009,Airapetian2010,Airapetian2011,Airapetian2012}, Jefferson Lab~\cite{Stepanyan2001,CLAS06_Pol,CLAS08_Asym,CLAS09_Asym,CLAS15_Asym,JoCLAS:2015,Defurne2015,Defurne2017}, and COMPASS~\cite{Akhunzyanov2019}. However, it does not include the unpolarized cross-section data from Hall-A used in this analysis.
A striking feature of our results is that the uncertainties are not uniformly larger than those obtained in fits that used both cross sections and asymmetries. Instead, we observe a nontrivial pattern: in certain regions of skewness $\xi$, the uncertainties widen, as expected given the reduced experimental input, while in other regions they are markedly smaller.  

This behavior underscores the effectiveness of the novel 
\xmi~local extraction procedure, which generates robust 
local constraints on $\Re e \mathcal{H}$, $\Re e \mathcal{E}$, and 
$\Re e \widetilde{\mathcal{H}}$ from the least-constraining 
observable. When these local CFFs are used as training input 
for the global DNN, they provide an intrinsic regularization 
mechanism that mitigates the degeneracies among CFFs typically 
encountered in global fits. As a result, the DNN learns to 
reproduce stable global behavior even in regions where 
traditional fits relying on local curvature around a 
$\chi^2$ minimum fail to converge reliably.

In particular, the extraction of $\Re e \mathcal{H}$ is found to 
be especially precise, with uncertainties matching or even 
surpassing previous global analyses over a broad range in $\xi$.  
Although $\Re e \mathcal{E}$ and $\Re e \widetilde{\mathcal{H}}$ are 
more difficult to constrain, as has been recognized in numerous 
studies, our results nevertheless achieve uncertainties of 
comparable magnitude to fits that had the benefit of 
beam-spin asymmetry inputs. This outcome is critical, 
demonstrating that even the least-constrained DVCS observable, 
when analyzed with the \xmi~approach, can yield sufficient 
information to provide nontrivial global constraints on three 
independent real CFFs using modern tools.  We save the extraction using additional observables to futher constrain for future work.

Taken together, these findings emphasize two important 
conclusions. First, the exclusive reliance on cross section 
measurements does not preclude a competitive global determination 
of the real parts of the CFFs. Second, the reduction of 
uncertainties in specific regions of $\xi$ demonstrates that 
our approach offers a complementary pathway to global DNN fits, 
capable of reducing model dependencies while retaining predictive 
power in unexplored kinematics. As additional observables become 
available in future measurements, these types of frameworks came be expected to 
achieve even greater precision, but the present results already 
establish that meaningful global constraints can be extracted 
from cross sections alone.

\section{summary}

In this work, we introduce a novel technique for extracting Compton Form Factors (CFFs) from experimental data by applying localized selections to the topology of two-dimensional $\chi^2$-map inference (\xmi) distributions. This approach reduces model bias and enables the extraction of three CFFs with reasonable accuracy using only a single observable—the least constrained one—namely, the unpolarized photon leptoproduction cross section. This is a remarkable result, since previous local extraction studies, such as Ref.~\cite{Dupre}, were only able to constrain two CFFs by simultaneously fitting two observables. Of particular importance, our method successfully extracts \ReE\, a CFF that has been notoriously difficult to access. Given its connection to the generalized parton distribution (GPD) $E$, this result has unique significance, as $E$ is directly tied to parton orbital angular momentum, a key component in the decomposition of the proton spin.

A comprehensive validation with pseudodata that mimicked real experimental conditions demonstrated the robustness of the method. We further compared our extracted CFF values with those reported in the literature that rely on multiple observables and found consistent agreement within uncertainties, despite our use of only the unpolarized cross section. This highlights the strong potential of our technique, particularly when additional observables are included. We also anticipate significant improvements when integrating advanced AI tools in the local extraction process, which is already underway at the University of Virginia (UVA). At the same time, the results of the present study underscore several limitations of traditional least-squares local fitting approaches.

Although our current application employs the BKM formalism, the method is not restricted to this framework and can be extended to other formalisms. Likewise, while the present study focuses on the unpolarized cross section, the approach can be naturally generalized to multiple observables in a simultaneous DNN-based extraction, which we plan to pursue in future work.

Beyond the local extraction, we constructed a global DNN fit trained solely on cross section data. Remarkably, the resulting global CFFs show uncertainties that are competitive with, and in some regions smaller than, those from previous DNN global fits that relied on both cross sections and beam asymmetries. This nontrivial behavior underscores the power of the \xmi-based local extraction as input to the DNN, which acts as an intrinsic regularization and mitigates parameter degeneracies. In particular, \ReH\ is determined with precision comparable to or better than earlier global analyses across much of the kinematic range, while \ReE\ and \ReHt\ are extracted with uncertainties of similar magnitude to fits that benefited from additional asymmetry constraints. This demonstrates that meaningful global information can be extracted from cross sections alone, a result with significant implications for future experiments where limited observables may be available.

Overall, this study presents a promising new strategy for CFF extraction from experimental data, leveraging \xmi-map contour selections to overcome challenges of model dependence, sparse data, and limited observables, while enabling robust global fits trained exclusively on cross section input. This approach opens the door to more comprehensive and unbiased extractions in the future and provides a foundation for incorporating additional observables as they become available at JLab and forthcoming facilities such as the EIC.

\section{Acknowledgments}
The Deep Neural Network models employed in this work were trained on the University of Virginia’s high-performance computing cluster, Rivanna. We acknowledge Research Computing at UVA for providing essential computational resources and technical support.
This work was supported by the U.S. Department of Energy (DOE) Contract No. DE-FG02-96ER40950 and by the DOE Office of Medium Energy Nuclear Physics.

\clearpage

\nocite{*}
\bibliographystyle{apsrev4-2}
%

\end{document}